\documentclass[twocolumn,aps,prx]{revtex4-1}

	\usepackage[usenames,dvipsnames]{xcolor}
	\usepackage{amsmath}
	\usepackage{amsfonts}
	\usepackage{amssymb}
	\usepackage[colorlinks=true,citecolor=blue,linkcolor=red]{hyperref}
	\usepackage{graphicx}
	\usepackage{bbold}					
	\usepackage[makeroom]{cancel}		
	\usepackage{multirow}				
	\usepackage{array}

	\newcolumntype{x}[1]{>{\centering\let\newline\\\arraybackslash\hspace{0pt}}p{#1}}

	\DeclareMathOperator{\sign}{sign}  	
	\DeclareMathOperator{\Pf}{Pf}  		
	\DeclareMathOperator{\tr}{tr}  		
	\DeclareMathOperator*{\ordprod}{\overline{\prod}}
	
	\DeclareMathAlphabet{\mathbbold}{U}{bbold}{m}{n}
	\def\abs#1{\left|{#1}\right|}      	
	\def\bs#1{\boldsymbol{#1}}			
	\def\imi{\mathrm{i}}				
	\def\e#1{\mathrm{e}^{#1}}				
	\def\de{\mathrm{d}}

	\def\mcH{\mathcal{H}}					
	\def\mcbsA{\boldsymbol{\mathcal{A}}}	
	\def\mcA{\mathcal{A}}					
	\def\mcbsF{\boldsymbol{\mathcal{F}}}	
	\def\mcF{\mathcal{F}}					
	\def\mcQ{\mathcal{Q}}					
	\def\mcT{\mathcal{T}}					
	\def\mcP{\mathcal{P}}					
	\def\mcC{\mathcal{C}}					
	\def\mcI{\mathcal{I}}					
	\def\mcK{\mathcal{K}}					
	\def\mcS{\mathcal{S}}
	\def\mcU{\mathcal{U}}
	\def\mcW{\mathcal{W}}					
	
	\def\mfT{\mathfrak{T}}					
	\def\mfP{\mathfrak{P}}					

	\def\intg{\mathbbold{Z}}					
	\def\ztwo{\mathbbold{Z}_2}					
	\def\triv{\mathbbold{0}}					
	\def\unit{\mathbbold{1}}					
	
	\def\bra#1{\left<{#1}\right|}				
	\def\ket#1{\left|{#1}\right>}				
	\def\braket#1#2{\left<{#1}|{#2}\right>}				
	\def\braketBig#1#2{\left<{#1}\Big|{#2}\right>}		
	
	\def\pdag{{\phantom{\dagger}}}
	
	\newcounter{subeqn} %
	\makeatletter
	\@addtoreset{subeqn}{equation}
	\makeatother

\begin{document}


\title{Robust doubly charged nodal lines and nodal surfaces in centrosymmetric systems}

\author{Tom\'{a}\v{s} Bzdu\v{s}ek$^{1,2}$}
\author{Manfred Sigrist$^{1}$}
\affiliation{$^{1}$Institut f\"{u}r Theoretische Physik, ETH Z\"{u}rich, 8093 Z\"{u}rich, Switzerland}
\affiliation{$^{2}$Department of Physics, Stanford University, Stanford, California 94305-4045, USA}
\date{\today}

\begin{abstract}
Weyl points in three spatial dimensions are characterized by a $\intg$-valued charge -- the Chern number -- which makes them stable against a wide range of perturbations. A set of Weyl points can mutually annihilate only if their net charge vanishes, a property we refer to as robustness. While nodal loops are usually not robust in this sense, it has recently been shown using homotopy arguments that in the centrosymmetric extension of the $\textrm{AI}$ symmetry class they nevertheless develop a $\mathbb{Z}_2$ charge analogous to the Chern number. Nodal loops carrying a non-trivial value of this $\mathbb{Z}_2$ charge are robust, i.e. they can be gapped out only by a pairwise annihilation and not on their own. As this is an additional charge independent of the Berry $\pi$-phase flowing along the band degeneracy, such nodal loops are, in fact, doubly charged. In this manuscript, we generalize the homotopy discussion to the centrosymmetric extensions of all Atland-Zirnbauer classes. We {\color{black} develop a taylored mathematical framework dubbed the AZ+$\mcI$ classification and} show that in three spatial dimensions such robust and multiply charged nodes appear in four of {\color{black} such centrosymmetric} extensions, namely {\color{black} AZ+$\mcI$} classes $\textrm{CI}$ and $\textrm{AI}$ lead to doubly charged nodal lines, while $\textrm{D}$ and $\textrm{BDI}$ support doubly charged nodal surfaces. We remark that no further crystalline symmetries apart from the spatial inversion are necessary for their {\color{black} stability}. We provide a description of the corresponding topological charges, and develop simple tight-binding models of various semimetallic and superconducting phases that exhibit these nodes. We also indicate how the concept of robust and multiply charged nodes generalizes to other spatial dimensions.
\end{abstract}

\maketitle




\section{Introduction}

The advent of topological aspects of condensed matter systems~\cite{Volovik:2003}, which has been initiated by the discovery of the Quantum Hall effects~\cite{Klitzing:1980,Thouless:1982,Laughlin:1981,Tsui:1982,Laughlin:1983,Moore:1991} and strongly amplified by the discovery of topological insulators~\cite{Kane:2005,Bernevig:2006,Bernevig:2006b,Konig:2007,Fu:2007,Fu:2007b,Hsieh:2008,Hasan:2010,Hughes:2011,Qi:2011,Shiozaki:2014,Bradlyn:2017}, has in recent years naturaly led to extensive research of \emph{gapless} topological phases. While electrons in such systems may emulate various high-energy physics particles like the massless Weyl~\cite{Murakami:2007,Wan:2011,Weng:2015,Huang:2015a,Lv:2015,Xu:2015,Xu:2015a} and Dirac~\cite{Novoselov:2005,Katsnelson:2006,Wang:2012,Young:2012,Neupane:2014,Liu:2014a,Liu:2014b,Wehling:2014,Steinberg:2014,Young:2015} fermions, it has been recognized that the lower degree of symmetry of condensed matter systems also enables the appearance of gapless spectra that have \emph{no} analog in high-energy physics. Examples include multi-Weyl points~\cite{Fang:2012,Tsirkin:2017}, ``tipped'' (or type-II) Weyl points~\cite{xu2015structured,Soluyanov:2015}, ``hyper-Dirac'' (or double Dirac) points~\cite{Wieder:2015}, effective spin-1 fermions~\cite{Bradlyn:2016}, nexus fermions \cite{Heikkila:2015b,Zhu:2016,Weng:2016,Weng:2016b,Chang:2016}, various Weyl and Dirac loops~\cite{Burkov:2011,Chen:2015,Yu:2015,Weng:2015b,Fang:2015,Youngkuk:2015,Bian:2015a,Bian:2015b,chen2015nanostructured,Liang:2016,Heikkila:2015,Hirayama:2017,Geilhufe:2017,Furusaki:2017} and even Weyl chains and Weyl nets~\cite{Bzdusek:2016,Rui:2017} and Dirac chains~\cite{wang2017hourglass}. All of these are associated with a topological protection, novel surface spectra, and some were found to exhibit unusual transport features, notably the chiral anomaly~\cite{Nielsen:1983,Son:2013,Hosur:2013,Huang:2015b}. In the present work we refer to a general touching of conduction and valence bands as a \emph{node}, and depending on its dimension we further specify it as a \emph{nodal point}, \emph{line} or \emph{surface}. To avoid confusion, by a \emph{nodal loop} we mean a nodal line that does not wind around the Brillouin zone (BZ).

The topological protection of many of the mentioned nodes requires the presence of certain \emph{crystalline} symmetries. For example, Dirac points require a rotational symmetry~\cite{Yang:2014,Yang:2015,Gao:2015}, most nodal line materials need a mirror or a glide symmetry~\cite{Chiu:2014,Bzdusek:2016}, and the hyper-Dirac points~\cite{Wieder:2015,Bradlyn:2016} only arise in certain non-symmorphic lattices. Once the relevant crystalline symmetry is removed (e.g. by applying strain), these nodes cease to be topologically protected and may be trivially gapped out. On the other hand, nodes protected solely by \emph{global} symmetries are more stable as they are \emph{not} susceptible to the removal of crystalline symmetries. By global symmetries we mean those of the Atland-Zirnbauer (AZ) classification~\cite{Atland:1997}, i.e. time-reversal symmetry ($\mcT$), particle-hole symmetry ($\mcP$) and chiral symmetry ($\mcC$). Examples of such highly stable nodes include nodal lines protected by $\mcC$ (in superconductors~\cite{Hayashi:2006} and in semimetals with sublattice symmetry~\cite{Burkov:2011}), and Weyl points (in semimetals~\cite{Wan:2011}, superfluids~\cite{Volovik:2003} and superconductors~\cite{Goswami:2013,Fischer:2014}) which do not require any symmetry {\color{black} (besides the translational symmetry of a crystalline system or a liquid)}. 

Nodes protected by global symmetries alone are still characterized by a differing degree of stability. For example, a Weyl point is characterized by a $\intg$-valued charge -- the Chern number -- and a set of Weyl points can mutually annihilate \emph{only} if their net charge vanishes. In this manuscript, we refer to this property as the \emph{robustness} of the nodes. An analogous $\intg$ charge can also be defined for nodal lines protected by $\mcC$ which wind around BZ. However, such a charge is \emph{absent} for nodal \emph{loops} of the same system -- these can be shrunk to a point and then gapped out on their own, hence we don't describe them as robust. A question therefore arises as whether there are other species of nodes that reach this high level of stability we call \emph{robustness}. The answer is unfavorable -- for systems with three spatial dimensions ($D=3$) and global symmetries only, the two examples above capture \emph{all} robust nodes. 

Interestingly, it was noticed in Ref.~\cite{Fang:2015} that the set of robust nodes becomes enlarged if one considers \emph{centrosymmetric} systems. Since inversion symmetry ($\mcI$) is typically not removed by straining (unless we encounter a structural phase transition~\cite{Bzdusek:2015}), nodes protected jointly by global symmetries and $\mcI$ are still very stable. In particular, Ref.~\cite{Fang:2015} used homotopy arguments to show that $D=3$ centrosymmetric {\color{black} electron bands} respecting $\mcT$ and $\textsf{SU}(2)$ spin-rotation symmetry (corresponding to AZ class $\textrm{AI}$) exhibit nodal loops with a $\ztwo$ charge analogous to the Chern number of Weyl points.~\footnote{Similar considerations using the notions of $K$-theory appeared recently also in Refs.~\cite{Zhao:2016,Zhao:2017}. {\color{black} The non-trivial second homotopy group has also been reported in Sec.~III.D of Ref.~\cite{morimoto2014weyl} although this work didn't study the consequences of this property. The non-trivial charge on a gapped 2D manifold has been also identified in Tab.~C.2 of Ref.~\cite{Ryu:2010}.}} A nodal loop with a non-trivial value of this charge cannot be gapped out on its own but only by an annihilation with another such a loop. Furthermore, since this charge is independent of the Berry $\pi$-phase flowing along the band degeneracy, such nodal loops are also \emph{doubly charged}. All these conclusions remain true in systems breaking both $\mcT$ and $\mcI$ provided that the composed $\mcT\mcI$ symmetry is preserved~\cite{Li:2017}. We refer to both described situations as belonging to the \emph{centrosymmetric extension} of the $\textrm{AI}$ class. Note that ``robust'' and ``multiply charged'' describe \emph{different} node attributes. We will argue that a single topological invariant protected by global symmetries and $\mcI$ is always sufficient to facilitate robust nodes, although they may have to wind around BZ for this to be the case. Being multiply charged is therefore a stronger and more peculiar feature.

In this manuscript, we generalize the homotopy arguments of Ref.~\cite{Fang:2015} to the centrosymmetric extension of \emph{all} AZ classes (termed ``AZ+$\mcI$ classes'' for brevity) of arbitrary spatial dimension $D$, and provide an exhaustive list of multiply charged nodes appearing in them. Specifically in $D=3$ we identify \emph{four} non-trivial instances: AZ+$\mcI$ classes $\textrm{CI}$ and $\textrm{AI}$ support doubly charged nodal loops, while we find doubly charged nodal \emph{surfaces} in classes $\textrm{D}$ and $\textrm{BDI}$. In $D=2$ the only class supporting doubly charged nodes is $\textrm{BDI}$, and there are no multiply charged nodes in $D=1$. On the other hand, all AZ+$\mcI$ classes support multiply charged nodes for $D\geq 6$. 

Focusing further on $D=3$, we work out a description of both topological charges and develop simple tight-binding (TB) models exhibiting doubly charged nodes for the four non-trivial instances. The presented TB models do not attempt to describe any specific material, but rather to supply the reader with a way of easily checking our claims. We nevertheless argue that all of these nodes can appear in realistic crystalline solids. Specifically, the doubly charged nodal loops of class $\textrm{CI}$ have natural realization in singlet superconducting (SC) phase of nodal line metals, nodal surfaces of class $\textrm{D}$ have very recently been discussed in the context of multi-orbital SCs~\cite{Agterberg:2017,Timm:2017b}, the nodal loops of class $\textrm{AI}$ apply to semimetals respecting $\mcT\mcI$ in the absence of spin-orbit coupling, and nodal surfaces of class $\textrm{BDI}$ are relevant to $\textrm{AI}$-like systems supplemented by the sublattice symmetry. {\color{black} Furthermore, the results for class $\textrm{AI}$ are naturally applicable to nodes formed by photonic~\cite{lu2013weyl}, phononic~\cite{po2016phonon} and magnonic~\cite{Li:2017} bands, since for bosonic excitations $\mcT$ naturally squares to $+1$.} We remark that while providing the complete topological classification of multiply charged nodes of the AZ+$\mcI$ classes, this manuscript does not investigate the possible associated surface states and transport signatures which we leave for a future work.

The manuscript is structured as follows: In Sec. \ref{sec:AZ+I-classes} we formalize the notion of AZ+$\mcI$ classes. In Sec.~\ref{sec:node-dim} we fix a canonical representation of the relevant symmetries in every AZ+$\mcI$ class, and use it to determine the dimensionality of stable nodes supported by the class in any spatial dimensions $D$. In Sec.~\ref{sec:charges} we use homotopy theory to determine the topological charges of these nodes. We further elaborate here on the distinction between \emph{robust} and \emph{multiply charged} nodes. Then in Sec.~\ref{sec:sym-class-rel} we discuss how the individual AZ+$\mcI$ classes relate to actual physical systems. Although such a discussion readily exists for AZ classes~\cite{Schnyder:2008}, the Cartan labels of the AZ and of the AZ+$\mcI$ class of a given centrosymmetric system can be \emph{different}, simply because $\mcI$ and $\mcP$ may be non-commuting. As this is a potentially confusing point, we dedicate significant space to carefully enumerate all the possibilities.

The sections introduced so far already contain our main findings, and the remaining pages mostly serve to strenghten our claims with examples and with derivations of the topological charges. We sequentially go through all four AZ+$\mcI$ classes supporting doubly charged nodes in $D=3$, and we develop a simple TB model on a hexagonal $\textrm{SrPtAs}$-like lattice~\cite{Fischer:2015} for each of them. For each model, we explicitly calculate both topological charges of the exhibited nodes. We gradually discuss AZ+$\mcI$ class $\textrm{D}$ exhibiting $\ztwo\oplus 2\intg$ nodal surfaces in Sec.~\ref{sec:nodes-D}, class $\textrm{BDI}$ with $\ztwo\oplus\ztwo$ nodal surfaces in Sec.~\ref{sec:nodes-BDI}, class $\textrm{CI}$ with $\intg\oplus\ztwo$ nodal lines in Sec.~\ref{sec:nodes-CI}, and finally class $\textrm{AI}$ with $\ztwo\oplus\ztwo$ nodal lines in Sec.~\ref{sec:nodes-AI}. While most of the formulated charges appear within the ten-fold way classification~\cite{Ryu:2010}, in three cases we encounter a somewhat unusual $\mathbb{Z}_2$ charge corresponding to a closed path winding inside Lie group $\mathsf{SO}(n)$~\cite{Fang:2015}, which we formalize using  Wilson loops~\cite{Soluyanov:2011,Alexandradinata:2016}. We conclude with some final remarks in Sec.~\ref{sec:conclusion}.

\begin{table*}[t]
	\begin{tabular}{|c|ccc||c|x{0.5cm}x{0.5cm}x{0.5cm}|x{0.5cm}x{0.5cm}x{0.5cm}x{0.5cm}x{0.5cm}x{0.5cm}|x{1.5cm}x{1.5cm}x{1.5cm}|}
	\hline
	\multirow{2}{*}{CL}					&	$\mfT$	
										&	$\mfP$	
										&	$\mcC$ 
										&	\multirow{2}{*}{classifying space $M_\textrm{CL}$}  
										&  	\multicolumn{9}{c|}{ homotopy group $\pi_p(M_\textrm{CL})$ for $p=\ldots$ }	
										&	\multicolumn{3}{c|}{ topological charge $c_{\textrm{CL}}(S^p)$} 			\\ 
										&	${\color{gray}(\mcT\phantom{)}}$	
										&	${\color{gray}\mcP}$	
										&	${\color{gray}\phantom{(}\mcC)}$	&				
										&	$0$		&	$1$		&	$2$		&	$3$		
										&	$4$		&	$5$		&	$6$		&	$7$		& $8$
										&	$p=0$	&	$p=1$	&	$p=2$
	\\ \hline 
	$\mathrm{A}$	&	$\times$	& 	$\times$	&	$\times$	&	
	$\textsf{U}(n+\ell)/\textsf{U}(n)\!\times\!\textsf{U}(\ell)$						
										&	${\color{gray} \triv}$	&	${\color{gray} \triv}$	&	$\pmb{\intg}$	&	$\triv$	
										&	$\intg$	&	$\triv$		&	$\intg$	&	$\triv$	&	$\intg$
										&	${\color{gray}-}$		&	${\color{gray}-}$			&	$\int\!\mcF$
	\\
	$\mathrm{AIII}$	&	$\times$	& 	$\times$	&	$1$			&	
	$\textsf{U}(n)$																
										&	${\color{gray} \triv}$	&	$\pmb{\intg}$	&	$\pmb{\triv}$	&	$\intg$	
										&	$\triv$	&	$\intg$	&	$\triv$	&	$\intg$	&	$\triv$
										& 	${\color{gray}-}$		&	$\int \! q^{-1}d q$	&	${\color{gray}-}$
	\\ \hline
	$\mathrm{AI}$	&	$+1$		& 	$\times$	&	$\times$	&	
	$\textsf{O}(n+\ell)/\textsf{O}(n)\!\times\!\textsf{O}(\ell)$						
										&	${\color{gray}\triv}$	&	$\pmb{\ztwo}$	&	$\pmb{\ztwo}$	&	$\triv$	
										&	$2 \intg$	&	$\triv$	&	$\triv$	&	$\triv$	&	$\intg$	
										& 	${\color{gray}-}$		& $\int\!\mcA$	& $\pi_1[\mathsf{SO}(n)]$
	\\
	$\mathrm{BDI}$	&	$+1$		& 	$+1$		&	$1$			&	
	$\textsf{O}(n)$														
										&	$\pmb{\ztwo}$	&	$\pmb{\ztwo}$	&	$\pmb{\triv}$	&	$2\intg$	
										&	$\triv$	&	$\triv$	&	$\triv$	&	$\intg$	&	$\ztwo$	
										&	$\sign\Pf$	&	$\pi_1[\mathsf{SO}(n)]$	& ${\color{gray}-}$
	\\
	$\mathrm{D}$	&	$\times$ 	& 	$+1$		&	$\times$	&	
	$\textsf{O}(2n)/\textsf{U}(n)$											
										&	$\pmb{\ztwo}$	&	$\pmb{\triv}$	&	$\pmb{2\intg}$	&	$\triv$	
										&	$\triv$	&	$\triv$	&	$\intg$	&	$\ztwo$	&	$\ztwo$	
										& 	$\sign\Pf$	&	${\color{gray}-}$		&	$\int\!\mcF$
	\\
	$\mathrm{DIII}$	&	$-1$	 	& 	$+1$		&	$1$			&	
	$\textsf{U}(2n)/\textsf{Sp}(n)$											
										&	${\color{gray}\triv}$	&	$\pmb{2\intg}$	&	$\pmb{\triv}$	&	$\triv$	
										&	$\triv$	&	$\intg$	&	$\ztwo$	&	$\ztwo$	&	$\triv$	
										&	${\color{gray}-}$		& $\int \! q^{-1}d q$	& ${\color{gray}-}$
	\\
	$\mathrm{AII}$	&	$-1$	 	& 	$\times$	&	$\times$	&	
	$\textsf{Sp}(n+\ell)/\textsf{Sp}(n)\!\times\!\textsf{Sp}(\ell)$						
										&	${\color{gray}\triv}$	&	${\color{gray}\triv}$	&	${\color{gray}\triv}$	&	${\color{gray}\triv}$	
										&	$\intg$	&	$\ztwo$	&	$\ztwo$	&	$\triv$	&	$2\intg$
										&	${\color{gray}-}$		& ${\color{gray}-}$			& ${\color{gray}-}$
	\\
	$\mathrm{CII}$	&	$-1$	 	& 	$-1$		&	$1$			&	
	$\textsf{Sp}(n)$												
										&	${\color{gray}\triv}$	&	${\color{gray}\triv}$	&	${\color{gray}\triv}$	&	$\intg$	
										&	$\ztwo$	&	$\ztwo$	&	$\triv$	&	$2\intg$	&	$\triv$	
										&	${\color{gray}-}$		& ${\color{gray}-}$			& ${\color{gray}-}$
	\\
	$\mathrm{C}$	&	$\times$	& 	$-1$		&	$\times$	&	
	$\textsf{Sp}(n)/\textsf{U}(n)$											
										&	${\color{gray}\triv}$	&	${\color{gray}\triv}$	&	$\pmb{\intg}$	&	$\ztwo$	
										&	$\ztwo$	&	$\triv$	&	$2\intg$	&	$\triv$	&	$\triv$	
										&	${\color{gray}-}$		& 	${\color{gray}-}$			&	$\int\!\mcF$
	\\
	$\mathrm{CI}$	&	$+1$	 	& 	$-1$		&	$1$			&	
	$\textsf{U}(n)/\textsf{O}(n)$							
										&	${\color{gray}\triv}$	&	$\pmb{\intg}$	&	$\pmb{\ztwo}$	&	$\ztwo$	
										&	$\triv$	&	$2\intg$	&	$\triv$	&	$\triv$	&	$\triv$	
										&	${\color{gray}-}$		& 	$\int \! q^{-1}d q$	&	$\pi_1[\mathsf{SO}(n)]$
	\\	\hline
	\end{tabular}
	\caption{Summary of the main findings of the manuscript. The first four columns list the ten possible symmetry classes together with their Cartan label (CL). The headers $\mcT,\mcP,\mcC$ (in gray) correspond to the AZ classification~\cite{Atland:1997}, while the headers $\mfT,\mfP,\mcC$ correspond to the AZ+$\mcI$ classification which we introduce in Sec.~\ref{sec:AZ+I-classes} and consider throughout the manuscript. The rest of the table applies to the AZ+$\mcI$ case only. The fifth column lists the classifying spaces $M_\textrm{CL}$~\cite{Kitaev:2009} relevant for each symmetry class as explained in Sec.~\ref{sec:node-dim}. Here, $n,\ell>0$ indicate the number of occupied and unoccupied bands ($n=\ell$ whenever $\mfP$ or $\mcC$ is present). The next block of the table lists the large $n,\ell$ limit homotopy groups of $M_\textrm{CL}$ which exhibit the Bott periodicity $\pi_{d+8}(M_\textrm{CL}) = \pi_d(M_\textrm{CL})$~\cite{Lundell:1992} with the exception of $\pi_0(M_\textrm{CL})$ which counts the number of connected components of $M_\textrm{CL}$. Note that in some cases we write $2\mathbb{Z}$ instead of $\mathbb{Z}$ as the naturally formulated topological invariant takes even values, and we type $\triv$ for a trivial (one-element) group. As discussed in Sec.~\ref{sec:charges}, the homotopy groups $\pi_p(M_\textrm{CL})$ that determine the charges of a node are those with $\delta_\textrm{CL} -1 \leq p \leq D-1$ where $\delta_\textrm{CL}$ is the node codimension determined in Sec.~\ref{sec:node-dim} and listed for all AZ+$\mcI$ classes in Tab.~\ref{tab:Hamiltonians}. The homotopy groups relevant in $D=3$ are typesetted in bold. We observe that AZ+$\mcI$ classes $\textrm{AI}$, $\textrm{BDI}$, $\textrm{D}$ and $\textrm{CI}$ support doubly charged nodes in this spatial dimension. The last three columns indicate the corresponding topological charges, namely $\int\!\mcF$ is the (first) Chern number, $\int \! q^{-1}d q$ is the winding of the off-diagonal block of $\mcQ(\bs{k})$, $\int\!\mcA$ is the Berry phase, $\sign\Pf$ is the sign of the Pfaffian of the Hamiltonian, and $\pi_1[\mathsf{SO}(n)]$ is the homotopy class of a closed path inside the orthogonal group. For the four classes with doubly charged nodes in $D=3$ we list in Tabs.~\ref{tab:Homotopies-small} and~\ref{tab:Homotopies-small-AI} the relevant homotopy groups for systems not reaching the large $n,\ell$ limit, which come to play in the simple models presented in Secs.~\ref{sec:nodes-D} to~\ref{sec:nodes-AI}.}
	\label{tab:Homotopies}
\end{table*}


\section{Symmetry classification}\label{sec:AZ+I-classes}

A classification of Hamiltonians according to their global symmetries was developed in Ref.~\cite{Atland:1997}. By these we mean time-reversal ($\mcT$), particle-hole ($\mcP$) and chiral ($\mcC$) symmetry . In momentum ($\bs{k}$-)space they fulfil
\begin{subequations}\label{eqn:sub-global}
\begin{eqnarray}
\mcT\mcH(\bs{k})\mcT^{-1}  = \mcH(-\bs{k})\,\,\,\,\,&\qquad& \mcT^2=\pm\unit\qquad \textrm{(AU)} \label{eqn:ordinary-TRS}\\
\mcP\mcH(\bs{k})\mcP^{-1}  = -\mcH(-\bs{k})&\qquad& \mcP^2=\pm\unit\qquad \textrm{(AU)} \label{eqn:ordinary-PHS}\\
\mcC\mcH(\bs{k})\mcC^{-1}  = -\mcH(\bs{k})\,\,\;\,&\qquad& \,\mcC^2=\unit \qquad \,\,\,\,\,\textrm{(U)}\label{eqn:CHS}
\end{eqnarray}
\end{subequations}
where (AU) indicates antiunitarity and (U) unitarity. There are ten possibilities which are listed alongside their Cartan label (CL) in the first four columns of Tab.~\ref{tab:Homotopies}.

Apart from the discrete set of $2^D$ time-reversal invariant momenta (TRIMs), $\mcT$ and $\mcP$ relate Hamiltonians at two \emph{different} $\bs{k}$-points. These non-local constraints have to be incorporated when developing the ten-fold way classification of gapped topological insulators and superconductors~\cite{Kitaev:2009,Schnyder:2009,Ryu:2010,wen2012symmetry,Ludwig:2016}. However, when characterizing \emph{nodes} of a gapless system, one only needs to know the Hamiltonian in the vicinity of the node, which is typically not constrained by conditions (\ref{eqn:ordinary-TRS}) and (\ref{eqn:ordinary-PHS}). Consequently, a $D=3$ system exhibits nodal points (lines) in the absence (presence) of $\mcC$, as is shown using homotopy theory in the subsequent sections. Additional nodal points imposed by conditions (\ref{eqn:ordinary-TRS}) and (\ref{eqn:ordinary-PHS}) may be fixed at TRIMs, but these are unremovable and thus not of relevance to us.

The conclusion that $\mcT$ and $\mcP$ don't affect the classification of nodes is changed in the presence of crystalline symmetries~\cite{Chiu:2014}. In the present work, we consider the presence of inversion symmetry ($\mcI$) which is typically robust against simple straining. It fulfils
\begin{equation}
\mcI\mcH(\bs{k})\mcI^{-1} = \mcH(-\bs{k})\qquad \mcI^2 = \unit \,\,\,\,\,\textrm{(U)}
\end{equation}
such that compositions 
\begin{subequations}\label{subeqn:AZ+I-symmetries}
\begin{equation}
\mcT\mcI=\mathfrak{T}\qquad\textrm{and}\qquad\mcP\mcI=\mathfrak{P}\label{eqn:def:newTP}
\end{equation}
impose \emph{local} antiunitary constraints in $\bs{k}$-space~\cite{Fang:2015,Zhao:2016} and become relevant for the characterization of nodes. Apart from the (non-)locality in $\bs{k}$-space, the set of operators $\mfT, \mfP, \mcC$ is mathematically equivalent to the set $\mcT, \mcP, \mcC$. It is therefore possible to define ten symmetry classes based on the presence of operators fulfilling
\begin{eqnarray}
\mfT\mcH(\bs{k})\mfT^{-1}  &=& \mcH(\bs{k})\qquad \,\,\,\;\,\mfT^2=\pm\unit \qquad\textrm{(AU)}\\
\mfP\mcH(\bs{k})\mfP^{-1}  &=& -\mcH(\bs{k})\qquad \mfP^2=\pm\unit\qquad\textrm{(AU)} \\
\mcC\mcH(\bs{k})\mcC^{-1}  &=& -\mcH(\bs{k})\qquad \;\mcC^2=\unit \qquad\;\;\,\textrm{(U)}.\label{eqn:AZIsyms-CHS}
\end{eqnarray}
\end{subequations}
We call {\color{black} such symmetry classes as} the \emph{centrosymmetric extensions of AZ classes} (or ``AZ+$\mcI$ classes'' for brevity) and list them together with their CL in the first four columns of Tab.~\ref{tab:Homotopies}.

We remark that CLs of the AZ class and of the AZ+$\mcI$ class corresponding to a given {\color{black} centrosymmetric} system may be \emph{different} -- a subtlety that is further detailed in Sec.~\ref{sec:sym-class-rel}. Furthermore, classification (\ref{subeqn:AZ+I-symmetries}) also applies to non-centrosymmetric systems. Especially, symmetries (\ref{eqn:def:newTP}) can be present even when $\mcT,\mcP,\mcI$ themselves are absent.

\begin{table*}[t]
\begin{tabular}{|c|ccc|c|c|c|c|}
\hline	
CL				&	$\mfT$					&	$\mfP$						&	$\mcC$				&
basis of $\mcH_{2\times2 }$																&
basis of $\mcH_{4\times 4}$																&
$\delta_\textrm{CL}$		&	nodes in $D=3$																\\	\hline
$\mathrm{A}$	&	$\times$				&	$\times$					&	$\times$			&
$\{\unit,\sigma_x,\sigma_y,\sigma_z\}$																&
${\color{gray}\{\unit,\sigma_x,\sigma_y,\sigma_z\}\otimes\{\unit,\tau_x,\tau_y,\tau_z\}}$			&
$3$		&	point																	\\
$\mathrm{AIII}$	&	$\times$				&	$\times$					&	$\sigma_z$			&
$\{\sigma_x,\sigma_y\}$																				&
${\color{gray}\{\sigma_x,\sigma_y\}\otimes\{\unit,\tau_x,\tau_y,\tau_z\}}$							&
$2$		&	line																	\\	\hline
$\mathrm{AI}$	&	$\mcK$					&	$\times$					&	$\times$			&
$\{\unit,\sigma_x,\sigma_z\}$																		&
${\color{gray}\{\unit,\sigma_x,\sigma_z\}\otimes\{\unit,\tau_x,\tau_z\}\cup\{\sigma_y\otimes \tau_y\} }$	&
$2$		& 	line~\cite{Fang:2015}									\\
$\mathrm{BDI}$	&	$\mcK$					&	$\sigma_z \mcK$				&	$\sigma_z$			&
$\{\sigma_x\}$																						&
${\color{gray} \sigma_x\otimes\{\unit,\tau_x,\tau_z\}\cup\{\sigma_y\otimes\tau_y\}}$				&
$1$		&	surface																					\\
$\mathrm{D}$	&	$\times$				&	$\mcK$						&	$\times$			&
$\{\sigma_y\}$																						&
${\color{gray}\sigma_y\otimes\{\unit,\tau_x,\tau_z\}\cup\{\unit,\sigma_x,\sigma_z\}\otimes\tau_y}$	&
$1$		&	 surface~\cite{Agterberg:2017}															\\
$\mathrm{DIII}$	&	$\imi\sigma_y\mcK$		&	$\sigma_x \mcK$				&	$\sigma_z$			&
${\color{gray}\varnothing}$																			&
$\{\sigma_x,\sigma_y\}\otimes\tau_y$																&
$2$		& 	line																					\\
$\mathrm{AII}$	&	$\imi\sigma_y\mcK$		&	$\times$					&	$\times$			&
${\color{gray}\{\unit\}}$																			&
$\unit\otimes\{\unit,\tau_x,\tau_z\}\cup\{\sigma_x,\sigma_y,\sigma_z\}\otimes\tau_y$				&
$5$ 	&	{\color{gray} (none)}																	\\
$\mathrm{CII}$	&	$\imi\tau_y\mcK$		&	$-\imi\sigma_z\otimes\tau_y\mcK$	&	$\sigma_z$	&
${\color{gray}\varnothing}$																			&
$\{\sigma_x\otimes\unit\}\cup\sigma_y\otimes\{\tau_x,\tau_y,\tau_z\}$								&
$4$ 	&	{\color{gray} (none)}																	\\
$\mathrm{C}$	&	$\times$				&	$\imi\sigma_y\mcK$			&	$\times$			&
$\{\sigma_x,\sigma_y,\sigma_z\}$																	&
${\color{gray}\{\sigma_x,\sigma_y,\sigma_z\}\otimes\{\unit,\tau_x,\tau_z\}\cup\{\unit\otimes\tau_y\}}$	&
$3$		&	point																					\\
$\mathrm{CI}$	&	$\mcK$					&	$\imi\sigma_y\mcK$			&	$\imi\sigma_y$		&
$\{\sigma_x,\sigma_z\}$																				&
${\color{gray}\{\sigma_x,\sigma_z\}\otimes\{\unit,\tau_x,\tau_z\}}$									&
$2$		&	line																					\\ \hline
\end{tabular}		
\caption{Determining the node codimension $\delta_\textrm{CL}$. The first four columns indicate for every AZ+$\mcI$ class our representation of $\mfT$, $\mfP$ and $\mcC$, if present. The fifth column lists the basis matrices $\sigma_i$ of a two-band Hamiltonian $\mcH_{2\times2}$ that are compatible with the listed operators, and the sixth column achieves the same for basis matrices $\sigma_i\otimes\tau_j$ of a four-band Hamiltonian $\mcH_{4\times 4}$. For classes $\mathrm{DIII}$ and $\mathrm{CII}$, Kramer's degeneracy associated with $\mfT^2=-\unit$ together with the presence of $\mfP$ necessitates a minimum of four bands, such that $\mcH_{2\times2}$ is non-existent ($\varnothing$). Similarly, $\mfT^2 =-\unit$ of class $\textrm{AII}$ implies a double degeneracy of bands for all $\bs{k}$, meaning that a minimum of four bands is necessary to obtain a node formed by touching valence and conduction bands. The column $\delta_\textrm{CL}$ indicates the node codimension for each AZ+$\mcI$ class. It is obtained by counting the number of traceless basis elements of the minimal model (typesetted in black). As explained in Sec.~\ref{sec:node-dim}, stable nodes {\color{black} are permitted} when $D\geq \delta_\textrm{CL}$ and their dimension is given by the difference $D - \delta_\textrm{CL}$. We list the node type appearing in $D=3$ explicitly in the last column.}
\label{tab:Hamiltonians}
\end{table*}


\section{Node dimensionality}\label{sec:node-dim}

In this section we systematically determine the dimension of nodes supported by the individual AZ+$\mcI$ classes for arbitrary $D$. To achieve this, we fix canonical representations of operators (\ref{subeqn:AZ+I-symmetries}) and use them to determine the \emph{codimension} $\delta_\textrm{CL}$ of the node, i.e. the number of conditions to be fulfilled to make two bands touch. Provided that $D\geq \delta_\textrm{CL}$, nodes of dimension $D-\delta_\textrm{CL}$ occur.

Our choice of representing operators $\mfT,\mfP,\mcC$ is shown in Tab.~\ref{tab:Hamiltonians} where $\mcK$ indicates complex conjugation, and $\sigma_i$ and $\tau_i$ are Pauli matrices corresponding to two different degrees of freedom. The choice is rather arbitrary, but this freedom does not influence the analysis of the node codimension and of the node charges in the subsequent text. Specifically, we apply the following rules:
\begin{itemize}
\item[$(i)$] Set $\mcC = \sigma_z$ to enforce a block-off-diagonal form~(\ref{eqn:Q-with-CHS}) of the Hamiltonian. 
\item[$(ii)$] Represent antiunitary operators squaring to $+1$ by $\mcK$, and those squaring to $-1$ by $\imi \sigma_y \mcK$. 
\end{itemize}
These requirements are incompatible if all $\mfT$, $\mfP$ and $\mcC$ are present, hence classes $\textrm{BDI}$, $\textrm{DIII}$, $\textrm{CII}$ and $\textrm{CI}$ contain certain exceptions to rules ($i$) and ($ii$). Finally, we always apply the rule:
\begin{itemize}
\item[$(iii)$] Set $\mfT\mfP=\mcC$ if all three symmetries are present. 
\end{itemize}

Having fixed the representations of $\mfT$, $\mfP$ and $\mcC$, the node codimension $\delta_{\textrm{CL}}$ is determined by considering the minimal model capturing degeneracies between the valence and the conduction bands. This is either a two-band Hamiltonian
\begin{subequations}\label{eqn:nxn-Hams}
\begin{equation}
\mcH_{2\times 2}(\bs{k}) = \sum_{i=0}^{3} f_i(\bs{k}) \sigma_i \label{eqn:2x2-Ham-f}
\end{equation}
where $\sigma_i$ are Pauli matrices, or a four-band model
\begin{equation}
\mcH_{4\times 4}(\bs{k}) = \sum_{i,j=0}^{3} g_{ij}(\bs{k}) \,\sigma_i \otimes \tau_j \label{eqn:4x4-Ham-g}
\end{equation}
\end{subequations}
expanded using a product of \emph{pairs} of Pauli matrices (i.e. \emph{Dirac} matrices) $\sigma_i\otimes\tau_j$, and $f_i$ and $g_{ij}$ are real-valued functions. Model~(\ref{eqn:nxn-Hams}) can be always locally obtained by projecting out bands not forming the node. By Kramer's theorem, the minimal model is~(\ref{eqn:4x4-Ham-g}) whenever the symmetry class contains $\mfT^2=-1$, and~(\ref{eqn:2x2-Ham-f}) otherwise.

Local constraints~(\ref{subeqn:AZ+I-symmetries}) forbid the presence of some of the basis matrices in expansions~(\ref{eqn:nxn-Hams}). For the representation of operators given in Tab.~\ref{tab:Hamiltonians} we systematically analyze these constraints, and we list the symmetry-compatible basis matrices and their number $\delta_{\textrm{CL}}$ further in Tab.~\ref{tab:Hamiltonians}. The unit matrix is not relevant for the node dimension (although it may be relevant for its experimental \emph{signatures}~\cite{Soluyanov:2015}) and is therefore not counted in $\delta_\textrm{CL}$. We observe that the remaining basis matrices of the minimal model are always anticommuting, which implies that a node occurs at $\bs{k}_0$ whenever the $\delta_{\textrm{CL}}$ function values $f_i(\bs{k}_0)$ [or $g_{ij}(\bs{k}_0)$] vanish. Solutions to $\delta_\textrm{CL}$ equations in $D$ dimensions generically occur on a $(D-\delta_\textrm{LC})$-dimensional manifold -- the node dimension of the corresponding AZ+$\mcI$ class. We provide the explicit outcome of the analysis for $D=3$ in the last column of Tab.~\ref{tab:Hamiltonians}. 

We remark that for classes with $\mfP$ or $\mcC$ the presented arguments only work for nodes located at zero energy (i.e. at the Fermi level). For nodes at a non-zero energy (i.e. ones formed entirely within the valence or within the conduction bands), symmetries $\mfP$ and $\mcC$ don't restrict the effective model~(\ref{eqn:nxn-Hams}) and can be dropped from the AZ+$\mcI$ description. This a rather trivial observation and we do not return to it again in the following sections. 


\section{Topological charges}\label{sec:charges}

\begin{figure}[t]
	\includegraphics[width=0.48\textwidth]{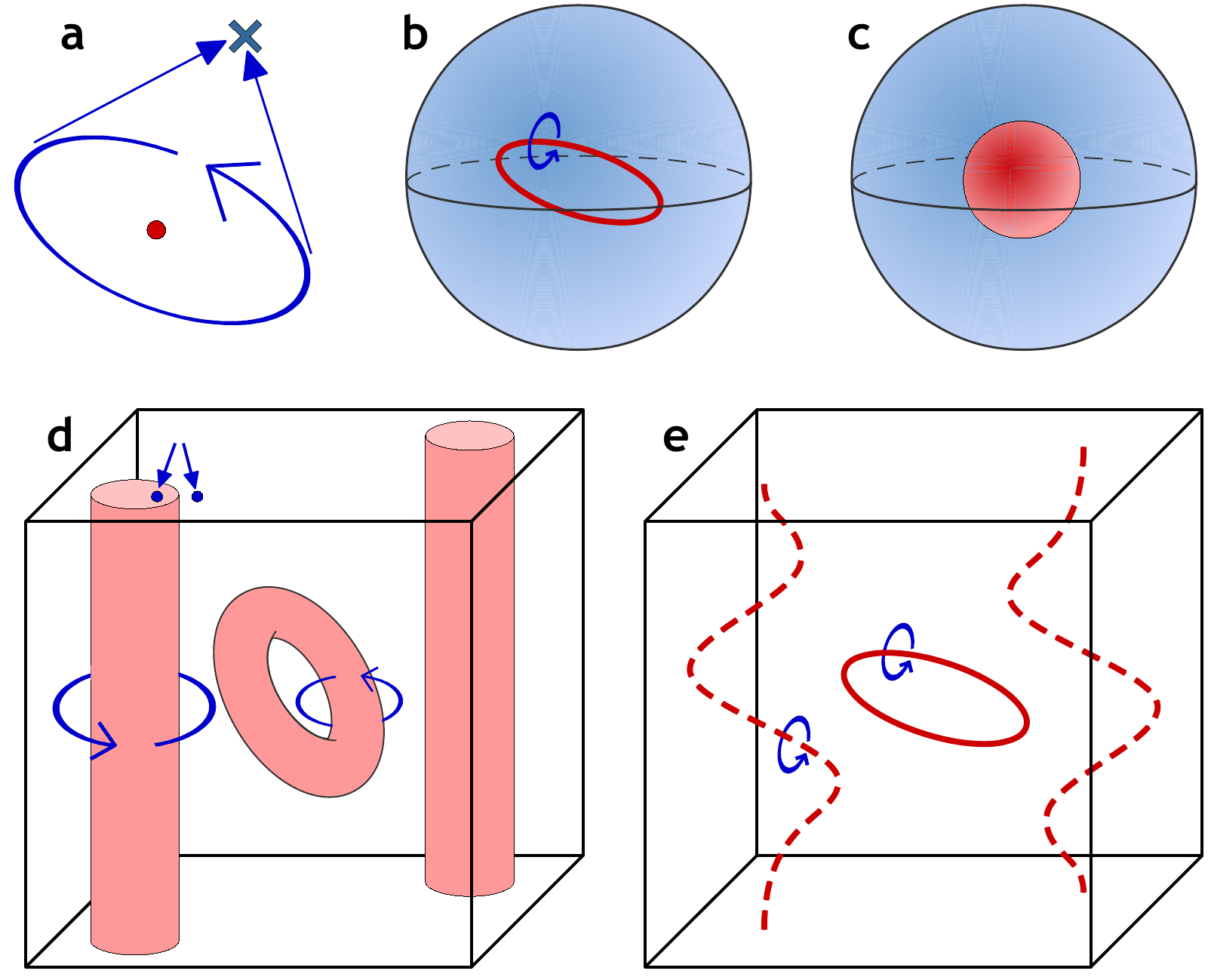}
	\caption{Enclosing nodes (red) by $p$-spheres $S^p$ (blue) for $D=3$. (a) The only non-contractible $S^p$ wrapping around a nodal point is $S^2$. Any $S^1$ would be trivially shrinkable to a point as indicated, and similarly for a pair of points $S^0$. (b) A nodal line can be wrapped by both $S^1$ and $S^2$, but not by $S^0$. It follows that nodal lines may be doubly charged. (c) A spherical nodal surface can be clearly encased by $S^2$ (shown), as well as by a pair of points $S^0$ located on the opposite sides of the surface. (d) If the nodal surface takes the form of a cylinder winding around BZ (black frame) or of a torus, it can be wrapped by $S^1$ too. If the nodal cylinder is characterized by a non-trivial charge on $S^1$, it becomes robust and a Nielsen-Ninomiya type of argument~\cite{Nielsen:1983} implies the presence of another such a cylinder. The same is \emph{not} true for torical nodal surface which can be gapped out by itself. The pair of points indicated by arrows is an example of $S^0$ ``enclosing'' the cylinder. (e) Two kinds of nodal lines. The dashed ones wind around BZ, while the solid one does not and is hence called a nodal \emph{loop}. Similar to the previous case, a non-trivial charge on $S^1$ makes only the winding loops robust.}
	\label{fig:nodal-objects}
\end{figure}

Knowing the \emph{dimensionality} of the nodes for all the AZ+$\mcI$ classes, we now determine the corresponding \emph{topological charges}. {\color{black} We generalize the method explained in Sec.~II.B of Ref.~\cite{fang2016topological}, and} consider $p$-spheres $S^p$ wrapping around the nodes. The largest sphere fitting into the Brillouin zone (BZ) is $S^{D-1}$. By the smalled sphere $S^0$ we mean a \emph{pair} of points (rather than just a single point). However, not all of these come into play. For example, all circles $S^1\subset\textrm{BZ}$ in a class supporting nodal points are continuously contractible to a single point without encountering a gap closing, and therefore cannot host any topological obstruction. The same is true for $S^0$ in such a system, and a few more similar examples are illustrated in Fig.~\ref{fig:nodal-objects}. Simple counting reveals that only $p$-spheres with $\delta_\textrm{CL}-1 \leq p \leq D -1$ can be \emph{non-contractible} and may therefore accommodate a topological charge.
\begin{table}[t]
	\begin{tabular}{|l|ccc|ccc|cc|}
	\hline
												&	
			\multicolumn{3}{c|}{$\textrm{BDI}$}	&
			\multicolumn{3}{c|}{$\textrm{D}$}	&
			\multicolumn{2}{c|}{$\textrm{CI}$}	\\ \cline{2-9}
												&
	$\pi_0$						&	$\pi_1$						&	$\pi_2$						&
	$\pi_0$						&	$\pi_1$						&	$\pi_2$						&
									$\pi_1$						&	$\pi_2$						\\ \hline
	$n=1$ (2 bands)				&
	{\color{gray}$\ztwo$}		&	$\pmb{\triv}$				&	{\color{gray}$\triv$}		&
	{\color{gray}$\ztwo$}		&	{\color{gray}$\triv$}		&	$\pmb{\triv}$				&
									{\color{gray}$\intg$}		&	$\pmb{\triv}$				\\ 
	$n=2$ (4 bands)				&
	{\color{gray}$\ztwo$}		&	$\pmb{\intg}$				&	{\color{gray}$\triv$}		&
	{\color{gray}$\ztwo$}		&	{\color{gray}$\triv$}		&	{\color{gray}$2\intg$}		&
									{\color{gray}$\intg$}		&	$\pmb{\intg}$				\\ 	
	$n\geq3$ 					&
	{\color{gray}$\ztwo$}		&	{\color{gray}$\ztwo$}		&	{\color{gray}$\triv$}		&
	{\color{gray}$\ztwo$}		&	{\color{gray}$\triv$}		&	{\color{gray}$2\intg$}		&
									{\color{gray}$\intg$}		&	{\color{gray}$\ztwo$}		\\ 	\hline
	\end{tabular}
	\caption{The relevant homotopy groups $\pi_p(M_\textrm{CL})$ for AZ+$\mcI$ symmetry classes supporting doubly charged nodes in $D=3$ for \emph{few-band} models. Class $\textrm{AI}$ contains more special cases and is treated separately in Tab.~\ref{tab:Homotopies-small-AI}. The exceptional values are typesetted in a bold font, while the values following the large $n$ result of Tab.~\ref{tab:Homotopies} are displayed in gray. Note that in all cases a minimum of four bands ($n=2$) is necessary to realize a node with a pair of non-trivial topological charges.}
	\label{tab:Homotopies-small}
\end{table}

\begin{table}[t]
	\begin{tabular}{|c|cc|}
	\hline
	$\pi_1(M_\textrm{AI})$	& $n=1$						& $n\geq 2$					\\ \hline
	$\ell=1$					& $\pmb{\intg}$				& ${\color{gray}\ztwo}$ 	\\
	$\ell\geq 2$				& ${\color{gray}\ztwo}$		& ${\color{gray}\ztwo}$		\\ \hline	
	\end{tabular}\hspace{0.2cm}
	\begin{tabular}{|c|ccc|}
	\hline
	$\pi_2(M_\textrm{AI})$	& $n=1$				& $n=2$						& $n\geq3$ 				\\ \hline
	$\ell=1$					& $\pmb{\triv}$		& $\pmb{2\intg}$ 			& $\pmb{\triv}$    		\\
	$\ell=2$					& $\pmb{2\intg}$	& $\pmb{\intg\oplus\intg}$	& $\pmb{\intg}$			\\
	$\ell\geq 3$				& $\pmb{\triv}$		& $\pmb{\intg}$				& ${\color{gray}\ztwo}$	\\ \hline	
	\end{tabular}
	\caption{The relevant homotopy groups $\pi_p(M_\textrm{AI})$ for few-band models contain multiple exceptions (displayed in bold) differing from the large $n,\ell$ limit (displayed in gray) of Tab.~\ref{tab:Homotopies}. We briefly tackle these exceptions in Sec.~\ref{sec:nodes-AI}.}
	\label{tab:Homotopies-small-AI}
\end{table}

To identify the topological charge supported by $S^p$, we first perform spectral flattening~\cite{Ryu:2010}. For a system with $n$ occupied and $\ell$ unoccupied bands, we decompose the Hamiltonian $\mcH(\bs{k})$ on $S^p$ using the eigensystem $\{\varepsilon^a(\bs{k}),\ket{u^a(\bs{k})}\}_{a=1}^{n+\ell}$ as 
\begin{subequations}\label{eqn:Ham-decomps}
\begin{equation}
\mcH(\bs{k}) = \sum_{a=1}^{n+\ell}\ket{u^a(\bs{k})}\varepsilon^a(\bs{k})\bra{u^a(\bs{k})}
\end{equation}
which can be continuously deformed without closing the gap on $S^p$ into a \emph{flat-band} Hamiltonian
\begin{equation}
\mcQ(\bs{k}) = \sum_{a=1}^{n+\ell}\ket{u^a(\bs{k})}\sign\left[\varepsilon^a(\bs{k})\right] \bra{u^a(\bs{k})}\label{eqn:flat-band-Ham-Q}.
\end{equation}
\end{subequations}
Note that the description using $\{\ket{u^a(\bs{k})}\}_{a=1}^{n+\ell}\in\mathsf{U}(n+\ell)$ is redundant because rotating the occupied (unoccupied) states by a $\mathsf{U}(n)$ [$\mathsf{U}(\ell)$] matrix leaves $\mcQ(\bs{k})$ invariant, meaning that $\mcQ(\bs{k}) \in \mathsf{U}(n+\ell)/\mathsf{U}(n)\!\times\!\mathsf{U}(\ell) \equiv M_\textrm{A}$ which is the classifying space in the absence of AZ+$\mcI$ symmetries. The decompositions~(\ref{eqn:Ham-decomps}) may not be achieved with a smooth gauge in the case of Chern bands, nevertheless, the flat-band Hamiltonian $\mcQ(\bs{k})\in M_\textrm{A}$ is smooth.

Conditions~(\ref{subeqn:AZ+I-symmetries}) of a given symmetry class CL constrain $\mcQ(\bs{k})$ to be an element of a \emph{smaller} classifying space $M_\textrm{CL}\subseteq M_\textrm{A}$. For example,  $\mcC=\sigma_z$ leads to
\begin{equation}
\mcQ(\bs{k}) = \left(\begin{array}{cc}
\triv					& q(\bs{k})			\\
q^\dagger(\bs{k})		& \triv				
\end{array}\right)\label{eqn:Q-with-CHS}
\end{equation}
with $q(\bs{k})\in \mathsf{U}(n)\equiv M_\textrm{AIII}$. All relevant classifying spaces are listed in Ref.~\cite{Kitaev:2009} and we reproduce them in Tab.~\ref{tab:Homotopies}.

Every $S^p\subset \textrm{BZ}$ with a gapped spectrum is associated with a continuous map $\mcF: S^p\to M_\textrm{CL}$. Continuous deformations of $S^p$ as well as of $\mcH(\bs{k})$ (such that the spectrum on $S^p$ is kept gapped) lead to continuous changes of $\mcF$. One may thus consider the \emph{equivalence class} $[\mcF]$ of maps continuously reachable from $\mcF$. Especially, if a constant map $\mcF_\mathfrak{m}: S^p \to \mathfrak{m}$ with a fixed element $\mathfrak{m}\in M_\textrm{CL}$ \emph{cannot} be reached from $\mcF$, then $S^p$ \emph{cannot} be shrunk to a single point $\bs{k}_0\in\textrm{BZ}$. {\color{black} This implies that $S^p$ accommodates a topological obstruction, i.e.} it contains an unremovable node. We therefore deduce a connection between the equivalence class $[\mcF]$ and the charge $c_\textrm{CL}(S^p)$ accommodated by the $p$-sphere. The \emph{order} of the charge for $p \geq 1$ corresponds to the number of distinct equivalence classes $[\mcF]$, which is captured by the \emph{homotopy group} $\pi_p(M_\textrm{CL})$~\cite{Nakahara:2003}. The charge on $S^p$ is then some element
\begin{subequations}
\begin{equation}
c_\textrm{CL}(S^p) \in \pi_p(M_\textrm{CL})\label{eqn:homotopy-charge-def}.
\end{equation}
Homotopy groups of spaces $M_\textrm{CL}$ in the large $n,\ell$ limit are listed in Ref.~\cite{Lundell:1992}, and we reproduce them in Tab.~\ref{tab:Homotopies}. The special $p=0$ case $\pi_0(M_\textrm{CL})$ counts the number of connected components of $M_\textrm{CL}$ and also happens to have a group structure (although this is not true about $\pi_0(X)$ for a general manifold $X$). \footnote{{\color{black} A similar table to our Tab.~\ref{tab:Homotopies} for symmetry classes $\textrm{A}$, $\textrm{AIII}$, $\textrm{D}$, $\textrm{C}$, $\textrm{DIII}$ and $\textrm{CII}$ also appears in Ref.~\cite{kobayashi2014topological} in the context of superconductors. However, this work assumes that a nodal object of a given AZ+$\mcI$ class can have arbitrary dimension and that it is characterized by a single topological charge -- contrary to the findings of the present work.}}

Collecting the charges supported by all $p$-spheres enclosing a node of a given AZ+$\mcI$ class, the complete topological charge of the node becomes
\begin{equation}
c_\textrm{CL}^{(D)} \in \bigoplus_{p\,=\,\delta_\textrm{CL}-1}^{D-1} \pi_p(M_\textrm{CL}).\label{eqn:TopoCharge}
\end{equation}
\end{subequations}
A node may become \emph{multiply charged} whenever more than one of the groups in direct sum~(\ref{eqn:TopoCharge}) are non-trivial. We read from Tabs.~\ref{tab:Homotopies} and~\ref{tab:Hamiltonians} that for $D=3$ in the large $n,\ell$ limit, doubly charged nodal lines appear in AZ+$\mcI$ classes $\textrm{AI}$ and $\textrm{CI}$, and doubly charged nodal \emph{surfaces} exist in classes $\textrm{BDI}$ and $\textrm{D}$.

More care is required if one studies \emph{few-band} models not reaching the large $n,\ell$ limit of Ref.~\cite{Lundell:1992}. In that case, the homotopy groups may differ from those listed in Tab.~\ref{tab:Homotopies}. If $M_\textrm{CL}$ is a Lie group (classes $\textrm{AIII}$, $\textrm{BDI}$ and $\textrm{CII}$), the homotopy groups can be readily found in various sources (e.g.~\cite{Ito:1993}), while if $M_\textrm{CL}$ is a fiber bundle $B=E/F$ (all other classes), one can determine $\pi_p(B)$ from the long exact sequence of homomorphisms~\cite{Hatcher:2002}
\begin{equation}
\hspace{-0.2cm}\ldots\to\pi_p(E)\to\pi_p(B)\to\pi_{p-1}(F)\to\pi_{p-1}(E)\to\ldots
\end{equation}
We carried out the analysis for the classes supporting doubly charged nodes in $D=3$, and we list the results in Tabs.~\ref{tab:Homotopies-small} and~\ref{tab:Homotopies-small-AI}. In particular, we find that in all four instances the minimal half-filled models exhibiting doubly charged nodes contain \emph{four bands}. 

Let us briefly discuss the concept of node \emph{robustness}, which is characterized by the presence of such a charge that a set of nodes can mutually annihilate only if their net charge vanishes. Clearly, the nodal loop in Fig.~\ref{fig:nodal-objects}(e) with a non-trivial $c(S^1)$ and trivial $c(S^2)$ is \emph{not} robust: It can be shrunk to a single point, and since at that stage $c(S^1)$ ceases to be defined (there is no ``loop interior''), nothing prevents us from gapping out the spectrum entirely. The result would be different if $c(S^2)$ were \emph{non-trivial}, because then the nodal point {\color{black}(corresponding now to the shrunk nodal loop)} would be enclosed by a sphere carrying a topological obstruction~\cite{Fang:2015}. This dichotomy should be contrasted with the case of \emph{winding} nodal lines in Fig.~\ref{fig:nodal-objects}(e) which \emph{cannot} be shrunk to a point, such that $c(S^1)$ \emph{never} ceases to be meaningful. Such winding nodal lines are robust regardless of the higher homotopy charge.

An analogous dichotomy exists for the nodal torus and the nodal cylinder with non-trivial $c(S^1)$ and a trivial higher charge that are illustrated in Fig.~\ref{fig:nodal-objects}(d): The cylinder is robust while the torus is not. In the same spirit, a nodal surface with a non-trivial $c(S^0)$ also becomes robust if \emph{both} of its dimensions wind around BZ. Generally, if the highest non-trivial charge in expansion (\ref{eqn:TopoCharge}) corresponds to $\tilde{p}$-sphere, the nodal object has to possess $D-\tilde{p}-1$ winding coordinates to become robust.


\section{Realizations of $\textrm{AZ}$+$\mcI$ classes}\label{sec:sym-class-rel}
We have already warned the reader that the Cartan labels of the AZ and of the AZ+$\mcI$ class corresponding to a given centrosymmetric system may be \emph{different}. In this section we explain that this disagreement occurs when $\mcI \mcP \neq \mcP \mcI$, and we further discuss how the individual AZ+$\mcI$ classes are realized as various semimetallic and superconducting phases. {\color{black}Our discussion} is focused on centrosymmetric systems only. The generalization to non-centrosymetric cases is straightforward, {\color{black} and we include it in the scheme of Fig.~\ref{fig:AZI-classification}.}

Before delving into the scrutiny of operators $\mcP$ and $\mfP$, let us show that time-reversal behaves ``nicely'' in the sense that always $\mcT^2 = \mfT^2$, thus not leading to a Cartan label difference. The reason is that $\mcT$ and $\mcI$ act on different degrees of freedom (flipping the sign of time vs. position), therefore the commutator $[\mcT,\mcI]=0$ and
\begin{equation}
\mfT^2 = \mcT\mcI\mcT\mcI = \mcT^2 \mcI^2 = \mcT^2
\end{equation}
where $\mcI^2=\unit$ is true for any system. {\color{black} This implies that in the absence of symmetries $\mcP$ and $\mcC$, the AZ label and the AZ+$\mcI$ label of a given centrosymmetric material \emph{do} coincide, and are either $\textrm{A}$, $\textrm{AI}$ or $\textrm{AII}$. Class $\textrm{A}$ exhibits nodal points, and corresponds to Weyl semimetals~\cite{Wan:2011} as well as to their photonic analogue~\cite{lu2013weyl}. Similarly, class $\textrm{AI}$ corresponds to nodal line semimetals in the absence of spin-orbit coupling (SOC) (including both ``Type A'' and ``Type B'' nodal line materials of Tab.~1 in Ref.~\cite{fang2016topological}) as well as their bosonic counterparts~\cite{lu2013weyl,po2016phonon,Li:2017}. Finally, class $\textrm{AII}$ does not exhibit any stable nodes in the absence of additional crystalline symmetries. These three scenarios correspond to the Fig.~\ref{fig:AZI-classification}(a).}

In the following two subsections we show that always $\mfP^2 = \pm \mcP^2$, where the sign depends on the specific system realization. The discussion is split into Subsec.~\ref{subsec:sublattice} dealing with non-superconducting systems with sublattice symmetry (SLS) which is summarized in Fig.~\ref{fig:AZI-classification}(b), and Subsec.~\ref{subsec:superconductivity} dealing with superconducting (SC) which is captured by Fig.~\ref{fig:AZI-classification}(c).


\subsection{$\mcP$ from sublattice symmetry} \label{subsec:sublattice}

The sublattice realization of $\mcC$ corresponds to acting on two sets of sites (i.e. ``sublattices'') with opposite sign, therefore $\mcC^2=\unit$. Furthermore, this operation is insensitive to spin, meaning that $\mcC$ and $\mcT$ commute. The composition $\mcP = \mcT\mcC$ fulfills
\begin{equation}
\mcP^2 = \mcT \mcC \mcT \mcC = \mcT^2 \mcC^2 = \mcT^2.
\end{equation}
This subsection is thus relevant only to $\textrm{AZ}$ classes $\textrm{BDI}$ and $\textrm{CII}$, which are themselves distinguished by the presence or absence of SOC. We want to find the associated $\mathrm{AZ}$+$\mcI$ classes.

We begin with the example of the graphene lattice which contains two sublattices [light red and dark blue in Fig~\ref{fig:BDI-vs-CI}(a)]. Within the nearest-neighbor (NN) TB description, electrons only hop \emph{between} the two sublattices (i.e. not \emph{within} them), meaning that the Hamiltonian is off-diagonal in the sublattice (orbital) basis captured by Pauli matrices $\tau_i$. Therefore, the Hamiltonian anticommutes with $\mcC = \tau_z$. On the other hand, inversion symmetry \emph{switches} the two sublattices, such that $\mcI_{\mathsf{(a)}} = \tau_x$ and $\mcI_{\mathsf{(a)}} \mcC = -\mcC \mcI_{\mathsf{(a)}}$. It follows that $\mfP = \mcP\mcI_{\mathsf{(a)}}$ obeys
\begin{subequations}\label{eqn:PHS-CHS-sublattice}
\begin{equation}
\mfP^2_{{\mathsf{(a)}}} = \mcT\mcC\mcI_{{\mathsf{(a)}}}\mcT\mcC\mcI_{{\mathsf{(a)}}} = -(\mcT\mcC)^2 \mcI_{{\mathsf{(a)}}}^2 = -\mcP^2\label{eqn:PHS-graphene}
\end{equation}
meaning that the $\textrm{AZ}+\mcI$ class relevant for graphene \emph{differs} from the underlying $\textrm{AZ}$ class. We refer to this situation as having $\mcI$-odd SLS. The Cartan labels are changed {\color{black} to $\textrm{CI}$ in the absence and to $\textrm{DIII}$ in the presence of SOC}.
\begin{figure}[t]
	\includegraphics[width=0.46\textwidth]{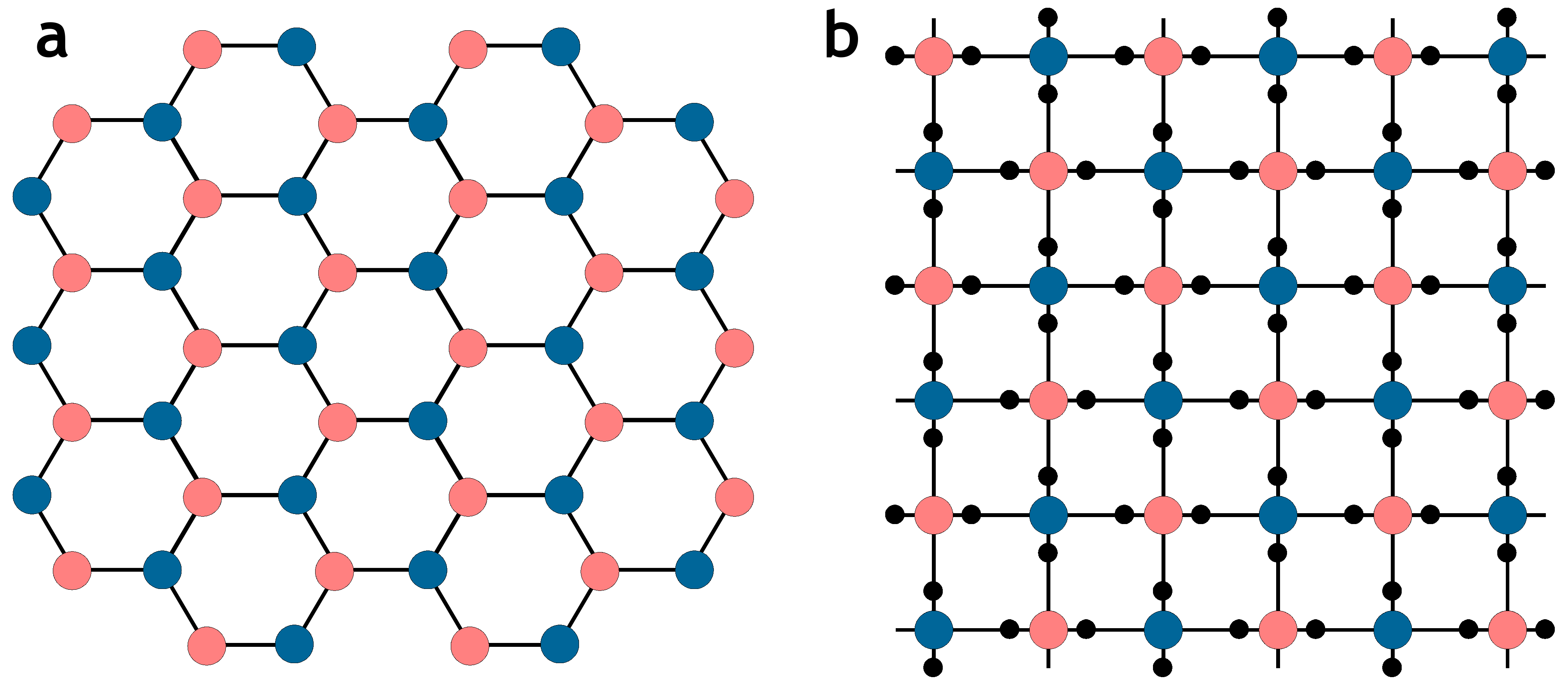}
	\caption{Two bipartite lattices with a different representation of the inversion operator $\mcI$. The red (light) and blue (dark) discs represent atoms on the two sublattices -- each unit cell contains one atom of every color. The black dots in example (b) indicate atoms that do not enter the effective TB model, but that nevertheless influence the hopping amplitudes by distorting the crystal field. The atoms of both colors are identical in all respects, such that the Hamiltonian is block-off-diagonal in the sublattice basis, leading to $\mcC = \tau_z$. While in (a) the inversion symmetry $\mcI_{\mathsf{(a)}}=\tau_x$ anticommutes with $\mcC$ ($\mcI$-odd SLS), in (b) the inversion operator $\mcI_{\mathsf{(b)}} = \unit_\tau$ commutes with it ($\mcI$-even SLS). In case (a), the Cartan labels of the relevant AZ and AZ+$\mcI$ classes are \emph{different}.}
	\label{fig:BDI-vs-CI}
\end{figure}
\begin{figure}[t]
	\includegraphics[width=0.48\textwidth]{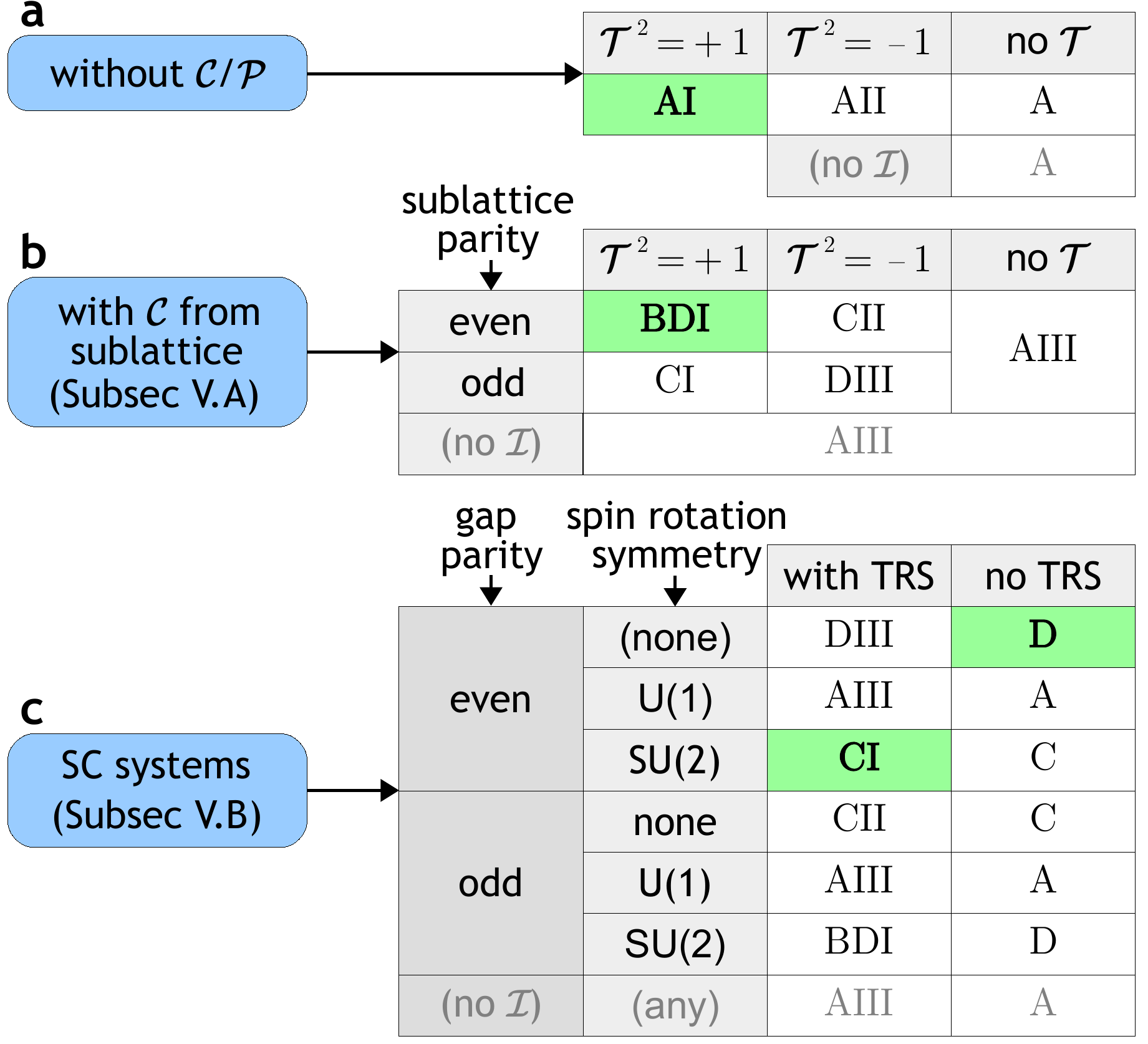}
	\caption{A schematic summary of Sec.~\ref{sec:sym-class-rel}. Here, $\mcT$, $\mcP$, $\mcC$ are the usual AZ symmetries, while the assigned Cartan labels correspond to the AZ+$\mcI$ classes characterized by symmetries $\mfT$, $\mfP$, $\mcC$. The even/odd sublattice parity corresponds to $\mcI \mcC = \pm \mcC\mcI$ (where $\mcI$ is the inversion operator), and the even/odd gap function corresponds to $\mcI_0\Delta_{\bs{k}} = \pm \Delta_{-\bs{k}}\mcI_0$ (where $\mcI_0$ is the \emph{normal state} inversion operator). {\color{black} With gray font we indicate the relevant AZ+$\mcI$ class for non-centrosymmetric systems (assuming for simplicity that $\mfT$ and $\mfP$ are also absent) which are not discussed explicitly in the text}. The cells with green background indicate the example models discussed explicitly in Secs.~\ref{sec:nodes-D} to~\ref{sec:nodes-AI}.}
	\label{fig:AZI-classification}
\end{figure}

On the other hand, the lattice in Fig.~\ref{fig:BDI-vs-CI}(b) has $\mcC = \tau_z$ and $\mcI_{\mathsf{(b)}} = \unit_\tau$ which \emph{commute} ($\mcI$-even SLS). In this case 
\begin{equation}
\mfP^2_{\mathsf{(b)}} = \mcT\mcC\mcI_{\mathsf{(b)}}\mcT\mcC\mcI_{\mathsf{(b)}} = (\mcT\mcC)^2 \mcI_{\mathsf{(b)}}^2 = \mcP^2,
\end{equation}
\end{subequations}
meaning that the $\textrm{AZ}$ and $\textrm{AZ}$+$\mcI$ Cartan labels coincide. Such a situation arises in various non-symmorphic lattices including various distorted perovskites with doubled unit cells~\cite{Shirane:1969,Dove:1997}.
\begin{table*}[t]
	\begin{tabular}{|c||c|c|c||c||c|}
	\hline  
	pair condensate material		& $\Delta_{\bs{k}}$ parity	&  	SRS 			&	$\mcT$ present	&	AZ+$\mcI$ class	&	nodes 
	\\ \hline
	SrPtAs $\textsf{(m)}$~\cite{Fischer:2015}, URu$_2$Si$_2$ $\textsf{(m)}$~\cite{Agterberg:2017}
									& even						& 	(none)			&	$\times$		&	D				&	surfaces
	\\	
	cuprates ($d$-wave)~\cite{scalapino1995case}, CeCoIn$_5$~\cite{higemoto2002musr}		
									& even				 	&	\textsf{SU}(2)		&	$\checkmark$	&	CI				&	lines
	\\
	SrPtAs $\textsf{(s)}$~\cite{Fischer:2014}, URu$_2$Si$_2$ $\textsf{(s)}$~\cite{kasahara2007exotic}
									& even					&	$\mathsf{SU}(2)$	&	$\times$		&	C				&   points
	\\
	$^3$He-B~\cite{vollhardt1990superfluid}				
									& odd						&	(none) 			&  	$\checkmark$	&	CII				&  	(none)
	\\
	PrOs$_4$Sb$_{12}$~\cite{aoki2007unconventional}
									& odd					&	(none) or $\mathsf{U}(1)$&	$\times$	& 	C or A			& 	points
	\\
	$^3$He-A~\cite{vollhardt1990superfluid}
									& odd					&	$\mathsf{U}(1)$		&	$\times$		&	A				&	points
	\\
	CePt$_3$Si~\cite{Hayashi:2006}, Li$_2$Pt$_3$B~\cite{yuan2006s}	
									& (no $\mcI$)				&	(none)			&	$\checkmark$	&  	AIII			& 	lines	
	\\
	LaNiC$_2$~\cite{hillier2009evidence}
									& (no $\mcI$)				&	(none)			& 	$\times$		& 	A				& 	points

	\\ \hline
	\end{tabular}
	\caption{{\color{black}List of selected unconventional superconducting (SC) and superfluid materials and their location within Fig.~\ref{fig:AZI-classification}(c), assuming the order parameters considered in the cited references. For certain entries, label $\textsf{(m)}$ indicates the explicit presence and $\textsf{(s)}$ the explicit absence of multi-orbital pairing, and for all materials we neglect the possible presence of SOC in $\Xi_{\bs{k}}$. The symmetry analysis of Sec.~\ref{subsec:superconductivity} predicts the dimensionality of stable nodes indicated in the last column. In some of the cases, further point group symmetries may impose additional nodes of a different dimension (e.g. the line node at $k_z = 0$ of URu$_2$Si$_2$ $\textsf{(s)}$~\cite{kasahara2007exotic} protected by a horizontal mirror symmetry). On the other hand, the vertical mirror symmetries of $d$-wave cuprates only serve to locate the nodal lines within high-symmetry planes, but are not essential for the  \emph{stability} of the NLs.}}
	\label{tab:SCmaterials}
\end{table*}


\subsection{$\mcP$ from superconductivity}\label{subsec:superconductivity}

The commutation relation of $\mcP$ and $\mcI$ in SC systems depends on the parity of the gap function $\Delta_{\bs{k}}$. Furthermore, the relevant AZ and AZ+$\mcI$ classes finely depend on the presence of time-reversal symmetry (TRS) and on the degree of spin-rotation symmetry (SRS). To systematically tackle all the possibilities, we largely follow Sec.~II.C of Ref.~\cite{Schnyder:2008} which similarly treats the case of AZ classification. For simplicity, we only consider SC with zero momentum and even frequency pairing.

The most general SC Hamiltonian takes the form~\cite{Atland:1997}
\begin{subequations}\label{eqn:BdG-big}
\begin{equation}
H(\bs{k}) = \frac{1}{2}\left(\begin{array}{cc}
c^{a\dagger}_{\bs{k}} 				& 	c^{a\pdag}_{-\bs{k}}
\end{array}\right)\mcH^{ab}_\textrm{BdG}(\bs{k})\left(\begin{array}{c}
c^{b\pdag}_{\bs{k}}					\\	c^{b\dagger}_{-\bs{k}}
\end{array}\right)
\end{equation}
where indices $a$ and $b$ encode both the spin and the orbital degree of freedom, and 
\begin{equation}
\mcH^{ab}_\textrm{BdG}(\bs{k}) = \left(\begin{array}{cc}
\Xi^{ab}_{\bs{k}}		&	\Delta^{ab}_{\bs{k}}	\\
-\Delta^{ab*}_{-\bs{k}}	&	-\Xi^{ba}_{-\bs{k}}
\end{array}\right) \label{eqn:BdG-4x4}
\end{equation}
\end{subequations}
is the Bololyubov-de Gennes (BdG) Hamiltonian, in which $\Xi_{\bs{k}}^\pdag = \Xi^\dagger_{\bs{k}}$ describes the underlying normal metal band structure {\color{black} which may or may not contain SOC}, and the gap function obeys $\Delta_{\bs{k}}^{\phantom{\top}} = -\Delta^\top_{-\bs{k}}$ due to the fermionic statistics. Hamiltonian (\ref{eqn:BdG-4x4}) is automatically furnished with particle-hole operator
\begin{equation}
\mcP_\textrm{t} = s_x \mcK \label{eqn:PHS-triplet}
\end{equation}
where Pauli matrices $s_i$ act on the particle-hole degree of freedom. Operator (\ref{eqn:PHS-triplet}) squares to $\mcP_\textrm{t}^2 = +\unit$.

The system described by~(\ref{eqn:BdG-big}) is assumed to be centrosymmetric. We can decompose the inversion operator of the normal metal state as $\mcI_0 = \mcI_\tau \otimes \unit_\sigma$ where $\mcI_\tau$ is the orbital component, Pauli matrices $\sigma_i$ represent the spin degree of freedom, and the $\unit_\sigma$ part follows because spin is an axial vector. The symmetry relates $\mcI_0 \Xi_{\bs{k}} \mcI_0^{-1} = \Xi_{-\bs{k}}$. Operator $\mcI_0$ is \emph{real} because it only permutes the orbitals. Furthermore, inversion symmetry does not mix electrons and holes, hence the inversion operator of (\ref{eqn:BdG-4x4}) is diagonal in $s$, leaving only two options 
\begin{equation}
\mcI_\textrm{BdG} = \mcI_0\oplus (\pm \mcI_0) \in\left\{\mcI_0\otimes \unit_s,\mcI_0\otimes s_z\right\} \label{eqn:SC-possible-Is}.
\end{equation}
These translate to a constraint on the gap function
\begin{subequations}
\begin{equation}
\pm \mcI_0 \Delta_{\bs{k}} \mcI_0^{-1} = \Delta_{-\bs{k}} = -\Delta_{\bs{k}}^\top \label{eqn:SC-INV-gen}
\end{equation}
where in the second step we used the fermionic statistics.

In \emph{single-orbital} SCs, Eq.~(\ref{eqn:SC-INV-gen}) contains $2\times 2$ matrices in the spin-degree of freedom with $\mcI_0 = \unit_\sigma$. Eq.~(\ref{eqn:SC-INV-gen}) therefore simplifies to
\begin{equation}
\pm\Delta_{\bs{k}} = -\Delta_{\bs{k}}^\top \label{eqn:SO-SC-Delta}
\end{equation}
\end{subequations}
where the $\pm$ sign is inherited from Eq.~(\ref{eqn:SC-INV-gen}). We see that single-orbital SCs have to follow one of two scenarios. Either ``$+$'' is realized in Eq.~(\ref{eqn:SO-SC-Delta}) (singlet, $\mcI$-even $\Delta_{\bs{k}}$), such that $\Delta_{\bs{k}} = \psi_{\bs{k}}(\imi\sigma_y)$ with a complex-valued scalar function $\psi_{\bs{k}}$ even in $\bs{k}$. In this case $\mcI_\textrm{BdG} = \mcI_0 \otimes \unit_s$ commutes with $\mcP_\textrm{t}$, such that $\mfP^2 = \mcP^2_\textrm{t} = +\unit$. Alternatively, the ``$-$'' sign in Eq.~(\ref{eqn:SO-SC-Delta}) is realized (triplet, $\mcI$-odd $\Delta_{\bs{k}}$), and the gap function takes the form $\Delta_{\bs{k}} = \left(\bs{d}_{\bs{k}}\cdot\bs{\sigma}\right)(\imi\sigma_y)$ with a complex-valued vector function $\bs{d}_{\bs{k}}$ odd in $\bs{k}$. In this case $\mcI_\textrm{BdG} = \mcI_0 \otimes s_z$ \emph{anti}commutes with $\mcP_\textrm{t}$, such that $\mfP^2 = -\mcP^2_\textrm{t} = -\unit$. However, as explained in Ref.~\cite{Schnyder:2008}, the presence of SRS renders description~(\ref{eqn:BdG-big}) redundant and a finer examination is necessary. We do so below after first commenting on the multi-orbital case.

In \emph{multi-orbital} SCs, the presence of inversion symmetry may be \emph{insufficient} to enforce the singlet/triplet separation. The gap function can be a \emph{mixture} of both, provided that the parity in the spin degree of freedom is compensated by the parity in the orbital one. Let us illustrate this on the graphene lattice of Fig.~\ref{fig:BDI-vs-CI}(a) with $\mcI_{0} = \tau_x \otimes \unit_\sigma$. The symmetry group permits two mixed representations. First, the \emph{even} one ($\mcI$-even $\Delta_{\bs{k}}$) with
\begin{subequations}
\begin{equation}
\Delta_{\bs{k}}^+ = \!\!\sum_{j=0,x,y}\psi_{\bs{k}}^j(\imi\sigma_y)\otimes \tau_j + \!\!\sum_{i=x,y,z}d_{\bs{k}}^{iz}\sigma_i(\imi \sigma_y)\otimes \tau_z \label{eqn:SrPtAs}
\end{equation}
where $\psi_{\bs{k}}^0,\psi_{\bs{k}}^x$ are even and $\psi_{\bs{k}}^y,d_{\bs{k}}^{iz}$ are odd functions of $\bs{k}$. In this case $\mcI_\textrm{BdG} = \mcI_0\otimes \unit_s$ commutes with $\mcP_\textrm{t}$ such that $\mfP^2=\mcP_\textrm{t}^2=+\unit$. The second options is the \emph{odd} representation ($\mcI$-odd $\Delta_{\bs{k}}$)
\begin{equation}
\Delta_{\bs{k}}^- = \psi_{\bs{k}}^z(\imi\sigma_y)\otimes \tau_z + \!\!\sum_{\substack{ i=x,y,z\\	j=0,x,y}}d_{\bs{k}}^{ij}\sigma_i(\imi \sigma_y)\otimes \tau_j \label{eqn:I-odd-Delta}
\end{equation}
\end{subequations}
where $\psi_{\bs{k}}^z,d_{\bs{k}}^{iy}$ are even and $d_{\bs{k}}^{i0},d_{\bs{k}}^{ix}$ are odd in $\bs{k}$. In this case $\mcI_\textrm{BdG} = \mcI_0\otimes s_z$ \emph{anti}commutes with $\mcP_\textrm{t}$ such that $\mfP^2 = -\mcP^2_\textrm{t}=-\unit$. Contrastingly, the lattice of Fig.~\ref{fig:BDI-vs-CI}(b) has a trivial $\mcI_0 = \unit_\tau\otimes \unit_\sigma$, such that the simplification to Eq.~(\ref{eqn:SO-SC-Delta}) becomes valid, and the singlet/triplet separation of the previous paragraph applies again.

We now combine the obtained information with the arguments of Ref.~\cite{Schnyder:2008}. We begin with systems without {\color{black} a continuous} SRS, {\color{black} meaning that there is no normalized combination of Pauli matrices $\mathbf{n}\cdot\bs{\sigma}\equiv {\sigma}_\mathbf{n}$ such that
\begin{subequations}\label{eqn:spin-rot-Paulis}
\begin{eqnarray}
\Xi_{\bs{k}}\sigma_\mathbf{n} - \sigma_\mathbf{n}\Xi_{\bs{k}} &=& 0 \\
\Delta_{\bs{k}}\sigma_\mathbf{n}^\top + \sigma_\mathbf{n}\Delta_{\bs{k}} &=& 0.
\end{eqnarray}
\end{subequations}
simultaneously for all momenta in BZ. Note that SRS can be removed either by the presence of SOC in $\Xi_{\bs{k}}$ or by the electron pairing encoded in $\Delta_{\bs{k}}$}. If {\color{black} such a} system \emph{breaks} TRS, the only AZ symmetry is $\mcP_\textrm{t}^2=+\unit$ which corresponds to $\textrm{AZ}$ class $\mathrm{D}$. Depending on the inversion-parity of $\Delta_{\bs{k}}$ in Eq.~(\ref{eqn:SC-INV-gen}) we obtain $\mfP^2=\pm\mcP_\textrm{t}^2=\pm\unit$. The positive sign ($\mcI$-even $\Delta_{\bs{k}}$) leads to $\textrm{AZ}$+$\mcI$ class $\textrm{D}$ while the negative sign ($\mcI$-odd $\Delta_{\bs{k}}$) moves us to $\textrm{AZ}$+$\mcI$ class $\textrm{C}$. On the other hand, a system \emph{respecting} TRS has additional 
\begin{equation}
\mcT =\imi\sigma_y \mcK
\end{equation}
squaring to $-\unit$, and belongs to $\textrm{AZ}$ class $\textrm{DIII}$. The corresponding $\textrm{AZ}$+$\mcI$ class for $\mcI$-even $\Delta_{\bs{k}}$ remains $\textrm{DIII}$, while $\mcI$-odd $\Delta_{\bs{k}}$ leads to $\textrm{AZ}$+$\mcI$ class $\textrm{CII}$. Note that {\color{black} in the absence of SOC} such realizations of $\textrm{AZ}$+$\mcI$ classes $\textrm{D}$ and $\textrm{DIII}$ are only possible in multi-orbital SCs, because $\mcI$-even $\Delta_{\bs{k}}$ in single-orbital case indicates a pure singlet state having the full $\textsf{SU}(2)$ SRS, which contradicts the original assumption.

We further consider SCs with $\textsf{U}(1)$ SRS {\color{black} which means that there is (up to the overall sign) a single matrix $\sigma_\mathbf{n}$ fulfilling Eqs.~(\ref{eqn:spin-rot-Paulis}) for all $\bs{k}\in\textrm{BZ}$}. As shown in Ref.~\cite{Schnyder:2008}, their Hamiltonians (\ref{eqn:BdG-4x4}) canonically decouple into two blocks, rendering the original description redundant. If we rotate the coordinates such that the conserved spin component is $\sigma_z$, then one of the blocks becomes
\begin{subequations}\label{subeqn:BdG-2x2-big}
\begin{equation}
H_{1/2}(\bs{k}) = \frac{1}{2}\left(\begin{array}{cc}
c^{\alpha\dagger}_{\bs{k}\uparrow} 				& 	c^{\alpha\pdag}_{-\bs{k}\downarrow}
\end{array}\right)\mcH^{\alpha\beta}_{1/2}(\bs{k})\left(\begin{array}{c}
c^{\beta\pdag}_{\bs{k}\uparrow}					\\	c^{\beta\dagger}_{-\bs{k}\downarrow}
\end{array}\right)
\end{equation}
where $\alpha,\beta$ stand for the orbital degree of freedom, and
\begin{equation}
\mcH^{\alpha\beta}_{1/2}(\bs{k}) = \left(\begin{array}{cc}
\Xi_{\bs{k},\uparrow\uparrow}^{\alpha\beta}			& \Delta_{\bs{k}\uparrow\downarrow}^{\alpha\beta}	\\
-\Delta_{-\bs{k}\downarrow\uparrow}^{\alpha\beta*}	& -\Xi_{-\bs{k}\downarrow\downarrow}^{\beta\alpha}
\end{array}\right)\label{eqn:BdG-2x2}
\end{equation}
\end{subequations}
is the \emph{reduced} BdG Hamiltonian. Operator $\mcP_\textrm{t}$ of Eq.~(\ref{eqn:PHS-triplet}) relates the two blocks, and in the \emph{absence} of TRS the single block~(\ref{eqn:BdG-2x2}) contains \emph{none} of symmetries (\ref{eqn:sub-global}). We thus end up in $\textrm{AZ}$ class $\textrm{A}$ which corresponds to the same AZ+$\mcI$ class. The additional \emph{presence} of TRS manifests within a single block as chiral symmetry $\mcC = \varsigma_y$ where Pauli matrices $\varsigma_i$ correspond to the particle-hole degree of freedom of Eqs.~(\ref{subeqn:BdG-2x2-big}). Such SCs correspond to $\textrm{AZ}$ and $\textrm{AZ}$+$\mcI$ class $\textrm{AIII}$, {\color{black} regardless of the parity of $\Delta_{\bs{k}}$}.

\begin{table}[t]
	\begin{tabular}{|x{1.25cm}|x{6.5cm}|}
	\hline
	symbol					& corresponding two-level degree of freedom														\\ \hline\hline
	$\sigma$				& spin ($\uparrow\,\leftrightarrow \, \downarrow$)							\\ \hline
	$s$						& particle-hole	($\textrm{p}\!\uparrow\,\,\leftrightarrow \,\textrm{h}\!\uparrow$)			\\ \hline
	$\varsigma$				& reduced particle-hole	($\textrm{p}\!\uparrow\,\,\leftrightarrow \,\textrm{h}\!\downarrow$)	\\ \hline 
	\multirow{2}{*}{$\tau$}	& orbital																\\
							& [same color \emph{inter}layer ($\textrm{A}\leftrightarrow \textrm{B}$) in Fig.~\ref{fig:SrPtAs-lattice}] 
							\\ \hline 
	\multirow{2}{*}{$r$}	& orbital														\\ 
							& (\emph{intra}layer $\textrm{red}\leftrightarrow \textrm{blue}$ in Fig.~\ref{fig:SrPtAs-lattice}) 
							\\ \hline
	\multirow{2}{*}{$\rho$}	& orbital														\\
							& (\emph{inter}layer $\textrm{red}\leftrightarrow \textrm{blue}$ in Fig.~\ref{fig:SrPtAs-lattice}) 
							\\ \hline	
	\end{tabular}
	\caption{Overview of the sets of Pauli matrices used in the manuscript. The last two rows are only relevant for Secs.~\ref{sec:nodes-BDI} to~\ref{sec:nodes-AI} when developing example TB models on a $\textrm{SrPtAs}$-like lattice. We use the words ``orbital'' and ``sublattice'' interchangeably if the orbitals reside at different lattice sites.}
	\label{tab:Paulis}
\end{table}

Finally, we tackle SCs with the complete $\textsf{SU}(2)$ SRS, {\color{black} meaning that Eqs.~(\ref{eqn:spin-rot-Paulis}) are fulfilled for every $\mathbf{n}$ and $\bs{k}$}. These are automatically pure singlet SCs {\color{black} with no SOC}. In this case, the block~(\ref{eqn:BdG-2x2}) develops particle-hole symmetry~\cite{Schnyder:2008} 
\begin{equation}
\mcP_\textrm{s} = \imi \varsigma_y \mcK \label{eqn:PHS-singlet}
\end{equation}
squaring to $-\unit$. In the \emph{absence} of TRS this corresponds to $\textrm{AZ}$ class $\textrm{C}$. Depending on the inversion-parity of $\Delta_{\bs{k}}$ in Eq.~(\ref{eqn:SC-INV-gen}), the inversion operator induced into the block~(\ref{eqn:BdG-2x2}) from $\mcI_\textrm{BdG}$ is 
\begin{equation}
\mcI_{1/2} = \mcI_\tau \oplus(\pm \mcI_\tau) \in \left\{\mcI_\tau\otimes \unit_\varsigma,\mcI_\tau\otimes \varsigma_z \right\}\label{eqn:INV-singlet-block}.
\end{equation}
The ``$+$'' sign of Eq.~(\ref{eqn:INV-singlet-block}) corresponds to $\mcI$-even $\Delta_{\bs{k}}$, when $\mcP_\textrm{s}$ and $\mcI_{1/2}$ commute and the $\textrm{AZ}$+$\mcI$ class remains $\textrm{C}$. This is the case of single-orbital singlet SCs, and of multi-orbital SC in which the three $d_{\bs{k}}^{iz}$ terms in Eq.~(\ref{eqn:SrPtAs}) vanish. On the other hand, the ``$-$'' sign relates to the $\mcI$-odd case $\Delta_{\bs{k}}$ with $\mfP^2 = -\mcP_\textrm{s}^2 = +\unit$ such that the $\textrm{AZ}$+$\mcI$ class switches to $\textrm{D}$. This situation corresponds to a very fine-tuned multi-orbital case when the nine $d_{\bs{k}}^{ij}$ of Eq.~(\ref{eqn:I-odd-Delta}) vanish, and as such may be relevant only to certain artificially engineered systems. Ultimately, if the singlet SC \emph{preserves} TRS, the block (\ref{eqn:BdG-2x2}) develops $\mcT_\textrm{s} = \mcK$ squaring to $+\unit$ leading to $\textrm{AZ}$ class $\textrm{CI}$. Under such circumstances, the $\mcI$-even $\Delta_{\bs{k}}$ case belongs to AZ+$\mcI$ class $\textrm{CI}$, while the $\mcI$-odd $\Delta_{\bs{k}}$ case corresponds to AZ+$\mcI$ class $\textrm{BDI}$.

{\color{black} We remark that the relevant AZ+$\mcI$ class of single-orbital singlet SCs is susceptible to the inclusion of SOC in $\Xi_{\bs{k}}$. In the presence of TRS, the gradual decrease of SRS due to the presence of SOC corresponds to AZ+$\mcI$ class evolution $\textrm{CI}\to\textrm{AIII}\to\textrm{DIII}$. While all three classes in 3D exhibit nodal lines, only class $\textrm{CI}$ nodal lines (corresponding to \emph{no} SOC) are according to Tab.~\ref{tab:Homotopies} characterized by a \emph{pair} of charges. On the other hand, in the absence of TRS the inclusion of SOC leads to AZ+$\mcI$ class evolution $\textrm{C}\to\textrm{A}\to\textrm{D}$. In 3D this implies that the removal of a continuous SOC inflates the Weyl points into doubly-charged nodal surfaces~\cite{Agterberg:2017}.}

{\color{black}  As a guide for the reader, we provide in Tab.~\ref{tab:SCmaterials} a list of some well-known examples of exotic superconductors with their relevant AZ+$\mcI$ classes. In several cases, there is no consensus on the symmetry of the SC order parameter, hence in the table we explicitly assume an order parameter suggested by the provided references. The focus of this manuscript, however, is not the classification of existing materials, but a study of doubly charged nodes permitted in 3D. To meet this goal, the subsequent sections focus on the realizations of AZ+$\mcI$ classes indicated with green background in Fig.~\ref{fig:AZI-classification}.}


\section{$\ztwo\oplus 2\intg$ nodal surfaces in class $\mathrm{D}$}\label{sec:nodes-D}

In the remainder of the manuscript we individually discuss each of the four AZ+$\mcI$ symmetry classes supporting doubly charged nodes in $D=3$. As shown in Tabs.~\ref{tab:Homotopies-small} and~\ref{tab:Homotopies-small-AI}, the minimal half-filled model {\color{black} capable to realize} doubly charged nodes always contains four bands, hence we begin each section by introducing the most general four-band Hamiltonian compatible with the AZ+$\mcI$ symmetries~(\ref{subeqn:AZ+I-symmetries}). We continue with an explanation of the two topological charges, and conclude by constructing a concrete TB model on a $\textrm{SrPtAs}$-like lattice.

The present section focuses on AZ+$\mcI$ class $\textrm{D}$. According to Tabs.~\ref{tab:Homotopies} and~\ref{tab:Hamiltonians}, its zero-energy nodes take the form of surfaces characterized by a pair of topological charges
\begin{equation}
c_\textrm{D}^{(3)} \in \pi_0(M_\textrm{D})\oplus\pi_2(M_\textrm{D}) = \ztwo\oplus 2\intg.
\end{equation}
We show that the $\pi_0$-charge can be understood as the sign of the Pfaffian of the Hamiltonian and the $\pi_2$-charge as the (first) Chern number, as is indicated in the rightmost block of Tab.~\ref{tab:Homotopies}. In the last subsection we discuss how such nodal surfaces naturally appear in TRS breaking multi-orbital superconductors.


\subsection{General $4$-band Hamiltonian}\label{subsec:D-Ham}

The six $4\times 4$ matrices compatible with $\mfP = \mcK$ which appear in Tab.~\ref{tab:Hamiltonians} can be arranged into a pair of vectors
\begin{subequations}\label{eqn:D-4matrices}
\begin{eqnarray}
\bs{v} &=& \left(\sigma_x\otimes \tau_y,\sigma_y\otimes\unit,\sigma_z\otimes\tau_y\right) \\
\bs{w} &=& \left(\sigma_y\otimes\tau_x,\unit\otimes\tau_y,\sigma_y\otimes\tau_z\right)
\end{eqnarray}
\end{subequations}
such that $\{v_i,v_j\}=\{w_i,w_j\}=2\delta_{ij}$ and $[v_i,w_j]=0$. The most general four-band Hamiltonian of this class is
\begin{subequations}
\begin{equation}
\mcH(\bs{k}) = \bs{a}(\bs{k})\cdot\bs{v}+\bs{b}(\bs{k})\cdot\bs{w}\label{eqn:Ham-D-class}
\end{equation}
where $\bs{a}(\bs{k})$ and $\bs{b}(\bs{k})$ are real-valued vector functions. The spectrum of (\ref{eqn:Ham-D-class}) is easily found to be
\begin{equation}
\varepsilon(\bs{k}) = \pm\lVert{\bs{a}(\bs{k})}\rVert \pm\lVert{\bs{b}(\bs{k})}\rVert.\label{eqn:eps-D-class}
\end{equation}
\end{subequations}
The gap closes whenever $\lVert{\bs{a}(\bs{k})}\rVert=\lVert{\bs{b}(\bs{k})}\rVert$. This is a single condition, manifesting the codimension $\delta_\textrm{D}=1$ listed in Tab.~\ref{tab:Hamiltonians}.

Although the $4\times 4$ matrices $v_i$ (as well as $w_i$) are anticommuting, they do \emph{not} form a Dirac basis because their algebra is closed under commutator, $[v_i, v_j] = 2 \imi \epsilon_{ijk} v_k$. Following Ref.~\cite{Zhao:2016}, we refer to it as a \emph{double Weyl basis}. An important characteristic is that a double Weyl Hamiltonian $\mcH_{\textrm{d.W.}}\propto \bs{k}\cdot\bs{v}$ describes two superimposed Weyl points of the \emph{same} chirality (which can be split using terms $\propto \!w_i$).~\footnote{{\color{black} We remark that our definition of a ``double Weyl point'' in a SC differs from that of Ref.~\cite{schnyder2015topological}}.} This is to be contrasted with a Dirac Hamiltonian $\mcH_\textrm{D.}\propto \bs{k}\cdot\bs{\Gamma}$ which creates two superimposed Weyl points of \emph{opposite} chirality (which can be split using a term $\propto \!\Gamma_{45}$). Consequently, a double Weyl point is a source of two Berry phase quanta, while the Chern number of a Dirac point vanishes.


\subsection{Interpretation of $\pi_0(M_\mathrm{D})=\ztwo$}\label{subsec:D-charge-0}

The six matrices (\ref{eqn:D-4matrices}) are antisymmetric and imaginary. The hermiticity of $\mcH(\bs{k})$ along with $\mfP = \mcK$ entail
\begin{equation}
\mcH^\top(\bs{k}) = \mcH^*(\bs{k}) = \mfP \mcH(\bs{k})\mfP^{-1} = -\mcH(\bs{k}).
\end{equation}
As a consequence, $\imi\mcH(\bs{k})$ is a skew-symmetric and even-dimensional matrix with real entries. One can therefore construct a non-vanishing Pfaffian $\Pf[\imi\mcH(\bs{k})]$ which is a real-valued function of $\bs{k}$. For (\ref{eqn:Ham-D-class}) specifically
\begin{equation}
\Pf[\imi\mcH(\bs{k})] = \bs{b}^2(\bs{k}) - \bs{a}^2(\bs{k})
\end{equation}
which changes sign at the nodal surface. 

This observation generalizes to an arbitrary class $\textrm{D}$ model: The presence of a node at $\bs{k}_0$ is revealed by a pair of zero energy states. This enforces $\det[\mcH(\bs{k})]=\prod_a \varepsilon^a(\bs{k})$ to vanish at $\bs{k}_0$ and to depend quadratically on $(\bs{k}-\bs{k}_0)$. By the identity
\begin{equation}
\det[\mcH(\bs{k})] = \Pf[\mcH(\bs{k})]^2 = (-1)^n \Pf[\imi\mcH(\bs{k})]^2\label{eqn:D-det-Pf-rel},
\end{equation}
the Pfaffian \emph{also} vanishes at $\bs{k}_0$, but it varies \emph{linearly} with $(\bs{k}-\bs{k}_0)$. This implies that $\Pf[\imi\mcH(\bs{k})]$ has a different sign at two points $\{\bs{k}_1,\bs{k}_2\}\cong S^0$ located on the opposite sides of the nodal surface. We can therefore formulate the zeroth homotopy charge as
\begin{equation}
c_\textrm{D}(S^0) = \sign\left\{\,\prod_{\bs{k}\in S^0} \Pf \left[\imi\mcH(\bs{k})\right]\,\right\} \in\left\{+1,-1\right\}.\label{eqn:D-pi0}
\end{equation}
This charge has been recently discussed in work~\cite{Agterberg:2017}.


\subsection{Interpretation of $\pi_2(M_\mathrm{D})=2\intg$}\label{subsec:D-pi2-charge}

A natural candidate for an integer topological charge on a $2$-sphere is the (first) Chern number. We show below that this is indeed the case here, and we explain why the particle-hole symmetry enforces it to be even.

The Chern number is formulated as~\cite{Ryu:2010}
\begin{equation}
c_\textrm{D}(S^2) = \frac{\imi}{2\pi}\oint_{S^2} \de^2 \bs{k}\cdot \tr \mcbsF(\bs{k})\in \intg \label{eqn:D-pi2-charge}
\end{equation}
where $\mcbsF(\bs{k})$ is the Berry curvature determined from 
\begin{subequations}
\begin{eqnarray}
\mcbsF &=& \bs{\nabla}_{\bs{k}}\times\mcbsA + \mcbsA\times\mcbsA \label{eqn:Hodge-dual-F} \\
\mathcal{A}^{ab}_i(\bs{k}) &=& \bra{u^a(\bs{k})}\partial_i\ket{u^b(\bs{k})} \label{eqn:Berry-connection-def}
\end{eqnarray}
\end{subequations}
where $\mcbsA(\bs{k})$ is the Berry-Wilczek-Zee (BWZ) connection~\cite{Berry:1984,Wilczek:1984}, $a,b$ label the \emph{occupied} bands, and $\partial_i \equiv \partial/\partial_{k_i}$. For non-degenerate bands, the second term in Eq.~(\ref{eqn:Hodge-dual-F}) vanishes after taking the trace such that we can decompose the integrand of Eq.~(\ref{eqn:D-pi2-charge}) into the contributions of individual bands,
\begin{equation}
\hspace{-0.1cm}\tr \mcbsF(\bs{k}) = \sum_{a} \bra{\bs{\nabla}_{\bs{k}} u^a(\bs{k})} \!\times\! \ket{\bs{\nabla}_{\bs{k}} u^a(\bs{k})} \equiv \sum_a \bs{F}^a(\bs{k}).\label{eqn:Curvature-decompo}
\end{equation}
One can similarly determine the Berry curvature and the Chern number of the \emph{unoccupied} bands.

The curvature of an occupied band $\ket{u^a(\bs{k})}$ and the curvature of a particle-hole related unoccupied band $\ket{\widetilde{u}^a(\bs{k})} = \mfP \ket{u^a(\bs{k})} {\color{black}= \ket{\widetilde{u}^a(\bs{k})}^*}$ differ only in the overall sign. To see this, first note that $\mcbsA(\bs{k})$ is skew-Hermitian,
\begin{equation}
0 = \bs{\nabla}_{\bs{k}}\delta^{ab} = \bs{\nabla}_{\bs{k}}\braket{u^a(\bs{k})}{u^b(\bs{k})} = \mcbsA^{ab}+(\mcbsA^{ba})^*,
\end{equation}
such that its diagonal terms are imaginary. Taking the curl in Eq.~(\ref{eqn:Hodge-dual-F}) preserves this property, hence the band curvatures in Eq.~(\ref{eqn:Curvature-decompo}) are imaginary too. It follows that
\begin{eqnarray}
\widetilde{\bs{F}}_a(\bs{k}) &=& \bra{\bs{\nabla}_{\bs{k}} \widetilde{u}^a(\bs{k})} \times \ket{\bs{\nabla}_{\bs{k}} \widetilde{u}^a(\bs{k})} \nonumber\\
&=& \left[\bra{\bs{\nabla}_{\bs{k}} u^a(\bs{k})} \times \ket{\bs{\nabla}_{\bs{k}} u^a(\bs{k})}\right]^* = -\bs{F}_a(\bs{k})\label{eqn:D-curvature-rel}
\end{eqnarray}
as we wanted to show.
\begin{figure}[t]
	\includegraphics[width=0.25\textwidth]{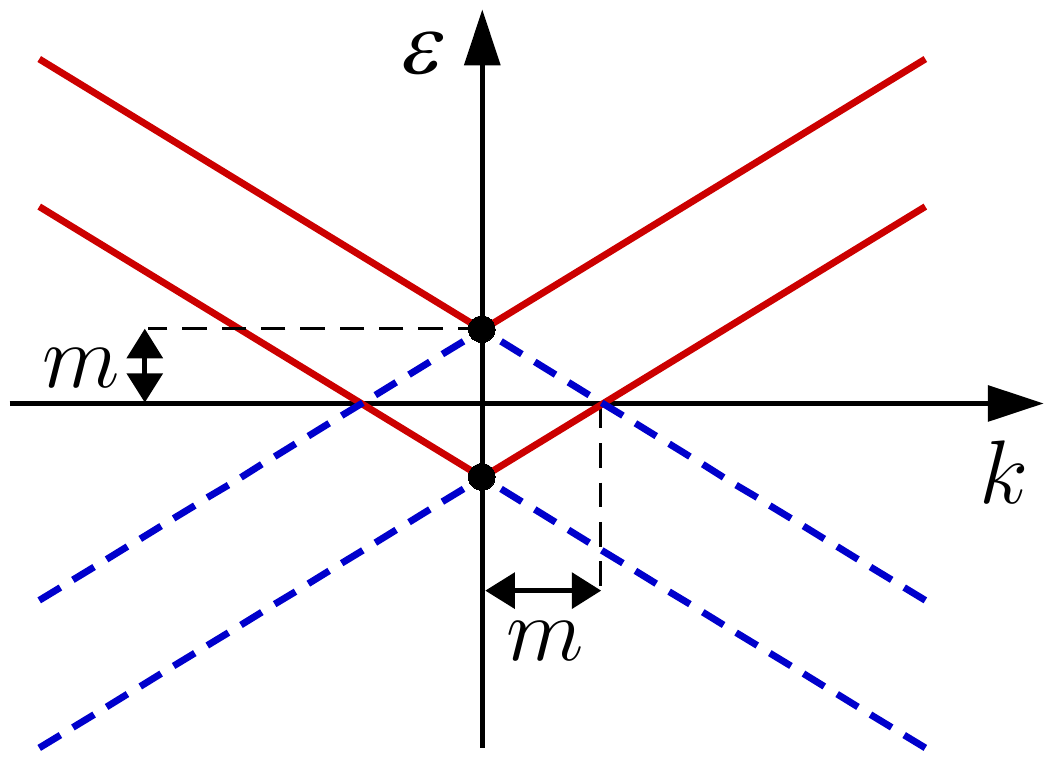}
	\caption{Spectrum of Hamiltonian (\ref{eqn:D-two-Weyls}) consists of two Weyl points (black dots) of the same chirality with energy offset $\pm m$, crossing on a nodal sphere with radius $\abs{m}$. The two species of lines indicate Berry curvatures of opposite signs, which integrate to opposite Chern numbers $\pm 1$. The total Chern number over the occupied states is zero inside and equals $\pm 2$ outside the nodal sphere, meaning that the nodal surface is a source of a pair of Berry phase quanta.}
	\label{fig:D-pi2}
\end{figure} 

Relation (\ref{eqn:D-curvature-rel}) readily explains why the Chern number (\ref{eqn:D-pi2-charge}) takes even values: Consider a nodal surface $S^2_\textrm{node}$ along with a surface $S^2_\textrm{in}$ inside of it and a surface $S^2_\textrm{out}$ enclosing it. Since $S^2_\textrm{in}$ can be trivially shrunk to a point without encountering a node, $c_\mathrm{D}(S^2_\textrm{in})=0$. The node $S^2_\textrm{node}$ is created by switching one occupied and one empty band, which by Eq.~(\ref{eqn:D-curvature-rel}) carry opposite Berry curvatures, which integrate to opposite Chern numbers $c$ and $-c$. Consequently, the charge $c_\textrm{D}(S^2_\textrm{out}) = c_\textrm{D}(S^2_\textrm{in}) + c - (-c) = 2c$ is indeed even. These considerations give the nodal surface a peculiar interpretation as an \emph{inflated double Weyl point}~\cite{Agterberg:2017}. The reason is that $S^2_\textrm{node}$ is a source of an even number of Berry phase quanta, reminiscent of double Weyl points. This is consistent with our discussion of the double Weyl basis in Subsec.~\ref{subsec:D-Ham}. Work~\cite{Agterberg:2017} also dubbed these objects in the context of SC as \emph{Bogolyubov Fermi surfaces}.

We demonstrate these observations by setting $\bs{a}(\bs{k}) = \left(0,m,0\right)$ and $\bs{b}(\bs{k}) =\bs{k}$ in the general $4$-band model~(\ref{eqn:Ham-D-class}). This produces a nodal sphere with radius $\abs{m}$. We further rotate the basis as $\sigma_x\mapsto \sigma_y\mapsto\sigma_z \mapsto\sigma_x$ using $\mcU_\sigma$, where
\begin{equation}
\mcU = \frac{1}{\sqrt{2}}\left(\begin{array}{cc}
1		&	-\imi	\\
1		&	\imi
\end{array}\right).\label{eqn:Pauli-forward-rotate}
\end{equation}
This is accompanied by a transformation $\mfP=\mcK \mapsto \sigma_x \mcK$. Then the Hamiltonian decouples into two blocks, $\mcH(\bs{k})=\mcH_1(\bs{k})\oplus\mcH_2(\bs{k})$, with
\begin{subequations} \label{eqn:D-two-Weyls}
\begin{eqnarray}
\mcH_1(\bs{k}) &=& + m\unit_\tau + k_x\tau_x + k_y \tau_y + k_z \tau_z \\
\mcH_2(\bs{k}) &=& -m\unit_\tau - k_x\tau_x + k_y \tau_y - k_z \tau_z. 
\end{eqnarray} 
\end{subequations}
Clearly, this is a pair of Weyl points of the \emph{same} chirality with energy offset $\pm m$, as illustrated in Fig.~\ref{fig:D-pi2}. We observe that indeed $c_\mathrm{D}(S^2_\textrm{in})=0$ and $\abs{c_\textrm{D}(S^2_\textrm{out})} =2$.


\subsection{Example class $\textrm{D}$ model}\label{subsec:D-model}
\begin{figure}[t]
	\includegraphics[width=0.46\textwidth]{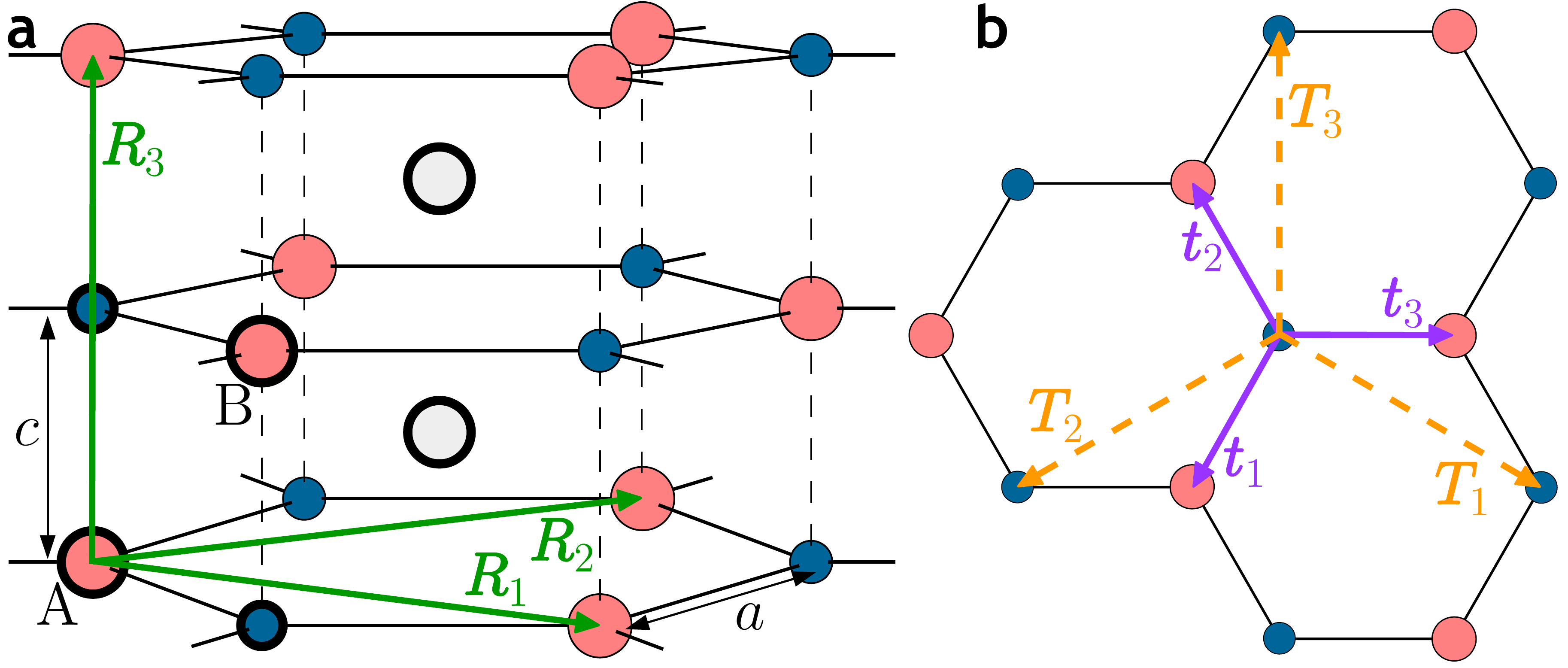}
	\caption{(a) Crystalline structure of $\textrm{SrPtAs}$ consists of three elements: Strontium (Sr -- large, bright gray), platinum (Pt -- large, light red), and arsenic (As -- small, dark blue). The TB model of the present Sec.~\ref{sec:nodes-D} considers $s$-like orbitals located at the $\textrm{Pt}$ sites, while the TB models of the subsequent Secs.~\ref{sec:nodes-BDI} to~\ref{sec:nodes-AI} capture $s$-like orbitals at \emph{both} the $\textrm{Pt}$ and the $\textrm{As}$ sites. The thick green arrows $\bs{R}_{1,2,3}$ indicate Bravais vectors~(\ref{eqn:SrPtAs-Bravais}), while the thin arrows $a,c$ indicate unit cell dimensions. Six atoms ($2\times\textrm{Sr},2\times\textrm{Pt},2\times\textrm{As}$) belonging to a single Bravais vector are highlighted with black thick circles in the bottom left part of the panel. (b) Top view of a single hexagonal layer of $\textrm{SrPtAs}$. The triplets of vectors ${\color{black} \mathbf{t}=\{\bs{t}_n\}_{n=1}^3}$ (solid purple) and ${\color{black} \mathbf{T}=\{\bs{T}_n\}_{n=1}^3}$ (dashed orange) are used to achieve a compact formulation of the TB Hamiltonians.}
	\label{fig:SrPtAs-lattice}
\end{figure}

We provide the reader with a simple way of testing the presented ideas by introducing a TB model on a $\textrm{SrPtAs}$-like lattice adapted from Ref.~\cite{Fischer:2015}. This {\color{black}reference} observed the formation of nodal surfaces in a multi-orbital $d\pm \imi d$ SC phase~\cite{Fischer:2014} with an admixed $p$-wave component, {\color{black} corresponding} to AZ+$\mcI$ class $\textrm{D}$. As illustrated in Fig.~\ref{fig:SrPtAs-lattice}(a), the centrosymmetric structure of $\textrm{SrPtAs}$ consists of graphene-like layers with platinum ($\textrm{Pt}$) and arsenic ($\textrm{As}$) atoms on the two sublattices, and of intercalated strontium ($\textrm{Sr}$) atoms. Our goal here is not to provide a realistic description of $\textrm{SrPtAs}$, but to demonstrate the possible appearance of the doubly charged nodal surfaces in this lattice for the properly set values of the TB parameters.

We assume that $s$-like orbitals located at the Pt sites enter the TB model. The primitive Bravais vectors are
\begin{subequations}
\begin{equation}
\bs{R}_{1,2} = \tfrac{3a}{2}\left(1,\mp\tfrac{1}{\sqrt{3}},0\right) \quad \textrm{and}\quad \bs{R}_3 = \left(0,0,2c\right)\label{eqn:SrPtAs-Bravais}
\end{equation}
and the positions of the two orbitals corresponding to a given Bravais vectors are
\begin{equation}
\bs{r}_\textrm{A} = \left(0,0,0\right) \quad\textrm{and} \quad \bs{r}_\textrm{B} = \left(\tfrac{a}{2},-\tfrac{a\sqrt{3}}{2},c\right).
\end{equation}
To write the Hamiltonian compactly, we use vectors
\begin{equation}
 \bs{t}_{1,2} =a\left(-\tfrac{1}{2},\mp \tfrac{\sqrt{3}}{2},0\right) \quad\textrm{and}\quad \bs{t}_3 = a\left(1,0,0\right)
\end{equation}
of Fig.~\ref{fig:SrPtAs-lattice}(b), and their differences $\bs{T}_i \!=\! \tfrac{1}{2}\!\sum_{jk}\!\epsilon_{ijk}\!\left(\bs{t}_k\!-\!\bs{t}_j\right)$. We further define $\omega_n = \e{\imi 2\pi n/3}\in\mathbb{C}$, and functionals
\begin{equation}
\mcS^{p,\mathbf{v}}_{f}(\bs{k}) = \sum_{n=1}^3 \left(\omega_n\right)^p f\left(\bs{k}\cdot\bs{v}_n\right)
\end{equation}
\end{subequations}
where $f$ denotes a function on $\mathbb{R}$, the first superscript $p\in\mathbb{Z}$, and the second superscript indicates a set of three vectors, $\mathbf{v}=\{\bs{v}_n\}_{n=1}^3$.  

The NN intra-layer hopping with amplitude $t_0/2$ and the intra-orbital hopping across two layers with amplitude $t_z'/2$ produce
\begin{subequations}\label{eqn:D-TB-Ham}
\begin{equation}
\mcH_1(\bs{k}) = \left[t_0\mcS^{0,\mathbf{T}}_{\cos}(\bs{k})+t_z'\cos(2k_z c)\right] \unit_\sigma \otimes \unit_\tau\label{eqn:SrPtAs-H1},
\end{equation}
while the inter-layer hopping with amplitude $t_z/2$ gives
\begin{equation}
\hspace{-0.1cm}\mcH_2(\bs{k}) = t_z \cos(k_z c)\unit_\sigma\otimes \left[\mcS^{0,\mathbf{t}}_{\cos}(\bs{k}) \tau_x +\mcS^{0,\mathbf{t}}_{\sin}(\bs{k})\tau_y\right]\!.\label{eqn:SrPtAs-H2}
\end{equation}
We further include the intra-layer SOC term
\begin{equation}
\mcH_3(\bs{k}) = \alpha_\textrm{so} \mcS^{0,\mathbf{T}}_{\sin}(\bs{k}) \sigma_z \otimes \tau_z.
\end{equation}
\end{subequations}
which reduces the SRS of $\mcH(\bs{k}) {\color{black} = \sum_i \mcH_i(\bs{k})}$ {\color{black} (which corresponds to $\Xi_{\bs{k}}$ in the notation of Subsec.~\ref{subsec:superconductivity})} to $\mathsf{U}(1)$. The meaning of Pauli matrices $\sigma_i,\tau_i$ follows Tab.~\ref{tab:Paulis}. 

We set the parameters of TB model~(\ref{eqn:D-TB-Ham}) to
\begin{equation}
\hspace{-0.3cm}t\!=\!1,\;\; t_z\!=\!-1,\;\; t_z'\!=\!1, \;\; \alpha_\textrm{so}\!=\!-1\;\;\textrm{and}\;\; \mu\!=\!-3.5 \label{eqn:D-model-params}
\end{equation}
where $\mu$ is the chemical potential. Then the Fermi surface consists of a pair of pockets centered on the corners of BZ as shown in Fig.~\ref{fig:SrPtAs-FS}(a). We further let the system develop a SC $d + \imi d$ order parameter
\begin{subequations}\label{eqn:D-example-gap-both}
\begin{equation}
{\Delta}^{d}_{\bs{k}}=\psi^0 \mcS^{1,\mathbf{T}}_{\cos}(\bs{k})(\imi\sigma_y)\otimes \unit_\tau \label{eqn:D-example-gap1}
\end{equation}
which is $\mcI$-even (more exactly, it belongs to the $E_{2\textrm{g}}$ representation of the $D_\textrm{6h}$ point group~\cite{Fischer:2015}). Order parameter~(\ref{eqn:D-example-gap1}) vanishes along the vertical edges of the BZ, meaning that nodal points are formed where these edges cross the Fermi pockets. This is compatible with the scheme in Fig.~\ref{fig:AZI-classification}(c): The complexity of the order parameter breaks TRS, and the combined Eqs.~(\ref{eqn:D-TB-Ham}) and (\ref{eqn:D-example-gap1}) preserve $\mathsf{U}(1)$ SRS. This locates the system in AZ+$\mcI$ class $\textrm{A}$, which according to Tab~\ref{tab:Hamiltonians} in $D=3$ indeed exhibits nodal points. In fact, because of the underlying spin degeneracy of the bands [i.e. due to the block reduction~(\ref{subeqn:BdG-2x2-big}) of the BdG Hamiltonian], these are precisely the previously discussed double Weyl points.
\begin{figure}[t]
	\includegraphics[width=0.48\textwidth]{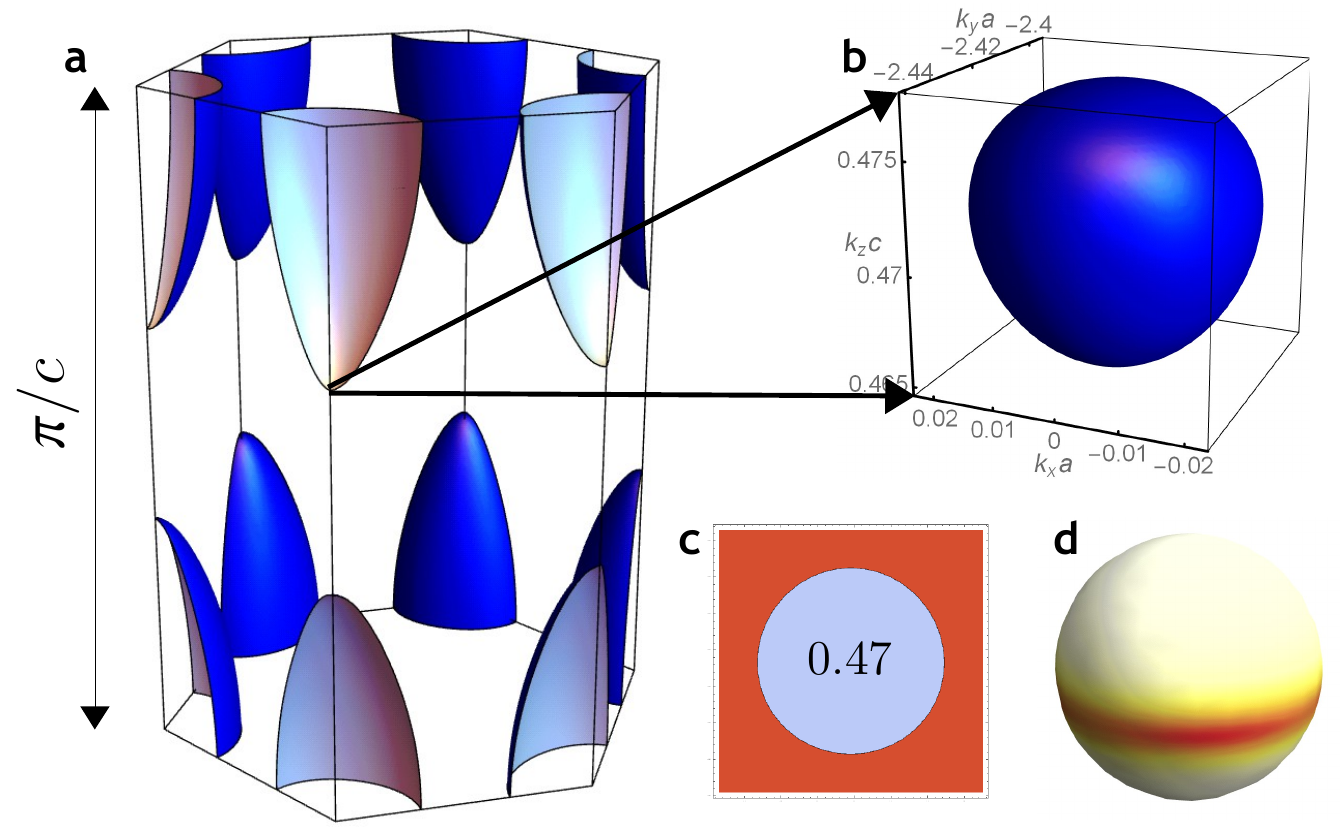}
	\caption{Doubly charged nodal surfaces of AZ+$\mcI$ class $\textrm{D}$. (a) Fermi surface of TB model~(\ref{eqn:D-TB-Ham}) with parameters~(\ref{eqn:D-model-params}) consists of a pair of pockets centered on the corners of BZ. (b) The $d + \imi d$ order parameter~(\ref{eqn:D-example-gap1}) creates double Weyl points where the vertical BZ edges cross the Fermi pockets. Admixing a $p$-wave order parameter~(\ref{eqn:D-example-gap2}) inflates them into tiny nodal surfaces. (c) Plot of $\sign\Pf\left[\imi \mcH(\bs{k})\right]$ for a horizontal slice through the nodal surface at the indicated value of $k_z c$. (d) Flow of the Berry curvature across an ellipsoid described in the text, which encloses the nodal surface. Integration reveals that the nodal surface is a source of two Berry phase quanta.}
	\label{fig:SrPtAs-FS}
\end{figure}

However, the multi-orbital character of the $\textrm{SrPtAs}$ lattice allows us to admix a $p$-wave order parameter~\cite{Fischer:2015}
\begin{equation}
{\Delta}^{p}_{\bs{k}}=d^{-z} \sin(2k_z c)\left(\sigma_x - \imi \sigma_y\right)(\imi \sigma_y)\otimes \tau_z \label{eqn:D-example-gap2},
\end{equation}
\end{subequations}
which preserves the even parity of $\Delta_{\bs{k}}$ (as well as the $E_{2\textrm{g}}$ representation) and that does \emph{not} vanish along the vertical BZ edges. Importantly,~(\ref{eqn:D-example-gap2}) breaks SRS altogether, such that the system is shifted to AZ+$\mcI$ class $\textrm{D}$ exhibiting doubly charged nodal surfaces. To check this, we set 
\begin{equation}
\psi^z = 0.2\qquad \textrm{and}\qquad d^{-z}=0.2
\end{equation} 
and find tiny nodal surfaces at the expected position, plotted in Fig.~\ref{fig:SrPtAs-FS}(b). Rotating the basis by $\mcU^\dagger_s\otimes \unit_\sigma \otimes \mcU^\dagger_\tau$ with matrix $\mcU$ from Eq.~(\ref{eqn:Pauli-forward-rotate}) leads to $\mfP=\mcK$, such that we can check the sign of $\Pf\left[\imi \mcH(\bs{k})\right]$ on a plane crossing the nodal surface, plotted in Fig.~\ref{fig:SrPtAs-FS}(c). Finally, we calculate the Berry curvature over the occupied states on an ellipsoid with radii $\frac{1}{10}(\frac{1}{a},\frac{1}{a},\frac{1}{c})$ centered on the nodal surface, plotted in Fig.~\ref{fig:SrPtAs-FS}(d). Numerical integration of~(\ref{eqn:D-pi2-charge}) reveals that indeed $\abs{c_\textrm{D}(S_\textrm{out}^2)}=2$.


\section{$\ztwo\oplus\ztwo$ nodal surfaces in class $\textrm{BDI}$}\label{sec:nodes-BDI}

The AZ+$\mcI$ class $\textrm{BDI}$ is the only one that supports doubly charged nodes already in $D=2$, explicitly
\begin{equation}
c_\textrm{BDI}^{(2,3)}\in \pi_0(M_\textrm{BDI})\oplus\pi_1(M_\textrm{BDI}) = \ztwo\oplus\ztwo.
\end{equation}
Most of our discussion applies equally well to both cases, although all the explicit examples are provided for $D=3$. In the following subsections we first construct the most general $4$-band Hamiltonian of this symmetry class and determine its spectrum. We show that the $\pi_0$-charge is again the Pfaffian invariant~(\ref{eqn:D-pi0}), although an alternative determinant formulation becomes possible too. On the other hand, the $\pi_1$-charge is new and corresponds to the winding of a closed path inside $\mathsf{SO}(n)$. We develop a way to determine this charge by plotting the matrix spectrum along the path. In the last subsection we develop a non-SC model on a lattice consisting of dimerized $\textrm{AAA}$-stacked graphene layers, which belongs to symmetry class $\textrm{BDI}$ and that exhibits {\color{black} doubly charged nodal cylinders. Nodal surfaces belonging to this symmetry class have been previously predicted in certain three-dimensional graphene networks~\cite{zhong2016towards}, although these didn't carry a non-trivial value of the higher $\pi_1$-charge.}


\subsection{General $4$-band Hamiltonian}

We order the four basis matrices of Tab.~\ref{tab:Hamiltonians} into a pair of two-component vectors
\begin{subequations}
\begin{eqnarray}
\bs{v} &=& (\sigma_x\otimes \unit,-\sigma_y \otimes \tau_y) \\
\bs{w} &=& (\sigma_x\otimes\tau_z,-\sigma_x \otimes \tau_x)
\end{eqnarray}
\end{subequations}
fulfilling $\{v_i,v_j\}=\{w_i,w_j\}=2\delta_{ij}$ and $[v_i,w_j]=0$. We encode the general $4$-band Hamiltonian using a pair of real-valued vector functions $\bs{p}=(p_1,p_2),\bs{r}=(r_1,r_2)$ as
\begin{subequations}
\begin{equation}
\mcH(\bs{k}) = \bs{p}(\bs{k})\cdot\bs{v} + \bs{r}(\bs{k})\cdot\bs{w}.\label{eqn:BDI-ham-form}
\end{equation}
Diagonalizing the Hamiltonian reveals the spectrum,
\begin{equation}
\varepsilon(\bs{k}) = \pm\lVert{\bs{p}(\bs{k})}\rVert \pm\lVert{\bs{r}(\bs{k})}\rVert.\label{eqn:BDI-spec}
\end{equation}
\end{subequations}
The gap closes whenever $\lVert{\bs{p}(\bs{k})}\rVert = \lVert{\bs{r}(\bs{k})}\rVert$. Since this is a single scalar constraint, we deduce the codimension $\delta_\textrm{BDI}=1$, thus confirming again that AZ+$\mcI$ class $\textrm{BDI}$ in $D=3$ supports nodal surfaces.


\subsection{Interpretation of $\pi_0(M_\mathrm{BDI}) = \ztwo$}

The presence of $\mcC = \sigma_z$ guarantees a block-off-diagonal form of $\mcH(\bs{k})$, while $\mfT = \mcK$ makes it real. Consequently, 
\begin{equation}
\mcH(\bs{k}) = \left(\begin{array}{cc}
\triv				&	A(\bs{k})			\\
A^\top(\bs{k})		&	\triv
\end{array}\right)\quad\!\!\textrm{with}\;A(\bs{k})\in\textsf{GL}(n,\mathbb{R}).
\end{equation}
For the $n=2$ model~(\ref{eqn:BDI-ham-form}) explicitly
\begin{equation}
A(\bs{k}) = p_1(\bs{k})\unit_\tau - r_2(\bs{k})\tau_x + p_2(\bs{k}) \imi\tau_y + r_1(\bs{k})\tau_z.
\end{equation}
Since $\mfP$ fixes nodes to zero energy, they are exposed by
\begin{equation}
0 = \det \mcH(\bs{k}) = \imi^{2n}[\det A(\bs{k})]^2.
\end{equation}
Nodal surfaces separate regions with opposite sign of $\det A(\bs{k})$. This implies that for $S^0 = \{\bs{k}_1,\bs{k}_2\}$ there is
\begin{equation}
c_{\textrm{BDI}}(S^0) = \sign\left[\prod_{\bs{k}\in S^0} \det A(\bs{k})\right]\in\{+1,-1\} \label{eqn:BDI-charge-0}.
\end{equation}
In fact, this is just the Pfaffian invariant~(\ref{eqn:D-pi0}) in disguise. Rotating the basis by $\mcU_\sigma$ leads to $\mfT = \sigma_x \mcK$, $\mfP = \mcK$ and $\mcC = \sigma_x$, such that the transformed Hamiltonian $\mcU_\sigma \mcH(\bs{k}) \mcU_\sigma^\dagger \equiv \widetilde{\mcH}(\bs{k})$ is antisymmetric and has a well-defined Pfaffian. It follows that
\begin{equation}
\left[\det A(\bs{k})\right]^2 = \Pf[\imi \widetilde{\mcH}(\bs{k})]^2
\end{equation}
such that formula~(\ref{eqn:D-pi0}) is applicable for calculating the charge $c_\textrm{BDI}(S^0)$ too.

{\color{black} Before moving on, we make a passing remark on the generality of the $\ztwo$ Pfaffian invariant~(\ref{eqn:D-pi0}) resp.~(\ref{eqn:BDI-charge-0}) for nodal surfaces occurring in an arbitrary systems with a \emph{spectral symmetry}, by which we mean that for any reason the eigenvalues of $\mcH(\bs{k})$ and $-\mcH(\bs{k})$ coincide. For AZ+$\mcI$ classes $\textrm{D}$ and $\textrm{BDI}$ of the last two sections, such a spectral symmetry was enforced by $\mfP$, but a very recent Ref.~\cite{Turker:2017} also considers systems with nodal surfaces protected by a global $\mathsf{U}(1)$ symmetry corresponding to $\left[\mcH(\bs{k}),\gamma^5\right]$ where $\gamma^5$ is the ``chiral'' Dirac matrix. The spectral symmetry in their example model follows from the additional chiral symmetry~(\ref{eqn:AZIsyms-CHS}) also present in the model.}

{\color{black} The spectral symmetry allows one to diagonalize the Hamiltonian as 
\begin{subequations}
\begin{equation}
\mcH(\bs{k}) = \mathcal{V}_{\bs{k}}\left[\mathcal{E}_{\bs{k}}\oplus(-\mathcal{E}_{\bs{k}})\right]\mathcal{V}_{\bs{k}}^\dagger
\end{equation}
where $\mathcal{E}_{\bs{k}}=\textrm{diag}\left[\varepsilon^1(\bs{k}),\varepsilon^2(\bs{k}),\ldots,\varepsilon^n(\bs{k})\right]$ is a collection of eigenvalues (not necessarily of the same sign) and the columns of unitary matrix $\mathcal{V}_{\bs{k}}$ are the corresponding eigenstates of $\mcH(\bs{k})$. Rotating the basis by $\mcU_s^\dagger \mathcal{V}_{\bs{k}}^\dagger$ brings the Hamiltonian to antisymmetric form
\begin{equation}
\widetilde{\mcH}(\bs{k}) = \mcU_s^\dagger \mathcal{V}_{\bs{k}}^\dagger \mcH(\bs{k}) \mathcal{V}_{\bs{k}} \mcU_s = \left(\begin{array}{cc}
\triv						&	-\imi\mathcal{E}_{\bs{k}}	\\
\imi\mathcal{E}_{\bs{k}}	&	\triv
\end{array}\right)
\end{equation}
such that we can consider the Pfaffian
\begin{equation}
\Pf\left[\imi\widetilde{\mathcal{H}}(\bs{k})\right] = \det[\mathcal{E}_{\bs{k}}] = \prod_{a=1}^n \varepsilon^a(\bs{k}).
\end{equation}
If one enforces the continuity of $\mathcal{V}_{\bs{k}}$ along a path in BZ, then one of the eigenvalues $\varepsilon^a(\bs{k})$ switches sign at the nodal surface. Therefore, Eq.~(\ref{eqn:D-pi0})
always defines a $\ztwo$ charge for spectral-symmetric systems exhibiting nodal surfaces. Reference~\cite{Turker:2017} shows that the $\mathsf{U}(1)$-symmetric nodal surfaces, in fact, exhibit a \emph{richer} $\intg$ classification, but this is not in contradiction with our statement. Instead, one can draw an analogy to nodal \emph{lines} when $\ztwo$ Berry phase can always be defined, but in the presence of certain symmetry a richer $\intg$ winding number also exists.
\end{subequations}}


\subsection{Interpretation of $\pi_1(M_\mathrm{BDI}) = \ztwo$}\label{subsec:BDI-pi1}

We first explain why $M_\textrm{BDI}=\mathsf{O}(n)$. Note that $\mcC = \sigma_z$ relates pairs of states with energies $\pm\varepsilon^a$,
\begin{subequations}\label{eqn:BDI-related-eigstates}
\begin{eqnarray}
\hspace{-0.4cm}\ket{u^{a,+}(\bs{k})} &=& \left(\begin{array}{c}
\ket{u^{a,1}_{\bs{k}}}_{\phantom{\dagger}}\!\!\!\!\!	\\	\ket{u^{a,2}_{\bs{k}}}^{\phantom{\dagger}}\!\!\!\!\!
\end{array}\right)\quad\;\;\textrm{with $\phantom{-}\varepsilon^a(\bs{k}) > 0$} \\
\hspace{-0.4cm}\ket{u^{a,-}(\bs{k})} &=& \sigma_z \ket{u^{a,+}(\bs{k})} \quad\textrm{with $-\varepsilon^a(\bs{k}) < 0$} 
\end{eqnarray}
\end{subequations}
where $(a,+)$ and $(a,-)$ label a pair of bands. It follows from the orthogonality of states (\ref{eqn:BDI-related-eigstates}) that
\begin{subequations}
\begin{eqnarray}
\braketBig{u^{a,1}_{\bs{k}}}{u^{b,1}_{\bs{k}}} &=& \braketBig{u^{a,2}_{\bs{k}}}{u^{b,2}_{\bs{k}}} = \tfrac{1}{2}\delta^{ab} \\
\sum_{a}\ket{u^{a,1}_{\bs{k}}}\bra{u^{a,1}_{\bs{k}}} &=& \sum_{a}\ket{u^{a,2}_{\bs{k}}}\bra{u^{a,2}_{\bs{k}}} = \tfrac{1}{2}\unit.
\end{eqnarray}
\end{subequations}
Furthermore, all the eigenstates are real because of $\mfT = \mcK$. It follows that the flat-band Hamiltonian $\mcQ(\bs{k})$ acquires the block-off-diagonal form~(\ref{eqn:Q-with-CHS}) with
\begin{equation}
q(\bs{k}) =  2\sum_a \ket{u^{a,1}(\bs{k})}\bra{u^{a,2}(\bs{k})}\in\mathsf{O}(n).
\end{equation}
Every $S^1\subset \textrm{BZ}$ with a gapped spectrum therefore traces a closed path image in $\mathsf{O}(n)$. The charge $c_\textrm{BDI}(S^1)$ corresponds to the homotopy equivalence class of this image.

The space $\mathsf{O}(n)$ consists of two separate components characterized by $\det q(\bs{k})=\pm 1$. This is just the $\pi_0$-charge~(\ref{eqn:BDI-charge-0}), i.e. the sign of $\det(\bs{k})$ is fixed unless one crosses a nodal surface, hence the image of a gapped $S^1\subset \textrm{BZ}$ lies entirely within one of the two components. If $\det[q(\bs{k})] = -1$, we replace $q(\bs{k})$ by its composition with a mirror symmetry with respect to the $n^\textrm{th}$ coordinate, while we keep it unchanged otherwise. Then the analysed $q(\bs{k})\in\mathsf{SO}(n)$, and we formally write down the $\pi_1$-charge as the homotopy equivalence class
\begin{equation}
c_\textrm{BDI}(S^1) = \left[q: S^1 \to \mathsf{SO}(n)\right].\label{eqn:BDI-charge-1}
\end{equation}
within the \emph{special} orthogonal group. In the subsequent sections we will encounter a similar charge appearing also in the case of nodal lines of AZ+$\mcI$ classes $\textrm{CI}$ and $\textrm{AI}$.

Determining the homotopy class~(\ref{eqn:BDI-charge-1}) can be achieved by tracking the eigenvalues of $q(\bs{k})$. For $n$ even, the eigenvalues come in complex conjugate pairs $\e{\pm\imi\alpha}$, while for $n$ odd there is an additional eigenstate (axis of rotation) with eigenvalue $1$. The charge $c_\textrm{BDI}(S^1)$ may be non-trivial if the eigenvalue phases $\alpha_i$ contain a non-trivial winding along $S^1$. More specifically, one has to count the number of crossings at $\alpha=\pm\pi$ which is a robust $\pi_1[\mathsf{SO}(2)]=\intg$ quantity for $n=2$, while for $n\geq 3$ only the parity $\pi_1[\mathsf{SO}(n)]=\ztwo$ is conserved since then a pair of $\pm\pi$ crossings is allowed to annihilate. For $n=1$ there is just the static unit eigenvalue and the topological charge $\pi_1[\mathsf{SO}(1)]=\triv$ is absent. These observations are in accord with the exceptional entries in Tab.~\ref{tab:Homotopies-small}.

We demonstrate this procedure on model~(\ref{eqn:BDI-ham-form}). The off-diagonal block $q\in\mathsf{SO}(2)$ becomes
\begin{equation}
q = \frac{1}{\lVert{\bs{p}}\rVert}\left(\begin{array}{cc}p_1 & p_2 \\ -p_2 & p_1 \end{array}\right)	\quad\textrm{or} \quad \frac{1}{\lVert{\bs{r}}\rVert}\left(\begin{array}{cc}r_1 & -r_2 \\ r_2 & r_1 \end{array}\right)	
\end{equation}
where the first expression applies if $\lVert{\bs{p}}\rVert > \lVert{\bs{r}}\rVert$, and the second one otherwise. Let us be more specific by setting $\bs{p}(\bs{k}) = (k_x,k_y)$ and $\bs{r}(\bs{k}) = (m,0)$. Then the spectrum $\varepsilon(\bs{k}) = \pm m \pm \sqrt{k_x^2+k_y^2}$ exhibits a zero-energy nodal cylinder at $k_x^2+k_y^2=m^2$, as wella as a pair of nodal lines with energies $\pm m$ coinciding at $k_x=0=k_y$. These nodal lines tie the (un)occupied bands in a way that makes the nodal cylinder robust: One can at best shrink the cylinder to a line by setting $m=0$, but it reappers for both $m>0$ and $m<0$. To check the topological charge, we consider a circular path $S^1:\bs{k}(\varphi) = (k\cos \varphi, k\sin\varphi,0)$ with $\varphi\in[0,2\pi]$. We find
\begin{equation}
q(\varphi) = \left(\begin{array}{cc} \cos \varphi & \sin\varphi \\ -\sin\varphi & \cos\varphi \end{array}\right) \quad\textrm{or}\quad \left(\begin{array}{cc} 1 & 0 \\ 0 & 1 \end{array}\right)
\end{equation}
where the first expression applies for $k>m$ ($S^1$ enclosing the nodal cylinder), and the second one otherwise. We observe that the invariant~(\ref{eqn:BDI-charge-1}) is trivial inside and non-trivial outside the nodal cylinder, as expected.
\begin{figure}[t]
	\includegraphics[width=0.48\textwidth]{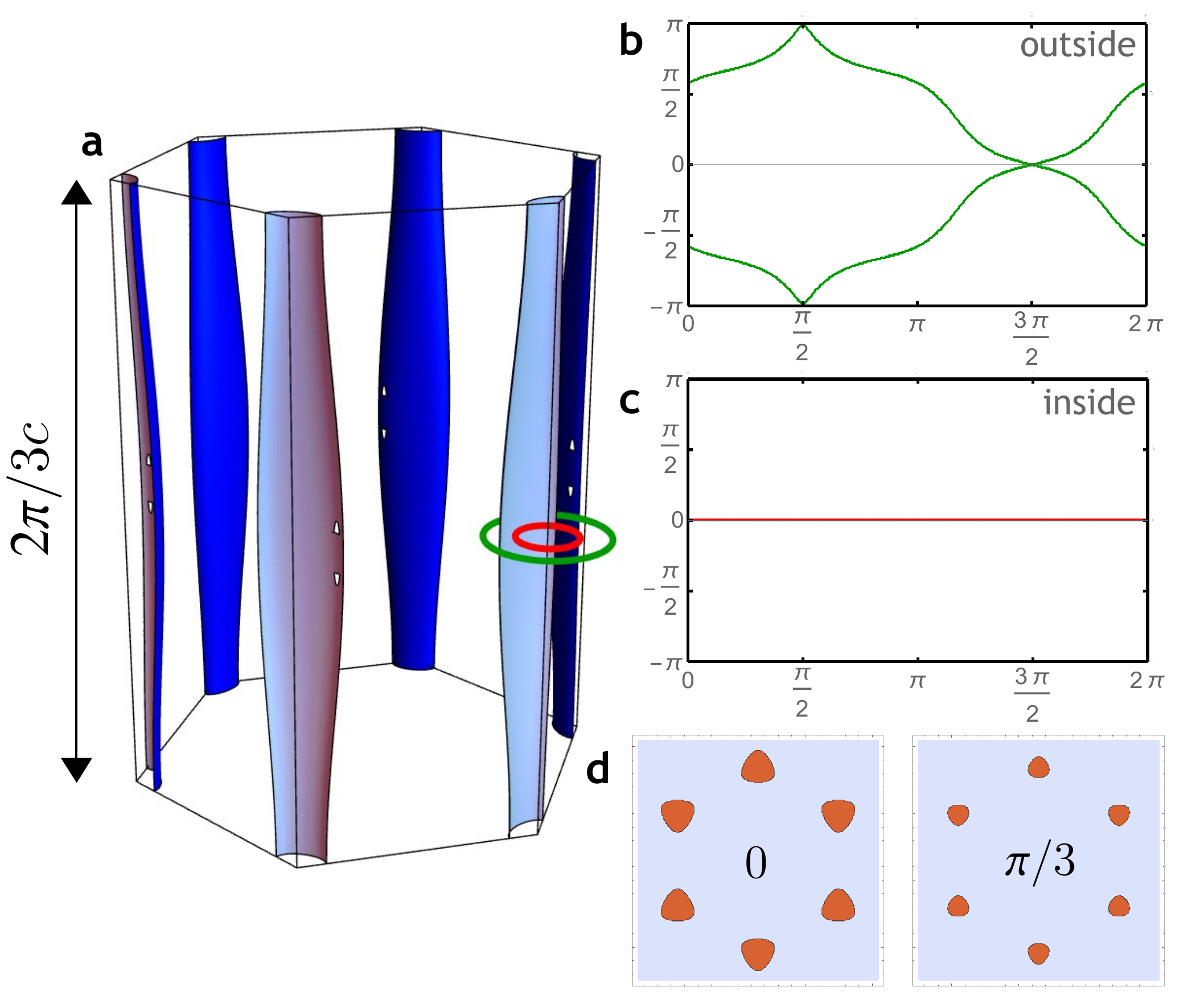}
	\caption{Doubly charged nodal surfaces of AZ+$\mcI$ class $\textrm{BDI}$. (a) Fermi surface of model~(\ref{eqn:BDI-tb-model}) with parameters~(\ref{eqn:BDI-2-params}) consists of a pair of nodal cylinders centered on the vertical BZ edges. (b,c) Evolution of the $q(\bs{k})\in\mathsf{SO}(2)$ eigenvalues along a path encircling the nodal surface (green) and a path inside of it (red) indicate a non-trivial $\pi_1$-charge~(\ref{eqn:BDI-charge-1}). (d) Plot of the sign of  $\det\left[q(\bs{k})\right]$ for horizontal planes at the indicated values of $k_z c$ demonstrate the non-trivial value of $\pi_0$-charge~(\ref{eqn:BDI-charge-0}).}
	\label{fig:BDI-model-2}
\end{figure}

We remark that invariant~(\ref{eqn:BDI-charge-1}) does \emph{not} correspond to the Berry phase. Although the nodal lines at energies $\pm m$ of the considered model both carry Berry $\pi$-phase, their pairwise appearance (imposed by $\mfP$) leads to net Berry $2\pi$-phase both outside and inside of the nodal cylinder. This doubling is similar to the presence of an \emph{even} Chern number in AZ+$\mcI$ class $\textrm{D}$. The difference here is that Berry phases $2\pi$ and $0$ are indistinguishable and therefore both trivial. 


\subsection{Example class $\textrm{BDI}$ model} \label{subsec:BDI-model}

As an example realization of the doubly charged nodal surfaces of $\textrm{AZ}$+$\mcI$ class $\textrm{BDI}$, we consider $\textrm{AAA}$-stacked graphene layers, obtained by identifying the $\textrm{Pt}$ and $\textrm{As}$ sites in Fig.~\ref{fig:SrPtAs-lattice}. The layers are further assumed dimerized into nearby pairs at distance $c$, the pairs being separated by larger gaps $r c$ with $r>1$. Importantly, we require the presence of a sublattice realization of $\mcC$ which is usually only approximately present in realistic systems.

Sticking to the Pauli matrix notation of Tab.~\ref{tab:Paulis}, we consider intralayer NN hoppings
\begin{subequations}\label{eqn:BDI-tb-model}
\begin{equation}
\mcH_1(\bs{k}) = t_1 \unit_\rho\otimes\left[\mcS^{0,\mathbf{t}}_{\cos}(\bs{k})r_x - \mcS^{0,\mathbf{t}}_{\sin}(\bs{k})r_y\right]
\end{equation}
and vertical hoppings of different amplitude within and between the dimerized layers
\begin{equation}
\mcH_2(\bs{k}) = \left(t_2 \e{\imi k_z c} + t_3 \e{-\imi k_z r c}\right) \rho_+\otimes\unit_r+\textrm{h.c.}
\end{equation}
\end{subequations}
where $\rho_{\pm} = \tfrac{1}{2}(\rho_x \pm \imi \rho_y)$, and $\textrm{h.c.}$ denotes the Hermitian conjugation. The spatial inversion of the system $\mcI = \rho_x \otimes r_x$ \emph{commutes} with the sublattice symmetry $\mcC = \rho_z \otimes r_z$, making this an $\mcI$-even SLS model. Since the Hamiltonian respects TRS and does not contain SOC, the scheme of Fig.~\ref{fig:AZI-classification}(b) locates us in $\textrm{AZ}$+$\mcI$ class $\textrm{BDI}$. 

To be specific, we consider parameter values
\begin{equation}
t_1 = 1,\quad t_2 = 0.5, \quad t_3 = 0.1 \quad\textrm{and}\quad r=2\label{eqn:BDI-2-params}
\end{equation}
which create a pair of nodal cylinders centred on the vertical BZ edges, see Fig.~\ref{fig:BDI-model-2}(a). Rotating the basis by
\begin{equation}
V = \frac{1}{\sqrt{2}}\left[\unit_\rho \otimes \left(\begin{array}{cc}
1		&	0	\\
\imi	&	0
\end{array}\right)_r + \rho_x \otimes \left(\begin{array}{cc}
0		&	1	\\
0		&	-\imi
\end{array}\right)_r\right]
\end{equation}
leads to $\mcC = \rho_z \otimes \unit_r$ and $\mfT = \mcK$, as has been required for the calculation of the topological charges. In Figs.~\ref{fig:BDI-model-2}(b,c) we check the $q(\bs{k})$ eigenvalue winding~(\ref{eqn:BDI-charge-1}) along a path that encloses (green) and a path that is enclosed by (red) the nodal cylinder. In Fig.~\ref{fig:BDI-model-2}(d) we plot the sign of $\det[q(\bs{k})]$ within the horizontal high-symmetry planes of the BZ. These two calculations confirm that both charges~(\ref{eqn:BDI-charge-0}) and~(\ref{eqn:BDI-charge-1}) of the nodal cylinders are non-trivial.


\section{$\intg\oplus\ztwo$ nodal lines in class $\mathrm{CI}$}\label{sec:nodes-CI}

The AZ+$\mcI$ class $\textrm{CI}$ in $D=3$ exhibits nodal lines characterized by a pair of topological charges
\begin{equation}
c_\textrm{CI}^{(3)} \in \pi_1(M_\textrm{CI})\oplus\pi_2(M_\textrm{CI}) = \intg\oplus\ztwo.
\end{equation}
According to Tab.~\ref{tab:Homotopies-small}, the minimal model supporting a non-trivial value of the higher charge contains four bands. In the following subsections we first introduce the general $4$-band Hamiltonian belonging to this symmetry class and determine its spectrum. We show that the $\pi_1$-charge corresponds to the winding of the determinant of $q(\bs{k})$ -- the off-diagonal block of the flat-band Hamiltonian $\mcQ(\bs{k})$. We further show that the $\pi_2$-charge corresponds to the homotopy equivalence class inside $\mathsf{SO}(n)$ just like for the $\pi_1$-charge of the AZ+$\mcI$ class $\mathrm{BDI}$. The difference is that the role of the Hamiltonian block $q(\bs{k})$ gets replaced by Wilson loop operators $\mcW(S^1)$. In the last subsection we show how this species of nodal lines may appear in TRS preserving singlet SC phase of nodal line metals without SOC, provided that the \emph{sign} of the gap function changes along the {\color{black} (cylindrical or torical)} Fermi surface. A viable route to realize such a phase experimentally might be through the SC proximity effect.


\subsection{General $4$-band Hamiltonian}

We order the six symmetry-compatible basis matrices of Tab.~\ref{tab:Hamiltonians} into a pair of three-component vectors
\begin{subequations}\label{subeqn:CI-matrices}
\begin{eqnarray}
\bs{v} &=& 	(\sigma_z\otimes\tau_z,-\sigma_z\otimes \tau_x,\sigma_x\otimes\unit) \\
\bs{w} &=&	(-\sigma_x\otimes\tau_z,\sigma_x\otimes \tau_x, \sigma_z\otimes\unit),
\end{eqnarray}
\end{subequations}
such that $\{v_i,v_j\}=\{w_i,w_j\}=2\delta_{ij}$ and $[v_i,w_j] = -\imi\delta_{ij} \, \sigma_y \otimes \unit_\tau$. A general $4$-band Hamiltonian is expressed using two real-valued vector functions $\bs{a}(\bs{k})$ and $\bs{b}(\bs{k})$ as
\begin{subequations}\label{eqn:CI-Ham-all}
\begin{equation}
\mcH(\bs{k}) = \bs{a}(\bs{k})\cdot\bs{v} + \bs{b}(\bs{k})\cdot\bs{w}.\label{eqn:CI-Ham-gen}
\end{equation}
and has spectrum
\begin{equation}
\varepsilon(\bs{k}) = \pm\sqrt{\bs{a}^2+\bs{b}^2\pm 2\lVert{\bs{a}\times\bs{b}}\rVert}.\label{eqn:CI-Ham-spec}
\end{equation}
\end{subequations}
The gap closes whenever simultaneously $\lVert{\bs{a}(\bs{k})}\rVert = \lVert{\bs{b}(\bs{k})}\rVert $ and $\bs{a}(\bs{k})\cdot\bs{b}(\bs{k})=0$. This is a \emph{pair} of scalar constraints, compatible with $\delta_\textrm{CI}=2$ that has been determined independently in Tab.~\ref{tab:Hamiltonians}. This means that the nodal objects in $D=3$ are one-dimensional \emph{lines}. 


\subsection{Interpretation of $\pi_1(M_\mathrm{CI})=\intg$}

The presence of $\mcC$ implies a natural interpretation of the $\intg$-valued $\pi_1$-charge as the usual winding number~\cite{Ryu:2010}, with $\mcI$ not playing a role. Rotating the basis of Tab.~\ref{tab:Hamiltonians} by $\mcU_\sigma$ of Eq.~(\ref{eqn:Pauli-forward-rotate}) leads to the canonical form $\mcC \propto \sigma_z$, when the off-diagonal block $q(\bs{k})$ of the flat-band Hamiltonian~(\ref{eqn:flat-band-Ham-Q}) acquires an integer winding
\begin{equation}
c_\textrm{CI}(S^1) = \frac{\imi}{2\pi} \oint_{S^1}\de \bs{k}\cdot\tr\left[q^\dagger(\bs{k})\bs{\nabla}_{\bs{k}} q(\bs{k})\right]\in\mathbb{Z}.\label{eqn:CI-pi1-charge}
\end{equation}
Formula~(\ref{eqn:CI-pi1-charge}) remains valid if we replace $q(\bs{k})$ by the off-diagonal block of the rotated $\mcH(\bs{k})$.


\subsection{Interpretation of $\pi_2(M_\mathrm{CI})=\intg$}\label{eqn:CI-pi2}

The $\pi_2$-charge of AZ+$\mcI$ class $\textrm{CI}$ can be understood in the way presented in the supplemental material to Ref.~\cite{Fang:2015}. Here we apply a slight modification to that procedure which replaces the gauge-dependent Wilson operators on open-ended paths by gauge-\emph{invariant} ones on closed loops. The invariant is also related to the $\pi_1$-charge of AZ+$\mcI$ class $\textrm{BDI}$. 

To determine the charge on a gapped $S^2$, we proceed as follows: We pick two arbitrary (but different) points $\textrm{N,S}\in S^2$ and consider a set of continuously varying closed paths $S^1(\theta)\subset S^2$ with $\theta\in[0,\pi]$ such that:
\begin{itemize} 
\item[$(i)$] $S^1(0)=\textrm{N}$ and $S^1(\pi)=\textrm{S}$ are single points,
\item[$(ii)$] for $0<\theta<\pi: S^1(\theta)$ are homeomorphic to a circle,
\item[$(iii)$] for $\theta_1\neq\theta_2: S^1(\theta_1)\cap S^1(\theta_2)=\varnothing$. 
\end{itemize}
One can visualize loops $S^1(\theta)$ as the parallels on a globe with $\textrm{N,S}$ the geographic poles, although any other choice is equally good. The union $\cup_\theta S^1(\theta)=S^2$ reproduces the original sphere. We further consider the Wilson operator~\cite{Wilczek:1984,Soluyanov:2011,Alexandradinata:2016}. For a path $\gamma$ from $\bs{k}_\textrm{i}$ to $\bs{k}_\textrm{f}$ this is
\begin{subequations}\label{subeqn:Wilson}
\begin{eqnarray}
\hspace{-0.9cm}\mcW^{ab}(\gamma)\! &=& \!\!\!\lim_{N\to\infty}\!\!\! \bra{u^a(\bs{k}_\textrm{f})}\!\!\left[\hspace{-0.45cm}\ordprod_{\hspace{0.4cm}\left\{\bs{k}_j\right\}_{j=1}^N\in \gamma}\hspace{-0.75cm}P_\textrm{occ.}\!(\bs{k}_j)\!\right]\!\! \ket{u^b(\bs{k}_\textrm{i})} \hspace{0.4cm}\label{eqn:Wilson-proj-def}\\
 \mcW(\gamma)\!&=&\!\!\! \lim_{N\to\infty}\!\! \hspace{-0.45cm} \ordprod_{\hspace{0.4cm}\left\{\bs{k}_j\right\}_{j=\imi}^\textrm{f}\in \gamma}\hspace{-0.65cm}\left[\unit-\de\bs{k}\cdot\mcbsA(\bs{k}_j)\right]\\
&\equiv & \overline{\exp}\left[-\int_{\gamma}\de\bs{k}\cdot \mcbsA(\bs{k})\right]
\end{eqnarray}
\end{subequations}
where $P_\textrm{occ.}(\bs{k})$ is the projector onto the occupied states of $\mcH(\bs{k})$ and the bar indicates path-ordering. Wilson loop operator $\mcW(\gamma)$ describes the adiabatic evolution of the occupied states along $\gamma$, and as such has to be unitary, $\mcW^\dagger \mcW = \unit$. For $\bs{k}_\textrm{i}\neq\bs{k}_\textrm{f}$ it depends on gauge, but for a closed path $S^1$
\begin{equation}
\mcW(S^1) = \overline{\exp}\left[-\oint_{S^1}\de\bs{k}\cdot \mcbsA(\bs{k})\right]
\end{equation}
it becomes gauge-\emph{invariant}. Additionally, the AZ+$\mcI$ class $\textrm{CI}$ respects $\mfT = \mcK$ which allows us to find a \emph{real} set of eigenvectors. It follows from the definition~(\ref{eqn:Wilson-proj-def}) that $\mcW(S^1)$ is \emph{real} unitary, i.e. an element of $\mathsf{O}(n)$.

To obtain the $\pi_2$-charge, we determine $\mcW[S^1(\theta)]\equiv \mcW(\theta)$ which depends continuously on $\theta$. Since $\mcW(0)=\mcW(\pi)=\unit$, the Wilson loop $\mcW(\theta)$ traces a closed path in $\mathsf{SO}(n)$, and $c_\textrm{CI}(S^2)$ is expressed as the homotopy equivalence class of loops in the special orthogonal group,
\begin{equation}
c_\textrm{CI}(S^2) = \left[\mcW: \theta \to\textsf{SO}(n)\right]\label{eqn:CI-charge-pi2}
\end{equation}
The topological classification for various $n$ follows
\begin{equation}
\pi_1[\mathsf{SO}(n)] = \left\{\begin{array}{ll}
\triv 	& \textrm{for $n=1$} 	\\
\intg 	& \textrm{for $n=2$}	\\
\ztwo	& \textrm{for $n=3$}
\end{array}\right.\label{eqn:SO-n-fund-groups}
\end{equation}
which agrees with Tab.~\ref{tab:Homotopies-small}.

\begin{figure}[t]
	\includegraphics[width=0.48\textwidth]{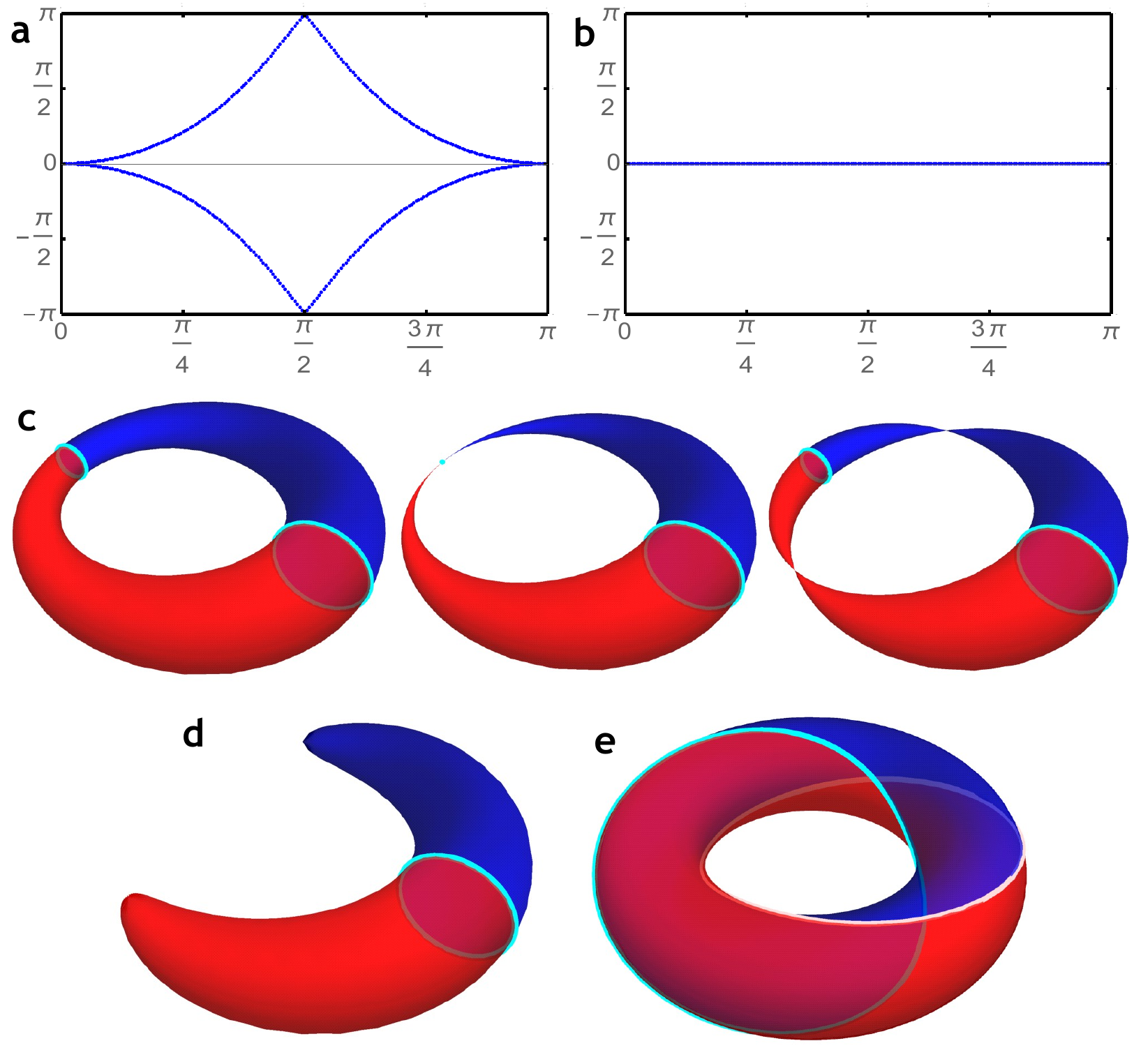}
	\caption{(a) Non-trivial eigenvalue winding of Wilson loop operators~(\ref{eqn:CI-charge-pi2}) for the nodal loop exhibited by model~(\ref{eqn:CI-illustrate-1}), and (b) the trivial winding for the nodal loop of model~(\ref{eqn:CI-illustrate-2}). Both spectra were calculated on a sphere with radius $2m$ centered at $\bs{k}=\bs{0}$. (c) Torical Fermi surface of a nodal loop metal. If the system develops a singlet SC order parameter with a gap function $\Delta_{\bs{k}}$ that changes sign along the torus (red and blue regions), a pair of SC nodal loops appear (cyan). By locally adjusting the energy of the metallic nodal loop to the chemical potential, it is possible to shrink the SC node to a single point. However, further energy variation of the metallic node leads to a regrowth of the SC nodal loop, thus manifesting a \emph{non-trivial} $\pi_2$-charge. (d) The same consideration for a torical Fermi surface \emph{without} an underlying nodal loop lead to \emph{removable} SC nodal loops with a \emph{trivial} $\pi_2$-charge. (e) Both torical Fermi surfaces admit a more complicated geometry of the SC nodal loops. Such {\color{black} \emph{linked} nodal} loops were argued to exhibit anomalous gravitomagnetoelectric response~\cite{Lian:2017,sun2017double}.}
	\label{fig:torical-FS}
\end{figure}

A more illustrative geometric interpretation of charge~(\ref{eqn:CI-charge-pi2}) exists for the $n=2$ case~(\ref{eqn:CI-Ham-all}). To keep the discussion simple, we consider a specific choice 
\begin{equation}
\bs{a}(\bs{k}) = \bs{k}\quad\textrm{and}\quad \bs{b}=(0,0,m)\label{eqn:CI-illustrate-1}.
\end{equation}
This creates a nodal loop at $k_x^2 + k_y^2 - m^2 = 0 = k_z$ which has a non-trivial $\pi_2$-charge~(\ref{eqn:CI-charge-pi2}) as is checked in Fig.~\ref{fig:torical-FS}(a). The nodal loop lies on a surface $S_{\bs{a},\bs{b}}^2=\left\{\bs{k}|k_x^2+k_y^2+k_z^2=m^2\right\}$ defined by $\lVert\bs{a}(\bs{k})\rVert = \lVert\bs{b}(\bs{k})\rVert$. Note that while vector field $\bs{b}(\bs{k})$ has a fixed direction, field $\bs{a}(\bs{k})$ has a hedgehog structure on $S_{\bs{a},\bs{b}}^2$. The map
\begin{equation}
S^2_{\bs{a},\bs{b}}\ni \bs{k}\longmapsto{\bs{n}}_{\bs{a}}(\bs{k}) = \frac{\bs{a}(\bs{k})}{\lVert{\bs{a}(\bs{k})}\rVert}\in S^2\label{eqn:CI-alternative-view}
\end{equation}
is characterized by the second homotopy group $\pi_2(S^2)=\intg$, and the hedgehog structure of the considered $\bs{a}(\bs{k})$ corresponds to a non-trivial element of $\pi_2(S^2)$. By virtue of Eq.~(\ref{eqn:CI-Ham-spec}), the nodal loop separates the ``northern hemisphere'' of $S_{\bs{a},\bs{b}}^2$ with $\bs{a}(\bs{k})\cdot\bs{b}(\bs{k})>0$ from the ``southern'' one with $\bs{a}(\bs{k})\cdot\bs{b}(\bs{k})<0$. Because of the non-trivial winding of map~(\ref{eqn:CI-alternative-view}), the two ``geographic poles'' with parallel vectors $\bs{a}(\bs{k})\times\bs{b}(\bs{k})=\bs{0}$ appear somewhere on $S^2_{\bs{a},\bs{b}}$. Since $\bs{a}(\bs{k})$ varies continuously on $S^2_{\bs{a},\bs{b}}$, the presence of the two ``poles'' makes the equator separating the two hemispheres -- i.e. the nodal loop -- robust.

We compare these observations to a model with
\begin{equation}
\widetilde{\bs{a}}(\bs{k})=(\sqrt{k_x^2 + k_y^2},\widetilde{m},k_z)\quad \textrm{and}\quad \bs{b}=(0,0,m)\label{eqn:CI-illustrate-2}
\end{equation}
which for $\widetilde{m}=0$ produces a nodal line at the same location as model~(\ref{eqn:CI-illustrate-1}) but with a \emph{trivial} $\pi_2$-charge~(\ref{eqn:CI-charge-pi2}) as checked in Fig.~\ref{fig:torical-FS}(b). In this case, the surface $S^2_{\widetilde{\bs{a}},\bs{b}}=S^2_{\bs{a},\bs{b}}$ remains unchanged, but the winding~(\ref{eqn:CI-alternative-view}) of $\bs{n}_{\widetilde{\bs{a}}}(\bs{k})$  is \emph{trivial} because $\bs{n}_{\widetilde{\bs{a}}}(\bs{k})$ lies on a circle with $n_y =0$. Indeed, increasing $\widetilde{m}$ from zero to $\pm m$ shrinks the nodal line to a point, and a gap opens for $\abs{\widetilde{m}} > \abs{m}$, thus manifesting the trivial value of its $\pi_2$-charge.


\subsection{Example class $\textrm{CI}$ model}

In Sec.~\ref{sec:sym-class-rel} we identified two qualitatively different realizations of AZ+$\mcI$ class $\textrm{CI}$. Here we focus on the case of a TRS-preserving singlet SC in the absence of SOC. We first show how such a class of systems relates directly to $\bs{k}\cdot\bs{p}$ models~(\ref{eqn:CI-illustrate-1}) and~(\ref{eqn:CI-illustrate-2}). We afterwards develop a concrete TB model of a nodal line metal on a $\textrm{SrPtAs}$-like lattice of Fig.~\ref{fig:SrPtAs-lattice}, and we show that its singlet SC phase exhibits doubly charged nodal loops whenever the gap function changes sign along the Fermi surface. The SC order parameter may either appear spontaneously at low enough temperatures, or it may be induced at an interface through the proximity effect.

Consider a two-orbital system with $\mcI = \unit_\tau$ {\color{black} without a sublattice realization of $\mcC$}. According to the discussion in Subsec.~\ref{subsec:superconductivity}, its singlet SC phase acquires AZ+$\mcI$ symmetries $\mfP = \imi \varsigma_y\otimes \unit_\tau \mcK$ and $\mfT = \mcK$. These are precisely the forms listed in Tab.~\ref{tab:Hamiltonians}, meaning that the symmetry-compatible basis matrices of the reduced $\textrm{BdG}$ Hamiltonian~(\ref{subeqn:BdG-2x2-big}) are exactly those organized in Eqs.~(\ref{subeqn:CI-matrices}) (with replaced $\sigma_i\mapsto \varsigma_i$). More specifically, model~(\ref{eqn:CI-illustrate-1}) corresponds to the (non-reduced) BdG Hamiltonian~(\ref{eqn:BdG-4x4}) with
\begin{subequations}
\begin{eqnarray}
\Xi_{\bs{k}} &=& \unit_\sigma \otimes \left(k_x \tau_z - k_y \tau_x + m\unit_\tau\right)\label{eqn:CI-connection-Xi}\\
\Delta_{\bs{k}} &=& k_z \left(\imi \sigma_y\right) \otimes \unit_\tau. \label{eqn:CI-connection-Delta}
\end{eqnarray}
The metallic state described by $\Xi_{\bs{k}}$ exhibits a cylindrical Fermi surface $\textrm{FS}=\left\{\bs{k}|k_x^2+k_y^2 = m^2\right\}$ connected to a nodal line at energy $m$, hence we call this system a \emph{nodal line metal}. Additionally, the gap function $\Delta_{\bs{k}}$ changes sign from negative for $k_z < 0$ to positive for $k_z > 0$, leading to a zero-energy nodal loop located at $k_z = 0$ in the SC phase. If one imagines warping the cylindrical $\textrm{FS}$ into a torus, then such SC nodal loops have to come in pairs as visible in Fig.~\ref{fig:torical-FS}(c). 

Notice that the underlying nodal line of the metallic band structure~(\ref{eqn:CI-connection-Xi}) makes nodal loops of the SC phase~(\ref{eqn:CI-connection-Delta}) robust: One can at best shrink the loop to a \emph{point} by setting $m=0$, but the loop reappears for $m\neq 0$ of both signs. The reason is that the underlying electron-like $\textrm{FS}$ (for $m<0$) evolves directly into a hole-like $\textrm{FS}$ (for $m>0$) via the nodal line which glues the bands together. This touching is protected by the Berry $\pi$-phase invariant. We illustrate such a band evolution and the robustness of such SC nodal loops in Fig.~\ref{fig:torical-FS}(c). 

We compare this to model~(\ref{eqn:CI-illustrate-2}) which arises from
\begin{equation}
\widetilde{\Xi}_{\bs{k}} = \unit_\sigma \otimes \left(\sqrt{k_x^2 + k_y^2} \tau_z - \widetilde{m} \tau_x + m\unit_\tau\right)\label{eqn:CI-connection-Xi2}.
\end{equation}
\end{subequations}
This leads to a cylindrical $\widetilde{\textrm{FS}}=\left\{\bs{k}|k_x^2+k_y^2 = m^2-\widetilde{m}^2\right\}$ \emph{without} an underlying nodal line (apart from the fine-tuned case $\widetilde{m}=0$ which is accidental and does not carry a topological charge). This makes it possible to remove $\widetilde{\textrm{FS}}$ by setting $\abs{\widetilde{m}}>\abs{m}$ which also eliminates the nodal loop of the SC phase~(\ref{eqn:CI-connection-Delta}) as illustrated in Fig.~\ref{fig:torical-FS}(d). We infer that the doubly charged nodes of the gap function are bound to SC phase of \emph{nodal line} metals. This observation supplements the already known unusual SC phases enabled by non-trivial Fermi surface topology~\cite{Li:2015,Nandkishore:2016}.

\begin{figure}[t]
	\includegraphics[width=0.472\textwidth]{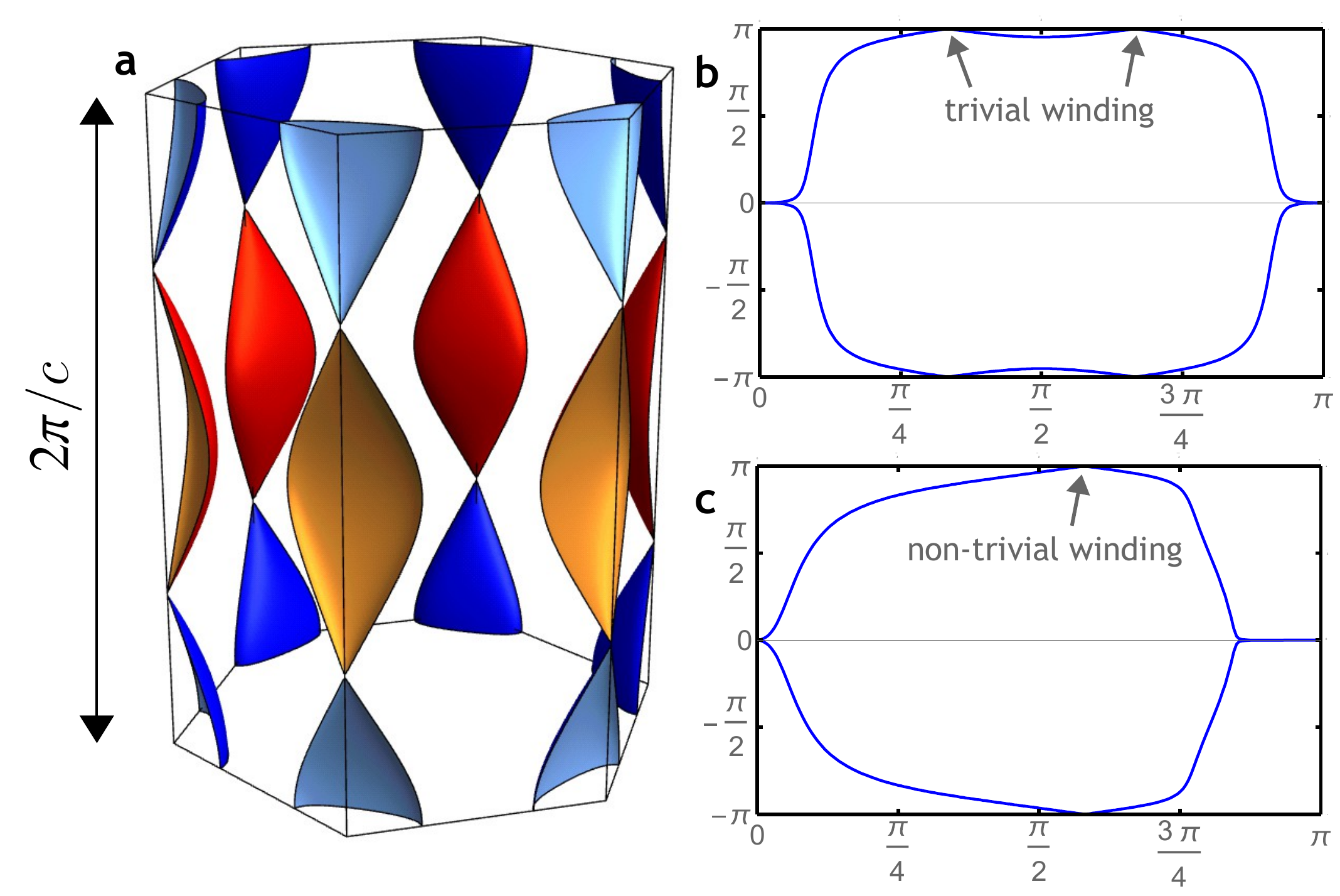}
	\caption{Doubly charged nodal lines of AZ+$\mcI$ class $\textrm{CI}$. (a) Fermi surface of the developed TB model with parameters~(\ref{eqn:CI-model-params}) consists of touching electron (blue) and hole (red) pockets. The SC order parameter~(\ref{eqn:CI-model-Delta}) creates two pairs of SC nodal loops when $\abs{\delta}<1$. (b) Winding of $\mcW(\theta)$ eigenvalues on an ellipsoid described in the text which contains a \emph{pair} of nodal loops, and (c) the same calculation on an ellipsoid containing a \emph{single} nodal loop reveal a non-trivial $\pi_2$-charge~(\ref{eqn:CI-charge-pi2}).}
	\label{fig:CI-TB-model}
\end{figure}

We now develop a concrete TB model belonging to this symmetry class to make our ideas more tangible. We consider again the lattice of Subsec.~\ref{subsec:BDI-model} but \emph{without dimerization}. There are two orbitals per unit cell that lead to $\mcI = r_x$ and $\mcC = r_z$. We consider parameters
\begin{equation}
t_1 = 1\quad\textrm{and} \quad t_2 = 0.4,\label{eqn:CI-model-params}
\end{equation}
and further $t_3=t_2$, $r=1$ and $\mu=0$. The hexagonal crystalline symmetry imposes nodal lines running along the vertical BZ edges. The chosen TB parameters~(\ref{eqn:CI-model-params}) lead to touching electron and hole Fermi pockets as shown in Fig.~\ref{fig:CI-TB-model}(a). We further assume that the system acquires a {singlet} SC order parameter 
\begin{equation}
\Delta_{\bs{k}} = \psi^0 \left[\delta + \cos (k_z c)\right](\imi\sigma^y)\otimes\unit_r.\label{eqn:CI-model-Delta}
\end{equation}
The AZ+$\mcI$ symmetries of this model are $\mfT = r_x \mcK$, $\mfP = \imi r_x \otimes \varsigma_y \mcK$ and $\mcC = \imi \varsigma_y$, which is modified to the choice of Tab.~\ref{tab:Hamiltonians} if one rotates the basis by $\mcU^\dagger_r$ of Eq.~(\ref{eqn:Pauli-forward-rotate}).

The developed model exhibits four SC nodal loops for $\abs{\delta}<1$ [one pair at heights $k_z c=\pm \arccos(-\delta)$ at both BZ edges] which move along the touching Fermi pockets when varying $\delta$. These nodes annihilate in pairs at $k_z=0$ ($k_z = \pi$) for $\delta=-1$ ($\delta=+1$). They shrink to points for $\delta=0$ when they coincide with the touching points of the Fermi pockets. To check the $\pi_2$-charge of these nodes, we set $\psi^0 = 0.2$ and $\delta=-\sqrt{3}/{2}$ which locates them at $k_z c = \pm \pi/4$. In Fig.~\ref{fig:CI-TB-model}(b) we plot the trivial eigenvalue winding on an ellipsoid with radii $(\tfrac{\pi}{3a},\tfrac{\pi}{3a},\tfrac{\pi}{3c})$ centered at $\bs{k}_\textrm{c}=(0,\frac{4\pi}{3\sqrt{3}a},0)$ which contains a \emph{pair} of SC nodal loops, while in Fig.~\ref{fig:CI-TB-model}(c) we plot the \emph{non-trivial} eigenvalue winding for an ellipsoid of the same dimensions centered at $\tilde{\bs{k}}_\textrm{c}=(0,\tfrac{4\pi}{3\sqrt{3}a},\tfrac{\pi}{4c})$ containing a \emph{single} SC nodal loop. These observations imply a non-trivial value of the $\pi_2$-charge~(\ref{eqn:CI-charge-pi2}). The $\pi_1$-charge (\ref{eqn:CI-pi1-charge}) is non-trivial for every (non-accidental) SC nodal line of this symmetry class.


\section{$\ztwo\oplus\ztwo$ nodal lines in class $\mathrm{AI}$}\label{sec:nodes-AI}

We finally discuss the AZ+$\mcI$ class $\textrm{AI}$ in $D=3$, which has been to varying degree considered in Refs~\cite{Fang:2015,Zhao:2016,Zhao:2017}. This symmetry class supports nodal lines with charge
\begin{equation}
c_\textrm{AI}^{(3)} \in \pi_1(M_\textrm{AI})\oplus\pi_2(M_\textrm{AI}) = \ztwo\oplus\ztwo.
\end{equation}
According to Tab.~\ref{tab:Homotopies-small-AI}, the minimal half-filled model supporting a non-trivial $\pi_2$-charge contains four bands. The nine basis matrices of Tab.~\ref{tab:Hamiltonians} can be organized into three vectors (called \emph{real Dirac basis} by Ref.~\cite{Zhao:2017})
\begin{subequations}\label{eqn:R-Dirac-bases}
\begin{eqnarray}
\bs{v}^{(1)} &=& 	(\sigma_z\otimes\tau_z,-\sigma_z\otimes \tau_x,\sigma_x\otimes\unit) \\
\bs{v}^{(2)} &=&  (\unit\otimes \tau_x,\unit\otimes\tau_z,\sigma_y\otimes\tau_y)\\
\bs{v}^{(3)} &=&	(-\sigma_x\otimes\tau_z,\sigma_x\otimes \tau_x, \sigma_z\otimes\unit)
\end{eqnarray}
\end{subequations}
fulfilling $\{v_i^{(a)},v_j^{(b)}\} = 2\left[\delta_{ij}\delta^{ab}\unit + \epsilon_{ijk}\epsilon^{abc}v_k^{(c)}\right]$. We failed to derive analytic conditions for the occurrence of a gap closing, so we proceed directly with the discussion of the topological charges. The $\pi_1$-charge is just the Berry phase, which is for closed paths quantized to $\{0,\pi\}$ by $\mfT$. The $\pi_2$-charge is again the $\pi_1[\mathsf{SO}(n)]$ homotopy equivalence class of Wilson loop operators that has been explained for AZ+$\mcI$ class $\textrm{CI}$ in Subsec.~\ref{eqn:CI-pi2}. In the last subsection we develop a TB model on a $\textrm{SrPtAs}$-like lattice that exhibits either singly or doubly charged nodal loops, depending on the chosen parameter values.


\subsection{Interpretation of $\pi_1(M_\mathrm{AI}) = \ztwo$}

The $\pi_1$-charge $c_\textrm{AI}(S^1)$ of AZ+$\mcI$ class $\textrm{AI}$ corresponds to the Berry phase acquired along $S^1$, which is obtained from the Wilson loop operator $\mcW(S^1)$ of Eq.~(\ref{subeqn:Wilson}) as
\begin{subequations}\label{subeqn:AI-pi1}
\begin{equation}
c_\textrm{AI}(S^1)=\frac{1}{\imi \pi}\log\det\mcW(S^1)\mod 2 \quad\in\ztwo\label{eqn:AI-charge-1}
\end{equation}
which is equivalent to
\begin{equation}
\widetilde{c}_\textrm{AI}(S^1) = \frac{\imi}{\pi}\oint_{S^1} \de \bs{k}\cdot\tr\mcbsA(\bs{k})\mod 2
\end{equation}
\end{subequations}
The Berry phase is quantized because $\mfT = \mcK$ makes it possible to find a \emph{real} set of eigenvectors $\ket{u^a(\bs{k})}$, such that $\mcW(S^1)\in\mathsf{O}(n)$ has determinant $\pm 1$. 

Note that according to Tab.~\ref{tab:Homotopies-small-AI} there is an exception to the \emph{order} of the $\pi_1$-charge in half-filed 2-band models. In this case, the basis of symmetry-compatible Hamiltonians is two-dimensional (spanned by $\sigma_x$ and $\sigma_z$) and thus allows for a richer $\intg$-valued winding number.


\subsection{Interpretation of $\pi_2(M_\mathrm{AI}) = \ztwo$}

The $\pi_2$-charge of this symmetry class is
\begin{equation}
c_\textrm{AI}(S^2) = [\mcW: \theta \to \mathsf{SO}(n)],\label{eqn:AI-charge-2}
\end{equation}
i.e. it corresponds to the homotopy equivalence class of Wilson loop operators $\mcW(\theta)\in\mathsf{SO}(n)$ for a set of closed paths $S^1(\theta)$ covering $S^2$, just like for the AZ+$\mcI$ class $\textrm{CI}$. However, we see from Tab.~\ref{tab:Homotopies-small-AI} that the \emph{order} of charge~(\ref{eqn:AI-charge-2}) is modified whenever $\min\{n,\ell\}\leq 2$. Such exceptions were absent in symmetry class $\textrm{CI}$. We use the rest of the subsection to clarify this complication.

The key observation is that the homotopy class (\ref{eqn:AI-charge-2}) can be determined for \emph{two} Wilson loop operators, $\mcW^\textrm{occ.}\in\mathsf{SO}(n)$ over the occupied and $\mcW^\textrm{un.}\in\mathsf{SO}(\ell)$ over the unoccupied bands. The presence of $\mfP$ in class $\textrm{CI}$ enforces the two Wilson operators to have identical spectra, so nothing is gained by considering both. On the other hand, particle-hole symmetry is absent in class $\textrm{AI}$ which allows the Wilson spectra to be different. As with the other topological charges, we expect the \emph{sum} of the two charges to be trivial, but the sum has to be perceived in the sense of $\pi_1[\textsf{SO}(n+\ell)]$, which may \emph{differ} from groups $\pi_1[\textsf{SO}(n)]$ and $\pi_1[\textsf{SO}(\ell)]$, cf.~(\ref{eqn:SO-n-fund-groups}). In fact, the latter two may be different themselves.

\begin{figure}[t]
	\includegraphics[width=0.48\textwidth]{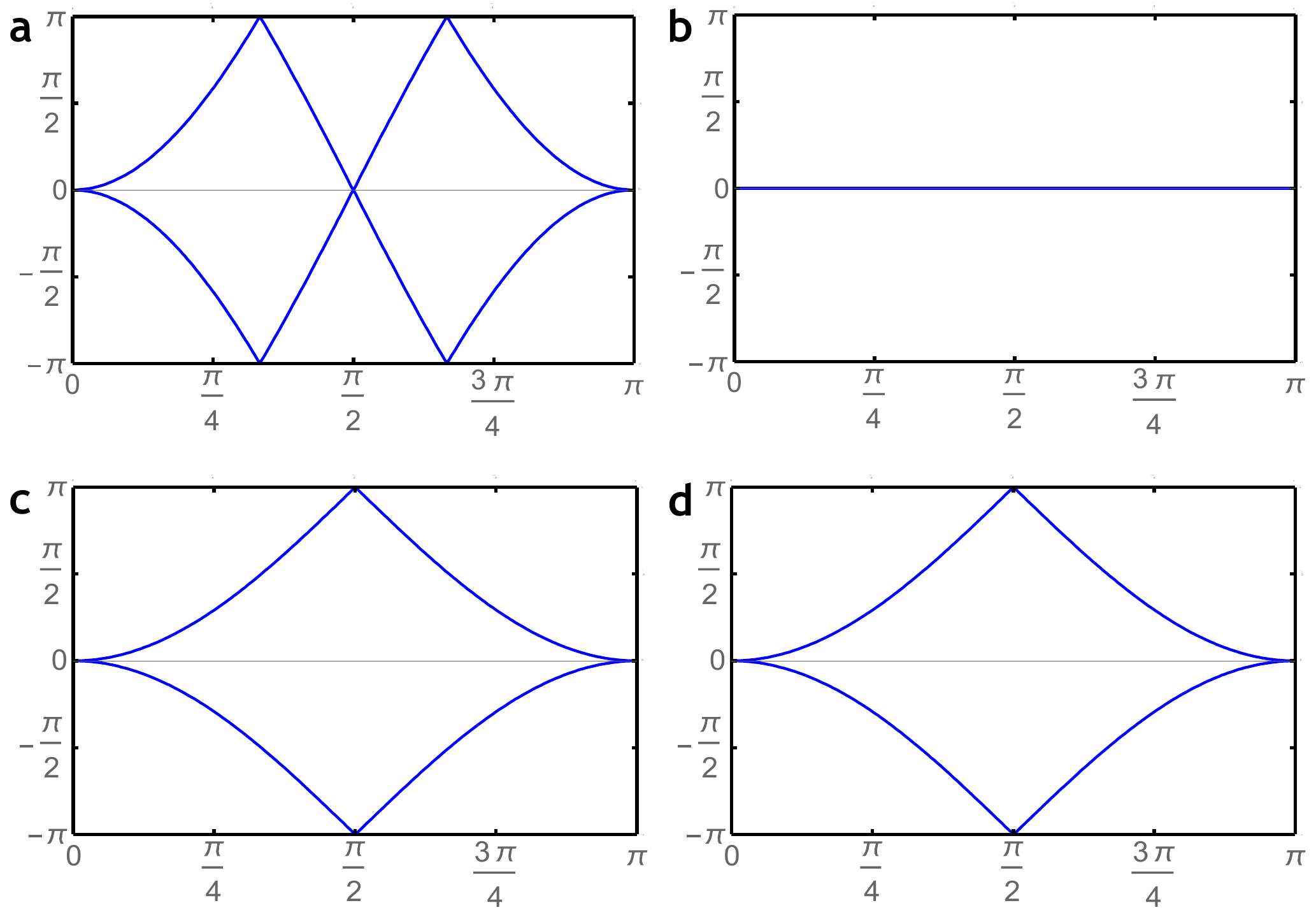}
	\caption{(a) Wilson loop spectrum over the occupied and (b) the unoccupied states of the (2+1)-band Hamiltonian~(\ref{eqn:AI-few-2-1}). The pair of plots looks identical for the (2+2)-band Hamiltonian~(\ref{eqn:AI-few-2-2-A}). (c,d) The corresponding Wilson loop spectra for the (2+2)-band Hamiltonian~(\ref{eqn:AI-few-2-2-B}).}
	\label{fig:few-band-AI}
\end{figure}

For example, for $n=2$ and $\ell=1$ the two occupied bands admit $c_\textrm{AI}^\textrm{occ.}(S^2)\in\mathsf{SO}(2)=\intg$, while the unoccupied band has a trivial $c^\textrm{un.}_\textrm{AI}(S^2)\in\mathsf{SO}(1)=\triv$. The sum of the two has to be trivial within $\pi_1[\textsf{SO}(2+1)]=\ztwo$ which only percieves the parity, hence $c^\textrm{occ.}_\textrm{AI}(S^2)$ must be an \emph{even} integer as indicated in Tab.~\ref{tab:Homotopies-small-AI}. An example Hamiltonian with a prescribed charge $2\nu\in2\intg$ is
\begin{equation}
\mcH^{(2,1)}_{2\nu}(\bs{k}) = k \left[2 \bs{n}_{(\theta,\nu\varphi)}\cdot\bs{n}^\top_{(\theta,\nu \varphi)} - \unit\right]\label{eqn:AI-few-2-1}
\end{equation}
where $\bs{k}=k\bs{n}_{(\theta,\varphi)}$ is expressed using spherical coordinates $(k,\theta,\varphi)$, and
\begin{equation}
\bs{n}^\top_{(\alpha,\beta)} = (\sin\alpha\cos\beta,\sin\alpha\sin\beta,\cos\alpha)
\end{equation}
with $\alpha\in[0,\pi]$ and $\beta\in[0,2\pi)$ is the unit vector in a specified direction, expanded in Cartesian coordinates. The Wilson loop spectra $\mcW^\textrm{occ.}$ and $\mcW^\textrm{un.}$ of model (\ref{eqn:AI-few-2-1}) with $2\nu = 2$ are plotted in Figs.~\ref{fig:few-band-AI}(a,b). This enrichment to $2\intg$ is in a stark contrast to models with $n\geq 3$ and $\ell=1$ when the \emph{only} consistent choice of charges summing to $0$ (mod $2$) is $c^\textrm{occ.}_\textrm{AI}(S^2) = 0\;(\in \ztwo)$ and $ c_\textrm{AI}^\textrm{un.}(S^2)=0\;(\in\triv)$, such that no topological classification remains. The same conclusion is also found for $n=\ell=1$. Clearly, the order of the charge is  unchanged if we exchange $n\leftrightarrow \ell$, which is manifested by the symmetry of Tab.~\ref{tab:Homotopies-small-AI}.

For $n=2$ and $\ell \geq 3$ we find $\intg$ topologically inequivalent classes: Charges $c^\textrm{occ.}_\textrm{AI}(S^2)\in\intg$ and $c_\textrm{AI}^\textrm{un.}(S^2)\in\ztwo$ have to sum up to $0$ (mod $2$). Clearly, there is a unique solution to $c^\textrm{un.}_\textrm{AI}(S^2)$ for any $c_\textrm{AI}^\textrm{occ.}(S^2)$. Finally, in the case $n=\ell=2$ we have to pick two integers that sum up to $0$ (mod $2$), meaning they are either both even or both odd. This corresponds to the $\intg\oplus\intg$ entry in the middle of Tab.~\ref{tab:Homotopies-small-AI}. As a non-trivial example, consider
\begin{equation}
\mcH^{(2,2)}_{(2\nu,0)}(\bs{k}) = \mcH^{(2,1)}_{2\nu}(\bs{k})\oplus (k)\label{eqn:AI-few-2-2-A}
\end{equation}
i.e. where we just increase the dimension of Hamiltonian~(\ref{eqn:AI-few-2-1}) by putting $k$ in its bottom-right corner, which produces an additional trivial unoccupied state. As indicated by the subscripts, Hamiltonian~(\ref{eqn:AI-few-2-2-A}) has charges $c^\textrm{occ.}_{\textrm{AI}} = 2\nu$ and $c_\textrm{AI}^\textrm{un.} = 0$, manifested again by the Wilson loop spectra of Fig.~\ref{fig:few-band-AI}(a,b). On the other hand, the two charges of~\cite{Zhao:2017}
\begin{eqnarray}
\hspace{-0.5cm}\mcH^{(2,2)}_{(\nu,-\nu)}&=&\textrm{Re}\left[(k_x+\sign(\nu)\,\imi k_y)^{\abs{\nu}}\right]v^{(a)}_1 \nonumber \\
&+& \textrm{Im}\left[(k_x + \sign(\nu)\,\imi k_y)^{\abs{\nu}}\right]v^{(a)}_2 + k_z v_3^{(a)}\label{eqn:AI-few-2-2-B}
\end{eqnarray}
with a real Dirac basis~(\ref{eqn:R-Dirac-bases}) are $\nu$ and $-\nu$. We plot the corresponding spectra of $\mcW^{occ.}$ and $\mcW^{un.}$ for $\nu=1$ in Fig.~\ref{fig:few-band-AI}(c,d). The somewhat unusual charge appearing for $n=\ell=2$ stems from the fact that {\color{black} the classifying space} $\mathsf{O}(4)/\mathsf{O}(2)\!\times\!\mathsf{O}(2)$ is homeomorphic to a double cover of $S^2\times S^2$~\cite{Gluck:1983,mathSE:2215495}.

We remark that Hamiltonians~(\ref{eqn:AI-few-2-1}),~(\ref{eqn:AI-few-2-2-A}) and~(\ref{eqn:AI-few-2-2-B}) are fine-tuned such that the node is contracted to a single point at $\bs{k}=\bs{0}$. The study of how the exceptional charges $c_\textrm{AI}^\textrm{occ.}(S^2)$ and $c_\textrm{AI}^\textrm{un.}(S^2)$ manifest themselves in the few-band spectra when the Hamiltonian becomes detuned is beyond the scope of the present manuscript.


\subsection{Example class $\textrm{AI}$ model}\label{subsec:AI-TB-model}

We consider a $\textrm{SrPtAs}$-like lattice with $s$-like orbitals at both the $\textrm{Pt}$ and the $\textrm{As}$ sites (assumed inequvalent) contributing to a TB model. In the absence of SOC, the hexagonal lattice symmetry enforces nodal lines running along the vertical BZ edges. A single unit cell extends over two layers and contains a pair of orbitals in each.

We stick to the Pauli matrix notation of Tab.~\ref{tab:Paulis}, such that the two orbitals of the same species [same color in Fig.~\ref{fig:SrPtAs-lattice}(a)] are exchanged by $\mcI = \rho_x \otimes r_x$. We consider intralayer hoppings with amplitude $t_1$
\begin{subequations}\label{eqn:AI-TB-model}
\begin{equation}
\mcH_1(\bs{k}) = t_1 \unit_\rho\otimes \left[\mcS^{0,\mathbf{t}}_{\cos}(\bs{k}) r_x - \mcS^{0,\mathbf{t}}_{\sin}(\bs{k}) r_y\right],
\end{equation}
and vertical interlayer hoppings with amplitude $t_2$
\begin{equation}
\mcH_2(\bs{k}) = 2t_2 \cos(k_z c) \rho_x \otimes \unit_r.
\end{equation}
We further consider vertical hoppings to the \emph{second nearest} layer with amplitudes $t_0\pm t_3$ and a staggered on-site potential $\pm m$ on the two elemental sublattices,
\begin{eqnarray}
\mcH_3(\bs{k}) &=& \left[m+2t_3\cos(2k_z c)\right]\rho_z \otimes r_z \nonumber \\
&\phantom{=}& +\, 2t_0 \cos(2k_z c) \unit_\rho \otimes \unit_r.
\end{eqnarray}
\end{subequations}
The representation $\mfT = \rho_x \otimes r_x \mcK$ in the employed basis differs from the choice of Tab.~\ref{tab:Hamiltonians}, but there is no need for a basis transformation as the Wilson loop formulation of charges~(\ref{subeqn:AI-pi1}) and (\ref{eqn:AI-charge-2}) applies to any basis.

\begin{figure}[t]
	\includegraphics[width=0.48\textwidth]{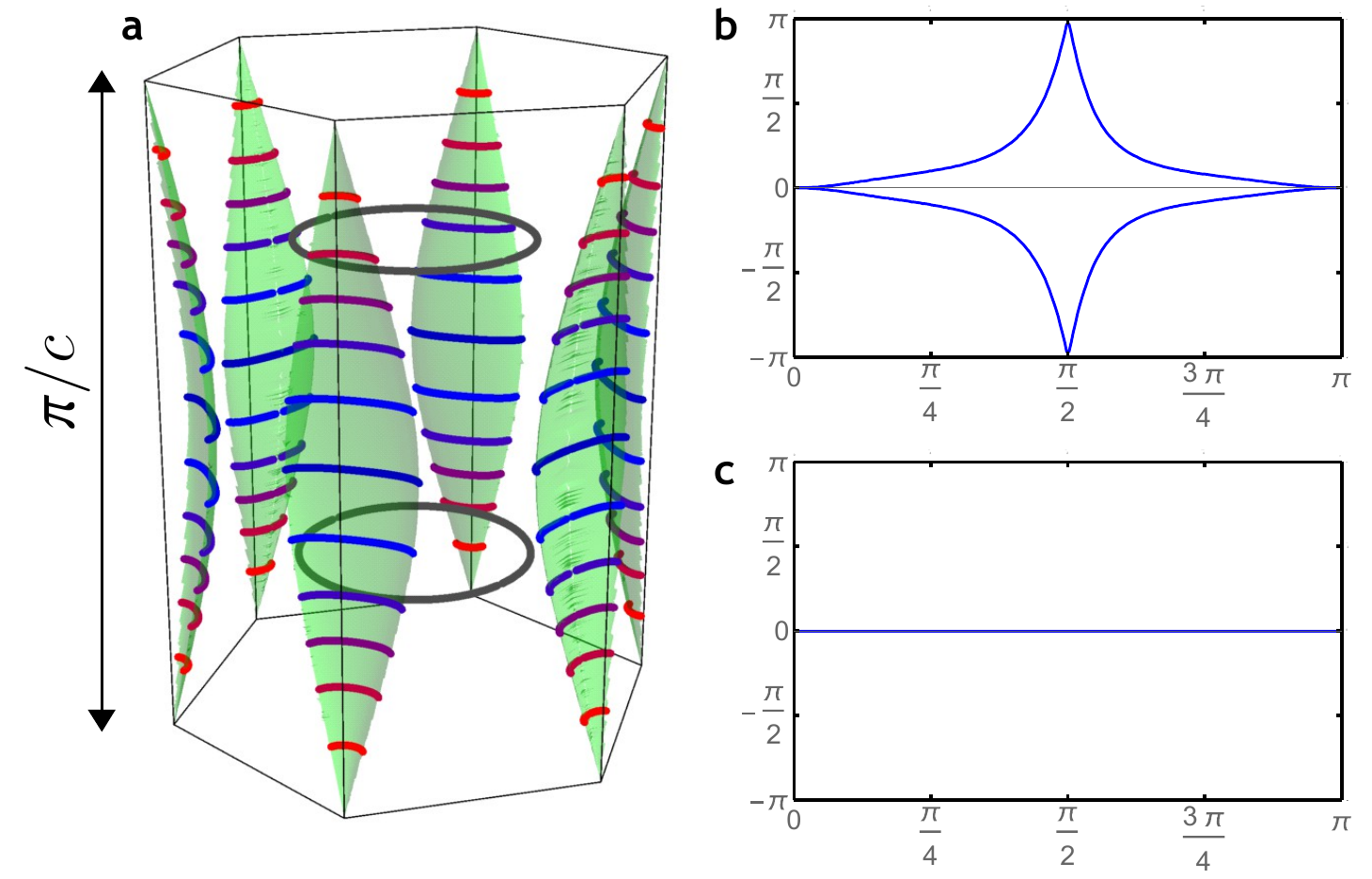}
	\caption{Doubly charged nodal loops of AZ+$\mcI$ class $\textrm{AI}$. (a) The transparent green sheets indicate the trajectory traced out by nodal loops of TB model~(\ref{eqn:AI-TB-model}) for parameters~(\ref{eqn:AI-param-vals-part1}) for $m\in[-2,2]$. Snapshots of these nodal loops for $m\in\{-2,-1.6,-0.8,0,0.8,1.6,2\}$ are shown in various shades of blue to red. The large circular gray loops inside the BZ correspond to nodes of the same model with $t_2 = 1.5$. (b) The Wilson Loop spectrum for one of the nodal lines for parameters~(\ref{eqn:AI-param-vals}) exposes the non-trivial value of their $\pi_2$-charge~(\ref{eqn:AI-charge-2}). This is related to the underlying nodal lines formed within the (un)occupied bands along the vertical BZ edges, running through the interior of the plotted nodal loops. (c) The $\pi_2$-charge is trivial for nodal loops with $t_2 = 1.5$ on the other side of the topological phase transition~(\ref{eqn:AI-topo-transition}).}
	\label{fig:TB-model-AI}
\end{figure}

Model~(\ref{eqn:AI-TB-model}) exhibits a gap closing at $\bs{k}$ with energy $-m t_0/t_3$ whenever the pair of conditions
\begin{subequations}\label{eqn:AI-model-NL-conds}
\begin{eqnarray}
2t_3\cos(2 k_z c) + m &=& 0 \\
\left[\mcS^{0,\mathbf{t}}_{\cos}(\bs{k})\right]^2 + \left[\mcS^{0,\mathbf{t}}_{\sin}(\bs{k})\right]^2 &=& 2\frac{t_2^2}{t_1^2}\left(1-\frac{m}{2t_3}\right).
\end{eqnarray}
\end{subequations}
are fulfilled. Nodal loops appear in the model for $\abs{m}<2$ and a small enough ratio $\abs{t_2/t_1}$. These loops annihilate in pairs at the $k_z c = 0$ plane for $m=-2$, and at the $k_z c = \pi/{2}$ plane for $m=2$, and we plot their trace for the intermediate values of $m$ and for parameters
\begin{subequations}\label{eqn:AI-param-vals}
\begin{equation}
t_1 = 1,\quad t_2 = 0.4,\quad\textrm{and}\quad t_3 = 1 \label{eqn:AI-param-vals-part1}
\end{equation}
in Fig.~\ref{fig:TB-model-AI}(a). We determine the $\pi_2$-charge~(\ref{eqn:AI-charge-2}) of these nodal loops for 
\begin{equation}
m=0
\end{equation}
\end{subequations} which sets their vertical location to $k_z c = \pi/4$. In Fig.~\ref{fig:TB-model-AI}(b) we plot the Wilson loop spectrum for an ellipsoid  with radii $\left(\tfrac{1}{a},\tfrac{1}{a},\tfrac{1}{c}\right)$ centered at $\bs{k}_\textrm{c}=\left(0,\tfrac{4\pi}{3\sqrt{3}a},\tfrac{\pi}{4c}\right)$ which contains a single nodal loop. The spectrum has a non-trivial winding, thus exposing the non-trivial value of the $\pi_2$-charge.

Interestingly, a topological transition occurs for 
\begin{equation}
1 = 2\frac{t_2^2}{t_1^2}\left(1-\frac{m}{2t_3}\right) \label{eqn:AI-topo-transition}
\end{equation} 
in which pairs of the doubly charged nodal loops merge together, thus forming nodal loops with a \emph{trivial} value of the $\pi_2$-charge. For the values of $t_1$, $t_3$, $m$ listen in Eqs.~(\ref{eqn:AI-param-vals}), the transition occurs for $t_{2\textrm{c}} = 1/\sqrt{2}$. Choosing $t_2 = 1.5 > t_{2\textrm{c}}$, we find the nearly circular nodal loops drawn in gray in Fig.~\ref{fig:TB-model-AI}(a). In Fig.~\ref{fig:TB-model-AI}(c) we plot the Wilson loop spectrum on an ellipsoid with radii $\left(\tfrac{3}{2a},\tfrac{3}{2a},\tfrac{3}{2c}\right)$ centred at $\tilde{\bs{k}}_\textrm{c}=\left(0,0,\tfrac{\pi}{4c}\right)$ which encloses a single such a nodal loop. The spectrum has a trivial winding, thus confirming the trivial value of the $\pi_2$-charge. Indeed, these loops disappear from the spectrum for $t_2 > 3/\sqrt{2}$, thus once again confirming their trivial nature.


\section{Summary}\label{sec:conclusion}

In this work we used homotopy theory to generalize the observation of doubly charged nodal lines in certain three-dimensional centrosymmetric semimetals by Ref.~\cite{Fang:2015} to the centrosymmetric extensions of all Atland-Zirnbauer (AZ) classes of arbitrary spatial dimension $D$. Our main results are summarized by {\color{black} Tab.~\ref{tab:Homotopies} and Fig.~\ref{fig:AZI-classification}}. After explaining our strategy in Secs.~\ref{sec:AZ+I-classes} to~\ref{sec:charges}, we treated in a greater detail all the symmetry classes supporting doubly charged nodes in $D=3$. This includes doubly charged nodal surfaces in the centrosymmetric extensions of the AZ classes (dubbed the ``AZ+$\mcI$ classes'') $\textrm{D}$ and $\textrm{BDI}$, and doubly charged nodal lines in the extensions of the classes $\textrm{CI}$ and $\textrm{AI}$. Both topological charges characterizing these nodes are protected solely by the inversion symmetry and the global symmetries, making them very stable against a wide range of perturbations. 

The doubly (and in higher spatial dimensions also ``multiply'') charged nodes are often naturally robust in the sense that they can't be gapped out on their own but only by annihilation in pairs, thus reaching the high degree of stability usually associated with Weyl points. In $D$ spatial dimensions, this is always the case for nodes with a non-trivial charge on $(D-1)$-spheres. On the other hand, if the largest $\tilde{p}$-sphere accommodating a non-trivial charge has only $\tilde{p}<(D-1)$ dimensions, the corresponding nodes need to possess $(D-1-\tilde{p})$ dimensions winding around the Brillouin zone torus to become robust in the same sense. This robustness is in a stark contrast with the case of the extensively studied nodes protected by crystalline symmetries (such as rotation axes or mirror planes) which are typically gapped out if the appropriate strain is applied.

We made a connection between the AZ+$\mcI$ symmetry classes and the physical systems in Sec.~\ref{sec:sym-class-rel}. Restating the results for the symmetry classes supporting doubly charged nodes in $D=3$, class $\textrm{D}$ nodal surfaces appear in time-reversal breaking multi-orbital superconductors (as has been recently also found in Ref.~\cite{Agterberg:2017}), robust nodal lines of class $\textrm{CI}$ appear in singlet superconducting phases of nodal line metals, nodal lines of class $\textrm{AI}$ are relevant for time-reversal symmetric semimetals without spin-orbit interaction, and nodal surfaces of class $\textrm{BDI}$ may appear in $\textrm{AI}$-like system respecting a sublattice symmetry. Other realizations are also possible and are listed in Fig.~\ref{fig:AZI-classification}. Although we didn't search for actual materials exhibiting doubly charged nodes, we provided simple and realistic tight-binding models for each of the four cases to support our claims. 

The present work leaves various questions open. For example, we didn't formalize the meaning of the topological charges in Tab.~\ref{tab:Homotopies} appearing on $p$-spheres with $p\geq 3$. Already for $p=1$ and $2$ we encountered an example of a charge that is absent in the tenfold way classification of gapped systems~\cite{Ryu:2010}, namely the homotopy equivalence class of closed paths in the special orthogonal group $\mathsf{SO}(n)$. We also didn't study the signatures of the multiple charges in the transport properties. For example, Ref.~\cite{Agterberg:2017} pointed out the unusual {\color{black} thermal conductivity} and specific heat associated with class $\textrm{D}$ nodal surfaces which are characterized by a $\intg$-valued Chern number. Since these nodal surfaces can be interpreted as inflated double Weyl points, it might be interesting to look for signatures of the chiral anomaly in such systems. As another example, very recent Ref.~\cite{Rui:2017b} showed that a small inversion-symmetry breaking perturbation in class $\textrm{AI}$ nodal line semimetals facilitates an anomalous transverse Hall-like current. Since the topological charges of class $\textrm{CI}$ nodal lines are closely related to those of class $\textrm{AI}$, we expect a related phenomenon to also exist in superconducting nodal line metals. Furthermore, doubly charged class $\textrm{AI}$ nodal lines were very recently predicted to exist in magnon spectra of certain realistic $\mfT$-symmetric antiferromangets~\cite{Li:2017}. The way the higher charges are manifested in the surface spectra has also been left out for future studies.


\section*{Acknowledgements}

The authors would like to acknowledge insightful discussion with D. F. Agterberg, A. Bouhon, M. H. Fischer, S. Moroz, T. Neupert, A. P. Schnyder, A. A. Soluyanov and F. Valach that helped shaping the final form of the manuscript. T.B. was supported by the Gordon and Betty Moore Foundation’s EPiQS Initiative, Grant GBMF4302. Both authors are grateful for the financial support through an ETH research grant and the Swiss National Science Foundation. 

\bibliography{bibliography}{}
\bibliographystyle{apsrev4-1}  
\end{document}